\documentclass[preprint,aps,12pt,showpacs,nofootinbib,tightenlines]{revtex4}
\usepackage{mathrsfs}
\usepackage{amsmath}
\usepackage{amssymb}
\usepackage{epsfig}
\usepackage{epstopdf}
\usepackage{graphicx}
\usepackage{subfigure}
\usepackage{booktabs}
\usepackage{float}
\usepackage{amsmath}
\usepackage{color}
\usepackage{multirow}
\usepackage{bm}

\def\be{\begin{eqnarray}}
\def\en{\end{eqnarray}}
\def\non{\nonumber\\}

\begin{document}
\title{Semileptonic and nonleptonic decays of  $B_{u,d,s,c}^{*}$ in the covariant light-front approach}
\author{Si-Yang Wang$^1$, You-Ya Yang$^2$, Zhi-Jie Sun$^2$, Hao Yang$^2$, Peng Li$^3$\footnote{lipeng@haut.edu.cn}  Zhi-Qing Zhang$^2$\footnote{zhangzhiqing@haut.edu.cn}} 
\affiliation{
 \it \small $^1$ Institute of Particle Physics and Key Laboratory of Quark and Lepton Physics (MOE),\\
  \small \it Central China Normal University, Wuhan, Hubei 430079, China\\
 \it \small $^2$ Institute of Theoretical Physics, School of Sciences, Henan University of Technology,
 Zhengzhou, Henan 450001, China \\
 \it \small $^3$ Institute for Complexity Science, Henan University of Technology,
 Zhengzhou, Henan 450001, China} 
\date{\today}
\begin{abstract}
The semileptonic and nonleptonic decays of the b-flavor vector mesons $B^{*}_{u,d,s}$ and $B_{c}^{*}$ are investigated within the covariant light-front quark model (CLFQM). By calculating the form factors of the transitions $B_{u, d, s, c}^{*}\to P$ under the CLFQM, with $P$ denoting a pseudoscalar meson, i.e., $\pi, K, \eta_c(1S,2S), D_{(s)}, B_{(s)}$, we predict and discuss several physical observables, including the branching ratios, polarization fractions $f_{L}, f_{\|}$, and forward-backward asymmetries $A_{FB}$. The total widths of the single-photon radiative decay channels for these b-flavor vector mesons are estimated using their partial widths. 
 In these considered decays, one can find that the semileptonic decays  $B_{s}^{*0}\to D_{s}^{-}\ell^{\prime+}{\nu}_{\ell^\prime}$ and $B_{c}^{*+}\to B_{s}^{0}\ell^{\prime+}{\nu}_{\ell^\prime}, \eta_{c}\ell^{\prime+}{\nu}_{\ell^\prime}$, with $\ell^\prime$ being $e$ or $\tau$, and the nonleptonic channels $B_{c}^{*+}\to B^0_{s} \pi^{+}, B^0_{s} \rho^{+}$ have the largest branching ratios, which can reach up to the $10^{-7}$ order, and are most likely to be accessible at the future high-luminosity LHCb and Belle-II experiments.
\end{abstract}

 \vspace{1cm}

\maketitle

\section{Introduction}\label{intro} 
The properties of b-flavored pseudoscalar mesons, namely $B_{u,d,s}$ and $B_{c}$, have been extensively studied through weak decays both theoretically and experimentally; they are important to precisely test the Standard Model (SM) and search for new physics (NP). However, the vector partners $B^{*}_{u,d,s}$ and $B^{*}_{c}$ still lack sufficient experimental measurements, owing to their low production rates and detection efficiencies compared to the pseudoscalar $B_{u,d,s}$ and $B_c$ mesons. With the increasing amount of data collected by LHCb and Belle II,  discovering  $B^*_c$ and precisely detecting the properties of $B^{*}_{u,d,s}$ have become possible. Consequently, we can expect these b-flavored vector mesons to provide another platform to study the heavy flavor meson weak decays in the future.

The vector mesons $B^*$ and $B^*_s$ have been found in experiments with masses $m_{B^*}=5324.71$ MeV and $m_{B^*_s}=5415.4$ MeV, respectively. Owing to the small mass difference between $m_{B^*_{(s)}}$
and $m_{B_{(s)}}$, $B^{*}_{(s)}$ cannot decay to the corresponding pseudoscalar partner and a light meson, such as $\pi, K$. This implies that the strong decays are forbidden and the radiative decays $B^*_{(s)}\to B_{(s)}\gamma$ become the dominant channels.  
Owing to the rapid development of heavy flavor experiments \cite{Belle-II:2010dht,R.A73,R.A153}, although $B^{*}_{(s)}$ decays are dominated by the electromagnetic processes $B^{*}_{(s)}\to B_{(s)}\gamma$, Belle-II may observe the branching ratios on the order of $10^{-8}$ \cite{Chang:2015jla,Q.767523} in the future. In addition, owing to the high production rates of $b\bar b$ pairs on $pp$ collisions \cite{RAIJ209.767523}, LHCb  may also provide comprehensive experimental information for the $B^{*}$ decays \cite{Chang:2016cdi}. Some theoretical studies on the $B^{*}_{(s)}$ weak decays in the SM \cite{Chang:2015jla,Grinstein:2015aua,Wang:2012hu,Chang:2016eto,Chang:2016cdi} and NP scenarios \cite{Chang:2018sud} have been performed.
 The $B_{c}^{*}$ meson comprises two different heavy quarks $\bar b$ and $c$, which can decay individually. Therefore, rich weak decay channels can be expected to exist for the $B_{c}^{*}$ meson. However, no experiment has confirmed their existence, and the corresponding theoretical investigations are limited. Similar to that for the $B^*_{(s)}$ mesons, the mass difference between $B_{c}^{*}$ and $B_c$ is much smaller ($\simeq 50$ MeV) than the pion mass. Therefore, the $B_{c}^{*}$ meson only decays through electromagnetic and weak interactions. Furthermore, finding $B^*_c$ at the LHC is possible by promoting the collider energy and luminosity.


The covariant light-front quark model (CLFQM) has some unique advantages compared with other quark-model approaches. First, the light-front wave functions used in this approach are independent of the hadron momentum and thus are manifestly Lorentz invariant. Moreover, the hadron spin is correctly constructed using the so-called Melosh rotation. Second, this model provides a relativistic treatment of the hadron. Since the final state meson at the maximum recoil point ($q^2=0$) or in heavy-to-light transitions can be highly relativistic, considering the relativistic effects is crucial. Consequently, the non-relativistic quark model may not work well. Finally, the spurious contribution in the CLFQM, which depends on the light-front orientation, is canceled by the zero-mode contribution, thereby restoring the covariance of the current matrix elements lost in the previous standard light-front quark model.
The CLFQM has been successfully extended to investigate the transition form factors and hadronic weak decays \cite{sunzj1,sunzj2,sunzj3,Yang}. The remainder of this paper is organized as follows. The formalism of the CLFQM, the hadronic matrix elements, and the helicity amplitudes combined via form factors are listed in Sec. \ref{form1}. In addition to the numerical results for the form factors of the transitions
$B^{*}\to D, \pi$, $B_{s}^{*}\to D_{s}, K$, and $B_{c}^{*}\to D, B, B_{s}, \eta_{c}, \eta_{c}(2S)$, the branching ratios, the forward-backward asymmetries $A_{FB}$, and the polarization fractions $f_{L, {\|}}$ for the corresponding decays are presented in Sec. \ref{numer}. Detailed comparisons with other theoretical values and relevant discussions are also included. Sec. IV provides the study summary. Some specific rules when performing the $p^-$ integration and the expression for each form factor are presented in Appendices A and B, respectively.
\section{Formalism}\label{form1}
\subsection{Form Factors}
The Bauer-Stech-Wirble (BSW) form factors for the transitions $B^* \to P$  are defined as follows:
\begin{footnotesize}
\be
\left\langle P \left(p^{\prime \prime}\right)\left|V_{\mu}-A_{\mu}\right| B^* \left(p^{\prime}, \epsilon\right)\right\rangle
&=&-\epsilon_{\mu \nu \alpha \beta} \epsilon_{B^*}^{\nu} q^{\alpha} p^{\beta} \frac{V\left(q^{2}\right)}{m_{B^*}+m_{P}}-i \frac{2 m_{B^*} \epsilon_{B^*} \cdot q}{q^{2}} q_{\mu} A_{0}\left(q^{2}\right) \non
&&-i \epsilon_{B^*, \mu}\left(m_{B^*}+m_{P}\right) A_{1}\left(q^{2}\right)-i \frac{\epsilon_{B^*} \cdot q}{m_{ B^*}+m_{P}}p_{\mu} A_{2}\left(q^{2}\right) \non
&&+i \frac{2 m_{B^*} \epsilon_{B^*} \cdot q}{q^{2}} q_{\mu} A_{3}\left(q^{2}\right),
\label{bsw}
\en
\end{footnotesize}
where $p=p'+p'', q=p'-p''$ ,and the convention $\epsilon_{0123}=1$ is adopted. In the above equations, $V_{\mu}$ and $A_{\mu}$ are the corresponding vector and axial-vector currents, which are dominant contributions in the weak decays. Here $p'(p'')$ is the four-momentum of the initial (final) meson. 

\begin{figure}[htbp]
	\centering \subfigure{
		\begin{minipage}{7cm}
			\centering
			\includegraphics[width=6cm]{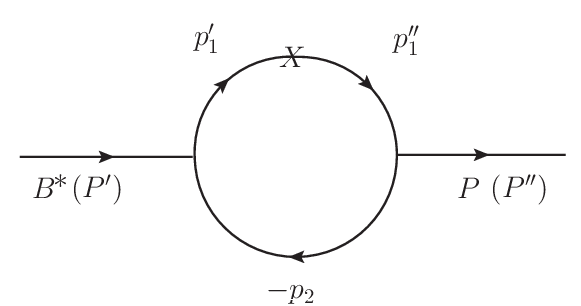}
	\end{minipage}}
	\caption{Feynman diagram for the $B^*\to P$ transitions,with $P$ representing a pseudoscalar meson, i.e., $\pi, K, \eta_c(1S,2S), D_{(s)}, B_{(s)}$, where $P^{\prime(\prime\prime)}$ is the
		$B^*$ ($P$) meson momentum, $p^{\prime(\prime\prime)}_1$
		is the quark momentum, $p_2$ is the antiquark momentum, and X
		denotes the vector or axial vector transition vertex.}
	\label{feyn}
\end{figure}
Following the convention and calculation rules for the form factors of the transition $J/\Psi\to D_{(s)}$ given in our previous work \cite{sunzj2},  
the decay amplitude can be given in the lowest order for the
$B^* \to P$ transitions, whose Feynman diagram is shown in Fig. \ref{feyn},
\be
\mathcal{B}_{\mu}^{B^*P}=-i^{3} \frac{N_{c}}{(2 \pi)^{4}} \int d^{4} p_{1}^{\prime} \frac{h_{B^*}^{\prime}\left(i h_{P}^{\prime\prime }\right)}{N_{1}^{\prime} N_{1}^{\prime \prime} N_{2}}
S_{\mu \nu}^{B^*P} \varepsilon^{*\nu},
\en
where the quark propagators $N^{\prime(\prime\prime)}, N_2$ and the trace $S_{\mu \nu}^{B^*P}$ can be obtained from those given in Ref. \cite{sunzj2} by the replacements $J/\Psi\to B^*, D_{(s)}\to P$. The covariant vertex function $h^\prime_{B^*}$ is defined as
\be
h_{B^*}^{\prime} &=&\left(M^{\prime 2}_{B^*}-M_{0B^*}^{\prime 2}\right) \sqrt{\frac{x_{1} x_{2}}{N_{c}}} \frac{1}{\sqrt{2} \widetilde{M}_{0}^{\prime}} \varphi^{\prime},
\en
where $M'_{0B^*}$ is the kinetic invariant mass of the initial $B^*$ meson and can be expressed as the energies
 $e^{(\prime)}_i$ of the constituent quark and anti-quark, with the masses (momentum fractions) being $m^\prime_1(x_1)$ and $m_2(x_2)$, respectively. Their definitions, including 
 the denominator $\widetilde{M}_{0}^{\prime}$, are given as follows:
\be
M_{0}^{\prime 2} &=&\left(e_{1}^{\prime}+e_{2}\right)^{2}=\frac{p_{\perp}^{\prime 2}+m_{1}^{\prime 2}}{x_{1}}
+\frac{p_{\perp}^{2}+m_{2}^{2}}{x_{2}}, \quad \widetilde{M}_{0}^{\prime}=\sqrt{M_{0}^{\prime 2}-\left(m_{1}^{\prime}-m_{2}\right)^{2}},\non
e_{i}^{(\prime)} &=&\sqrt{m_{i}^{(\prime) 2}+p_{\perp}^{\prime 2}+p_{z}^{\prime 2}} \;\;(i=1,2), x_1+x_2=1,
\en
where $p_{z}^{\prime}=\frac{x_{2} M_{0}^{\prime}}{2}-\frac{m_{2}^{2}+p_{\perp}^{\prime 2}}{2 x_{2} M_{0}^{\prime}}$.
The light-front wave functions (LFWFs) can be derived in principle by solving the relativistic Schr$\ddot{o}$dinger equation; however, obtaining their exact solutions is challenging in many cases. Here, the harmonic oscillator wave function with an exponential term is used, and its expression for the S-wave mesons is written as
\be
\varphi^{\prime} &=&\varphi^{\prime}\left(x_{2}, p_{\perp}^{\prime}\right)=4\left(\frac{\pi}{\beta^{\prime 2}}\right)^{\frac{3}{4}}
\sqrt{\frac{d p_{z}^{\prime}}{d x_{2}}} \exp \left(-\frac{p_{z}^{\prime 2}+p_{\perp}^{\prime 2}}{2 \beta^{\prime 2}}\right),
\en
where $\beta^\prime$ is a phenomenological parameter describing the momentum distribution and can be fixed by fitting the corresponding decay constant. For example, the shape parameter $\beta^\prime_{B^*}$ can be obtained through
\be
f_{B^*}&=&\frac{N_{c}}{4 \pi^{3} M^{\prime}_{B^*}} \int d x_{2} d^{2} p_{\perp}^{\prime} \frac{h_{B^*}^{\prime}}{(1-x_{2}) x_{2}\left(M^{\prime 2}_{B^*}-M_{0B^*}^{\prime 2}\right)}\non &&
\times\left[(1-x_{2}) M_{0B^*}^{\prime 2}-m_{b}^{\prime}\left(m_{b}^{\prime}-m_{q}\right)-p_{\perp}^{\prime 2}+\frac{m_{b}^{\prime}+m_{q}}{w_{B^*}^{\prime}} p_{\perp}^{\prime 2}\right],
\en
where $w_{B^*}^{\prime}=M_{0B^*}^{\prime}+m_{b}^{\prime}+m_{q}$, $M^{\prime}_{B^*}$ and $m_b^\prime (m_q)$ are the $B^*$ meson and $b, (q=u,d)$ quark masses, respectively. The expressions of the vertex functions $h^{\prime\prime}_P$ for our considered pseudoscalar mesons are similar.
After expanding the trace $S_{\mu \nu}^{B^*P}$ using the Lortentz contraction, the form factors $V(q^2), A_0(q^2), A_1(q^2)$, and $A_0(q^2)$ can be obtained by matching their coefficients given in Eq. (\ref{bsw}). Their specific expressions are listed in Appendix B.

The differential decay widths of the semileptonic $B^*$ decays can be obtained by the combinations of the helicity amplitudes via the form factors, which are listed as follows:
 \begin{footnotesize}
\begin{eqnarray}
 \frac{d\Gamma_L(B^*\to P\ell\nu_\ell)}{dq^2}&=&(\frac{q^2-m_\ell^2}{q^2})^2\frac{ {\sqrt{\lambda(m_{B^*}^2,m_{P}^2,q^2)}} G_F^2 |V_{CKM}|^2} {384m_{B^*}^3\pi^3}
 \times \frac{1}{q^2} \left\{ 3 m_\ell^2 \lambda(m_{B^*}^2,m_{P}^2,q^2) A_0^2(q^2)\right.\nonumber\\
 && +\frac{m_\ell^2+2q^2}{4m^2_{P}}\left.\left|
 (m_{B^*}^2-m_{P}^2-q^2)(m_{B^*}+m_{P})A_1(q^2)-\frac{\lambda(m_{B^*}^2,m_{P}^2,q^2)}{m_{B^*}+m_{P}}A_2(q^2)\right|^2
 \right\},\label{eq:decaywidthlon}\;\;\non\\
\frac{d\Gamma_\pm(B^*\to P \ell\nu_\ell)}{dq^2}&=&(\frac{q^2-m_\ell^2}{q^2})^2\frac{ {\sqrt{\lambda(m_{B^*}^2,m_{P}^2,q^2)}} G_F^2 |V_{CKM}|^2} {384m_{B^*}^3\pi^3}
  \nonumber\\
 &&\;\;\times \left\{ (m_\ell^2+2q^2) \lambda(m_{B^*}^2,m_{P}^2,q^2)\left|\frac{V(q^2)}{m_{B^*}+m_{P}}\mp
 \frac{(m_{B^*}+m_{P})A_1(q^2)}{\sqrt{\lambda(m_{B^*}^2,m_{P}^2,q^2)}}\right|^2
 \right\},\label{eq:widthlon2}
\end{eqnarray}
\end{footnotesize}
where $\lambda(q^2)=\lambda(m^{2}_{B^*},m^{2}_{P},q^{2})=(m^{2}_{B^*}+m^{2}_{P}-q^{2})^{2}-4m^{2}_{B^*}m^{2}_{P}$ and $m_{\ell}$ is the mass of the lepton $\ell$, with $\ell=e,\mu, \tau$ \footnote{From now on, we use $\ell$ to represent $e,\mu, \tau$  for simplicity. }. Notably, despite the electron and nuon being very light, we do not ignore their masses in our calculations so as to observe the mass effects.
The combined transverse and total differential decay widths are defined as
\be
\frac{d \Gamma_{T}}{d q^{2}}=\frac{d \Gamma_{+}}{d q^{2}}+\frac{d \Gamma_{-}}{d q^{2}}, \quad \frac{d \Gamma}{d q^{2}}=\frac{d \Gamma_{L}}{d q^{2}}+\frac{d \Gamma_{T}}{d q^{2}}.
\en

For the $B^*$ decays, defining the polarization fraction is important owing to the existence of different polarizations
\be
f_{L}=\frac{\Gamma_{L}}{\Gamma_{L}+\Gamma_{+}+\Gamma_{-}}. \label{eq:fl}
\en
The analytical expression of the forward-backward asymmetry is defined as \cite{Sakaki:2013bfa}
\be
A_{FB} = \frac{\int^1_0 {d\Gamma \over dcos\theta} dcos\theta - \int^0_{-1} {d\Gamma \over dcos\theta} dcos\theta}
{\int^1_{-1} {d\Gamma \over dcos\theta} dcos\theta} = \frac{\int b_\theta(q^2) dq^2}{\Gamma_{B^*}},\label{eq:AFB}
\en
where $\theta$ is the angle between the three-momenta of the lepton $\ell$ and the initial meson in the $\ell\nu_\ell$ rest frame. The function $b_{\theta}(q^2)$ represents the angular coefficient, which can be written as \cite{Sakaki:2013bfa}
\be
b_\theta(q^2) &=& {G_F^2 |V_{CKM}|^2 \over 128\pi^3 m_{B^*}^3} q^2 \sqrt{\lambda(q^2)}
\left( 1 - {m_\ell^2 \over q^2} \right)^2 \left[ {1 \over 2}(H_{V,+}^2-H_{V,-}^2)+ {m_\ell^2 \over q^2} ( H_{V,0}H_{V,t} ) \right],
\label{eq:btheta2}
\en
where the helicity amplitudes
\be
H_{V,\pm}\left(q^{2}\right)&=&\left(m_{B^*}+{m_{P}}\right) A_{1}\left(q^{2}\right) \mp \frac{\sqrt{\lambda\left(q^{2}\right)}}{m_{B^*}+m_{P}} V\left(q^{2}\right), \non
H_{V,0}\left(q^{2}\right)&=&\frac{m_{B^*}+m_{P}}{2 m_{B^*} \sqrt{q^{2}}}\left[-\left(m_{B^*}^{2}-m_{P}^{2}-q^{2}\right) A_{1}\left(q^{2}\right)+\frac{\lambda\left(q^{2}\right) A_{2}\left(q^{2}\right)}{\left(m_{B^*}+m_{P}\right)^{2}}\right],\non
H_{V,t}\left(q^{2}\right)&=&-\sqrt{\frac{\lambda\left(q^{2}\right)}{q^{2}}} A_{0}\left(q^{2}\right),
\en
with the subscript $V$ in each helicity amplitude referring  to the $\gamma_\mu(1-\gamma_5)$ current.

Based on the effective Hamiltonian,
the amplitudes of the decays $B^{*} \rightarrow P M_{1} $ with $M_{1}= \pi, K, D_{(s)}$ can be expressed as
\be
A( B^{*} \rightarrow P M_{1})=\left\langle P M_{1}\left|\mathcal{H}_{e f f}\right| B^{*}\right\rangle\approx\left\langle M_{1}\left|J^{\mu}\right| 0\right\rangle\left\langle P\left|J_{\mu}\right| B^{*}\right\rangle,
\en 
Notably,the second equality is an approximation. Here, we only consider the dominant 
contributions from the tree-emission diagrams. In some cases, W-exchange and W-annihilation diagrams might give sizeable deviations to the results.
The amplitudes of the decays $B^{*} \to \pi D_{(s)}$ are given as
\be
\mathcal{A}\left( B^{*+} \rightarrow \pi^{+} \bar{D}^{0}\right) &=& \sqrt{2}G_{F} V_{c b} V_{u d}^{*}m_{B^{*}}
\left(\epsilon \cdot p_{D}\right) (a_{2}f_{D} A_{0}^{ B^{*} \pi}+a_1f_\pi A_{0}^{ B^{*} D}),\\
\mathcal{A}\left( B^{*0} \rightarrow \pi^{-} D^{+}\right) &=& \sqrt{2}G_{F} V_{u b} V_{c d}^{*} a_{1} m_{B^{*}}\left(\epsilon \cdot p_{D}\right) f_{D} A_{0}^{B^{*} \pi},\\
\mathcal{A}\left( B^{*0} \rightarrow \pi^{-} D_{s}^{+}\right) &=& \sqrt{2}G_{F} V_{u b} V_{c s}^{*} a_{1} m_{B^{*}}\left(\epsilon \cdot p_{D_{s}}\right) f_{D_{s}} A_{0}^{ B^{*} \pi},\\
\mathcal{A}\left( B^{*0} \rightarrow  D^{-}\pi^{+}\right) &=& \sqrt{2}G_{F} V_{c b} V_{u d}^{*} a_{1} m_{B^*}\left(\epsilon \cdot p_{\pi}\right) f_{\pi} A_{0}^{ B^{*} D},
\en
where $\epsilon$ is the polarization four vector of the $B^*$ meson. The combinations of the Wilson coefficients are $a_1=C_2+C_1/3$ and $a_2=C_1+C_2/3$.
Similarly, the amplitudes of the decays $ B^{*0}_{(s)} \to D_{(s)} K$ are listed as
\be
\mathcal{A}\left( B^{*0} \rightarrow D^{-} K^{+} \right) &=& \sqrt{2}G_{F} V_{c b} V_{u s}^{*} a_{1} m_{B^{*}}\left(\epsilon \cdot p_{K}\right) f_{K} A_{0}^{ B^{*} D},\\
\mathcal{A}\left( B_{s}^{*0} \rightarrow D_{s}^{-}K^{+} \right) &=& \sqrt{2}G_{F} V_{c b} V_{u s}^{*} a_{1} m_{B_{s}^{*}}\left(\epsilon \cdot p_{K}\right) f_{K} A_{0}^{ B_{s}^{*} D_{s}},\\
\mathcal{A}\left( B_{s}^{*0} \rightarrow K^{-} D^{+}\right) &=& \sqrt{2}G_{F} V_{u b} V_{c d}^{*} a_{1} m_{B_{s}^{*}}\left(\epsilon \cdot p_{D}\right) f_{D} A_{0}^{ B^{*}_s K},\\
\mathcal{A}\left( B_{s}^{*0} \rightarrow K^{-} D_{s}^{+}\right) &=& \sqrt{2}G_{F} V_{u b} V_{c s}^{*} a_{1} m_{B_{s}^{*}}\left(\epsilon \cdot p_{D_{s}}\right) f_{D_{s}} A_{0}^{ B^{*}_s K}.
\en

The amplitudes of the decays $B_{c}^{*+} \to B_{(s)}^{0}\pi(K), \eta_{c}(1S,2S)\pi(K)$ are written as
\be
\mathcal{A}\left( B^{*+}_{c} \rightarrow B_{s}^{0} \pi^{+}\right) &=& \sqrt{2}G_{F} V_{u d} V_{c s}^{*} a_{1} m_{B^{*}_{c}}\left(\epsilon_{B^{*}_{c}} \cdot p_{B_{s}}\right) f_{\pi} A_{0}^{ B^{*}_{c} B_{s}},\\
\mathcal{A}\left( B^{*+}_{c} \rightarrow B_{s}^{0} K^{+}\right) &=& \sqrt{2}G_{F} V_{u s} V_{c s}^{*} a_{1} m_{B^{*}_{c}}\left(\epsilon_{B^{*}_{c}} \cdot p_{B_{s}}\right) f_{K} A_{0}^{ B^{*}_{c} B_{s}},\\
\mathcal{A}\left( B^{*+}_{c} \rightarrow B^{0} \pi^{+}\right) &=& \sqrt{2}G_{F} V_{u d} V_{c d}^{*} a_{1} m_{B^{*}_{c}}\left(\epsilon_{B^{*}_{c}} \cdot p_{B^{0}}\right) f_{\pi} A_{0}^{ B^{*}_{c} B^{0}},\\
\mathcal{A}\left( B^{*+}_{c} \rightarrow B^{0} K^{+}\right) &=& \sqrt{2}G_{F} V_{u s} V_{c d}^{*} a_{1} m_{B^{*}_{c}}\left(\epsilon_{B^{*}_{c}} \cdot p_{B^{0}}\right) f_{K} A_{0}^{ B^{*}_{c} B^{0}},\\
\mathcal{A}\left( B^{*+}_{c} \rightarrow \eta_{c} \pi^{+}\right) &=& \sqrt{2}G_{F} V_{u d} V_{c b}^{*} a_{1} m_{B^{*}_{c}}\left(\epsilon_{B^{*}_{c}} \cdot p_{\eta_{c}}\right) f_{\pi} A_{0}^{ B^{*}_{c} \eta_{c}},\\
\mathcal{A}\left( B^{*+}_{c} \rightarrow \eta_{c} K^{+}\right) &=& \sqrt{2}G_{F} V_{u s} V_{c b}^{*} a_{1} m_{B^{*}_{c}}\left(\epsilon_{B^{*}_{c}} \cdot p_{\eta_{c}}\right) f_{K} A_{0}^{ B^{*}_{c} \eta_{c}},\\
\mathcal{A}\left( B^{*+}_{c} \rightarrow \eta_{c}(2S) \pi^{+}\right) &=& \sqrt{2}G_{F} V_{u d} V_{c b}^{*} a_{1} m_{B^{*}_{c}}\left(\epsilon_{B^{*}_{c}} \cdot p_{\eta_{c}(2S)}\right) f_{\pi} A_{0}^{ B^{*}_{c} \eta_{c}(2S)},\\
\mathcal{A}\left( B^{*+}_{c} \rightarrow \eta_{c}(2S) K^{+}\right) &=& \sqrt{2}G_{F} V_{u s} V_{c b}^{*} a_{1} m_{B^{*}_{c}}\left(\epsilon_{B^{*}_{c}} \cdot p_{\eta_{c}(2S)}\right) f_{K} A_{0}^{ B^{*}_{c} \eta_{c}(2S)}.
\en

The hadronic matrix elements for the decays $B^{*}\rightarrow P M_{2}$, with $M_2$ being $\rho, K^*$, can be expressed as
\be
\mathcal{A}\left( B^{*} \rightarrow P M_{2} \right)=\left\langle P M_{2}\left|\mathcal{H}_{\mathrm{eff}}\right|  B^* \right\rangle=\frac{G_{F}}{\sqrt{2}} V_{cq}^{*} V_{ud(s)} a_{1} H_{\lambda},
\en
where $\lambda=0, \pm$ denotes the helicity of vector meson, $q$ stands for $s,d$, and $b$ quarks corresponding to the
transitions $B^*_c\to B, B^*_c\to B_s$, and $B_c\to \eta_c(1S,2S), B^*_{(s)}\to D_{(s)}$, respectively. $\mathcal{H}_{\lambda}=\left\langle M_{2}\left|J^{\mu}\right| 0\right\rangle\left\langle P \left|J_{\mu}\right|  B^{*}\right\rangle$  is given as follows:
\be
H_{0} &\equiv&  \left\langle M_2 \left(\varepsilon_{0}^{\prime}, p_{M_2}\right)\left|\bar{u} \gamma^{\mu} d(s)\right| 0\right\rangle\left\langle P\left(p_{P}\right)\left|\bar{c} \gamma_{\mu}\left(1-\gamma_{5}\right) q\right|  B^{*} \left(\varepsilon, p_{ B^{*}}\right)\right\rangle \non
&=&\frac{i f_{M_2}}{2 m_{ B^{*}}}\left[\left(m_{ B^{*}}^{2}-m_{P}^{2}+m_{M_2}^{2}\right)\left(m_{ B^{*}}+m_{P}\right) A_{1}^{ B^{*} P}\left(m_{M_2}^{2}\right)\right. \non&&
 \left.+\frac{4 m_{ B^{*}}^{2} p_{c}^{2}}{m_{ B^{*}}+m_{P}} A_{2}^{ B^* P}\left(m_{M_2}^{2}\right)\right], \\
H_{\pm} &\equiv& \left\langle M_2\left(\varepsilon_{\pm}^{\prime}, p_{M_2}\right)\left|\bar{u} \gamma^{\mu} d(s)\right| 0\right\rangle\left\langle P\left(p_{P}\right)\left|\bar{c} \gamma_{\mu}\left(1-\gamma_{5}\right) q\right|  B^{*}\left(\varepsilon_{\pm}, p_{ B^{*}}\right)\right\rangle \non
&=&i f_{M_2} m_{M_2}\left[-\left(m_{B^{*}}+m_{P}\right) A_{1}^{B^{*}  P}\left(m_{M_2}^{2}\right) \pm \frac{2 m_{B^{*}} p_{c}}{m_{B^{*}}+m_{P}} V^{ B^{*} P}\left(m_{M_2}^{2}\right)\right].
\en
The polarization and branching fraction, the polarization fractions are defined as
\be
f_{L,\parallel,\perp}=\frac{H_{0,\parallel,\perp}}{H_{0}+H_{\parallel}+H_{\perp}}, \label{eq:fl}
\en
where $H_{\parallel}$ and $H_{\perp}$ are parallel and perpendicular amplitudes, respectively, and can be obtained through $H_{\parallel,\perp}=\frac{(H_{-}\pm H_{+})}{\sqrt{2}}$.
\section{Numerical results and discussions} \label{numer}
\subsection{Transition Form Factors}
\begin{tiny}
\begin{table}[H]
\caption{Values of input parameters \cite{sunzj1,Workman,Ebert:2002pp,damir,chiu,Wingate}. }
\label{tab:constant}
\begin{tabular*}{16.5cm}{@{\extracolsep{\fill}}l|cccccc}
  \hline\hline
\textbf{Mass(\text{GeV})} &$m_{b}=4.8$
&$m_{c}=1.4$&$m_{s}=0.37$&$m_{u,d}=0.25$&$m_{e}=0.000511$   \\[1ex]
&$m_{\pi}=0.140$&$m_{K}=0.494$&$m_{\rho}=0.775$&$m_{K^{*}}=0.892$& $m_{\mu}=0.106$\\[1ex]
& $m_{\eta_c}=2.9839$& $m_{\eta_c(2S)}=3.6377 $  &$ m_{D}=1.86966$ & $m_{D_{s}}=1.96835 $& $m_{\tau}=1.78 $ \\[1ex]
& $m_{B}=5.27965 $ & $m_{B_{s}}=5.36688 $& $m_{B^{*}}=5.32470 $& $m_{B^{*}_{s}}=5.4154 $& $m_{B^{*}_{c}}=6.332 $\\[1ex]
\hline
\end{tabular*}
\begin{tabular*}{16.5cm}{@{\extracolsep{\fill}}l|ccccc}
  \hline
\multirow{2}{*}{{\textbf{CKM}}}&$V_{ub}=(0.00382\pm0.0002)$&$V_{cb}=(0.0408\pm0.0014)$\\[1ex]
& $V_{cd}=(0.221\pm0.004)$&$V_{us}=(0.2243\pm0.0008)$ \\[1ex]
& $V_{ud}=(0.97373\pm0.00031)$&$V_{cs}=(0.975\pm0.006)$ \\[1ex]
\hline
\end{tabular*}
\begin{tabular*}{16.5cm}{@{\extracolsep{\fill}}l|ccccc}
\hline
\textbf{ Decay constants(\text{GeV})} & $f_{\pi}=0.130\pm0.002$ & $f_{K}=0.16\pm0.004$
& $f_{\rho}=0.209\pm0.002$\\[1ex]  &$f_{K^{*}}=0.217\pm0.005$
& $f_{\eta_c}=0.335\pm0.075$& $f_{\eta_c(2S)}=0.243^{+0.079}_{-0.111}$\\[1ex]
& $f_{D}=0.235\pm0.016$  &$f_{D_s}=0.290\pm0.046$
& $f_{B^{*}}=0.210\pm0.020$\\[1ex]  &$f_{B^{*}_{s}}=0.260\pm0.020$&$f_{B^{*}_{c}}=0.5355\pm0.0578$&$f_{B}=0.190\pm0.020$\\[1ex]
&$f_{B_{s}}=0.230\pm0.030$\\[1ex]
\hline\hline
\end{tabular*}
\begin{tabular*}{16.5cm}{@{\extracolsep{\fill}}l|ccccc}
\textbf{Shape parameters(\text{GeV})}&$\beta^{'}_{\eta_{c}}=0.814^{+0.092}_{-0.086}$&$\beta^{'}_{\eta_{c}(2S)}=0.488^{+0.140}_{-0.187}$&$\beta^{'}_{B^{*}_{c}}=0.954^{+0.065}_{-0.069}$\\[1ex]
& $\beta^{'}_{B^{*}}=0.528^{+0.033}_{-0.034}$&$\beta^{'}_{B^{*}_{s}}=0.599^{+0.033}_{-0.032}$&$\beta^{'}_{B}=0.555^{+0.048}_{-0.048}$\\[1ex]
& $\beta^{'}_{D}=0.464^{+0.011}_{-0.014}$&$\beta^{'}_{D_{s}}=0.497^{+0.032}_{-0.028}$\\[1ex]
\hline\hline
\end{tabular*}
\end{table}
\end{tiny}
The input parameters, including the masses of the initial and final mesons, the CKM matrix elements, and the shape parameters fitted by the decay constants are listed in Table \ref{tab:constant}.

Although the $B^{*}$ and $B_{s}^{*}$ mesons have been experimentally discovered \cite{Workman}, experimental and theoretical information on their decay widths is still limited. The total decay width is essential for evaluating the branching ratio. Given the prohibition of the strong decays in phase space, the total decay widths of these vector b-flavored mesons can be estimated by the partial widths of the single-photon decays. This study adopts the result of Refs. \cite{Priyadarsini:2016tiu,Patnaik:2017cbl} for the decay width $\Gamma(B^{*+}\to B^+\gamma)$ and considers the most recent results \cite{Issadykov:2023pwl} for the $\Gamma(B^{*0}_{s}\to B^0_s\gamma)$ and $\Gamma(B^{*0}\to B^0\gamma)$ to evaluate the branching fractions,
\be
\Gamma_{tot}(B^{*+})&\simeq& \Gamma(B^{*+}\to B^{+}\gamma)=(0.577\pm0.120)\; \mathrm{keV},\\
\Gamma_{tot}(B^{*0})&\simeq&\Gamma(B^{*0}\to B^{0}\gamma)=(0.117\pm0.022)\; \mathrm{keV},\\
\Gamma_{tot}(B^{*0}_{s})&\simeq& \Gamma(B^{*0}_{s}\to B^{0}_{s}\gamma)=(0.094\pm0.018)\; \mathrm{keV}.
\en
An obvious difference can be observed between the charged and neutral $B^*$ mesons, whose weak decays will be calculated later. The difference will 
induce a significant discrepancy in the branching ratios between their decay channels. Several other theoretical predictions on the single-photon decay widths can be found in Table \ref{tab:3}.

For the $B^{*}_{c}$ meson, which has not been experimentally discovered, we take the theoretical value $m_{B^{*+}_{c}}=6.332$ GeV predicted by the relativistic quark model \cite{Ebert:2002pp}. Predictions of its single-photon decay width have been given in many different theoretical models, which are also listed in Table \ref{tab:3}. This study uses an intermediate value \cite{Fulcher:1998ka},
\be
\Gamma_{tot}(B^{*+}_{c})&\simeq&\Gamma(B^{*+}_{c}\to B^{+}_{c}\gamma)=(0.059\pm0.012)\; \mathrm{keV}.
\en
\begin{table}[H]
\caption{Widths of radiative decays of $B^{*}_{u,d,s,c}$ mesons in units of keV.
}\label{tab:3}
	\begin{center}
\scalebox{0.75}{
\begin{tabular}{ccccccccccccc}
\hline\hline
References&\cite{Issadykov:2023pwl}&\cite{Ebert:2002pp,Ebert:2002xz}&\cite{Simonis:2018rld}&\cite{Jena:2002is}&\cite{Chang:2020xvu}&\cite{Priyadarsini:2016tiu,Patnaik:2017cbl}&\cite{Lahde:1999ih,Lahde:2002wj}&\cite{Choi:2007se,Choi:2009ai}&\cite{Eichten:1994gt}&\cite{Gershtein:1994dxw}&\cite{Nobes:2000pm}\\
\hline
$\Gamma(B^{*+}\to B^{+}\gamma)$&$0.362\pm0.072$&$0.19$&$0.520$&$0.52$&$0.349\pm0.018$&$0.577$&$0.0674$&$0.4$&$-$&$-$&$-$\\
\hline
$\Gamma(B^{*0}\to B^{0}\gamma)$&$0.117\pm0.022$&$0.070$&$0.165$&$0.14$&$0.116\pm0.006$&$0.181$&$0.0096$&$0.13$&$-$&$-$&$-$\\
\hline
$\Gamma(B^{*0}_{s}\to B^{0}_{s}\gamma)$&$0.094\pm0.018$&$0.054$&$0.115$&$0.06$&$0.084^{+0.011}_{-0.009}$&$0.119$&$0.148$&$0.068$&$-$&$-$&$-$\\
\hline
$\Gamma(B^{*+}_{c}\to B^{+}_{c}\gamma)$&$0.045\pm0.009$&$0.033$&$0.039$&$0.030$&$0.049^{+0.028}_{-0.021}$&$0.023$&$0.034$&$0.022$&$0.135$&$0.060$&$0.050$\\
 \hline\hline
\end{tabular}
}
\end{center}
\end{table}

The form factor, as a nonperturbative hadronic parameter, is essential for evaluating the hadronic matrix element. We adopt the CLFQM \cite{41,44,60,Y. Cheng} to calculate the form factor values.  All computations are performed within the $q^+=0$ reference frame, where the form factors can only be obtained at spacelike momentum transfers $q^2=-q^2_{\bot}\leq0$. Determining the form factors in the timelike region for the physical decay processes is necessary. Hence, in this study, we use the
following double-pole approximation to parametrize the form factors obtained in the spacelike region and then extend to the timelike region,
\be
F\left(q^{2}\right)=\frac{F(0)}{1-a q^{2} / m^{2}+b q^{4} / m^{4}},
\en
where $m$ represents the initial meson mass and $F(q^{2})$ denotes the different form factors, such as $V,A_{0},A_{1}$, and $A_{2}$.
The values of $a$ and $b$ can be obtained by performing a three-parameter fit to the form factors in the range $-(m_{B^*}-m_P)^2\; \text{GeV}^2\leq q^2\leq0$, with
the subscript $P$ representing the final states, such as $\pi$, $K$, $B_{(s)}$, $D_{(s)}$, and $\eta_{c}(1S,2S)$. The results are presented in Tables \ref{tab:1}-\ref{tab:2}, with uncertainties arising from the decay constants of the initial and final mesons.

\begin{table}[H]
\caption{Form factors of the transitions $B^{*}\to D,\pi, B_{s}^{*}\to D_{s},K$ in the
CLFQM. The uncertainties are from the decay constants of the initial and final mesons, respectively.
}\label{tab:1}
\begin{center}
	\scalebox{1.1}{
\begin{tabular}{|cccccc|}
\hline\hline
 & $F$ & $F(0)$ &$F(q^2_{\rm {max}})$ & $a$ & $b$ \\
 \hline
 & $V^{B^{*} D}$ & $0.75^{+0.00+0.00}_{-0.00-0.01}$  & $0.79^{+0.08+0.16}_{-0.07-0.14}$   &  $0.70^{+0.30+0.25}_{-0.27-0.25}$ & $1.39^{+0.07+0.07}_{-0.06-0.07}$
 \\
 & $A_0^{B^{*} D}$ & $0.63^{+0.00+0.00}_{-0.00-0.00}$  & $0.69^{+0.06+0.09}_{-0.05-0.08}$   & $0.43^{+0.30+0.27}_{-0.27-0.27}$  & $0.59^{+0.06+0.08}_{-0.07-0.06}$
 \\
 & $A_1^{B^{*} D}$ & $0.66^{+0.00+0.00}_{-0.01-0.00}$  & $0.72^{+0.02+0.05}_{-0.01-0.04}$   & $0.35^{+0.28+0.24}_{-0.25-0.23}$  & $0.43^{+0.06+0.06}_{-0.04-0.05}$
 \\
 & $A_2^{B^{*} D}$ & $0.57^{+0.00+0.00}_{-0.00-0.00}$  & $0.61^{+0.01+0.07}_{-0.01-0.06}$   & $0.62^{+0.26+0.23}_{-0.24-0.23}$  & $1.16^{+0.04+0.06}_{-0.04-0.05}$\\
 \hline
 & $V^{B^{*} \pi}$ & $0.29^{+0.00+0.00}_{-0.00-0.00}$  & $0.29^{+0.07+0.16}_{-0.07-0.14}$   &  $1.21^{+0.23+0.21}_{-0.21-0.21}$ & $1.27^{+0.04+0.06}_{-0.04-0.06}$
 \\
 & $A_0^{B^{*} \pi}$ & $0.25^{+0.00+0.00}_{-0.00-0.00}$  & $0.43^{+0.06+0.08}_{-0.05-0.08}$   & $0.78^{+0.23+0.22}_{-0.21-0.23}$  & $0.36^{+0.06+0.06}_{-0.04-0.06}$\\
 & $A_1^{B^{*} \pi}$ & $0.30^{+0.00+0.01}_{-0.00-0.00}$  & $0.46^{+0.01+0.04}_{-0.01-0.04}$   & $0.58^{+0.21+0.19}_{-0.19-0.19}$  & $0.22^{+0.03+0.04}_{-0.03-0.04}$
 \\
 & $A_2^{B^{*} \pi}$ & $0.21^{+0.00+0.00}_{-0.00-0.00}$  & $0.23^{+0.01+0.06}_{-0.02-0.06}$   & $1.04^{+0.19+0.18}_{-0.18-0.18}$  & $0.97^{+0.03+0.04}_{-0.04-0.04}$\\
 \hline
 & $V^{B^{*}_{s} D_{s}}$ & $0.76^{+0.01+0.01}_{-0.01-0.01}$  & $0.78^{+0.03+0.09}_{-0.04-0.08}$   &  $0.74^{+0.13+0.10}_{-0.13-0.10}$ & $1.62^{+0.02+0.03}_{-0.03-0.03}$
 \\
 & $A_0^{B^{*}_{s} D_{s}}$ & $0.63^{+0.00+0.01}_{-0.00-0.01}$  & $0.67^{+0.02+0.07}_{-0.02-0.07}$   & $0.47^{+0.13+0.11}_{-0.13-0.11}$  & $0.72^{+0.02+0.02}_{-0.03-0.02}$\\
 & $A_1^{B^{*}_{s} D_{s}}$ & $0.66^{+0.00+0.01}_{-0.00-0.01}$  & $0.70^{+0.00+0.05}_{-0.01-0.05}$   & $0.39^{+0.12+0.12}_{-0.12-0.12}$  & $0.56^{+0.01+0.02}_{-0.02-0.02}$
 \\
 &$A_2^{B^{*}_{s} D_{s}}$ & $0.56^{+0.00+0.01}_{-0.00-0.00}$  & $0.59^{+0.02+0.05}_{-0.03-0.04}$   & $0.66^{+0.11+0.11}_{-0.10-0.11}$  & $1.36^{+0.01+0.02}_{-0.01-0.02}$\\
  \hline
 & $V^{B^{*}_{s} K}$ & $0.34^{+0.00+0.01}_{-0.00-0.01}$  & $0.36^{+0.03+0.09}_{-0.04-0.08}$   &  $1.23^{+0.13+0.10}_{-0.13-0.10}$ & $1.43^{+0.02+0.03}_{-0.03-0.03}$
 \\
 & $A_0^{B^{*}_{s} K}$ & $0.27^{+0.00+0.01}_{-0.00-0.01}$  & $0.42^{+0.02+0.07}_{-0.02-0.07}$   & $0.77^{+0.13+0.11}_{-0.13-0.11}$  & $0.42^{+0.02+0.02}_{-0.03-0.02}$\\
 & $A_1^{B^{*}_{s} K}$ & $0.31^{+0.00+0.01}_{-0.00-0.01}$  & $0.45^{+0.00+0.05}_{-0.01-0.05}$   & $0.60^{+0.12+0.12}_{-0.12-0.12}$  & $0.28^{+0.01+0.02}_{-0.02-0.02}$
 \\
 &$A_2^{B^{*}_{s} K}$ & $0.22^{+0.00+0.00}_{-0.00-0.01}$  & $0.26^{+0.02+0.05}_{-0.03-0.04}$   & $1.02^{+0.11+0.11}_{-0.10-0.11}$  & $1.02^{+0.01+0.02}_{-0.01-0.02}$\\
 \hline\hline
\end{tabular}
}
\end{center}
\end{table}

\begin{table}[H]
\caption{Form factors of the transitions $B^{*}_c\to D, B_{(s)},\eta_c(1S, 2S)$  in the
CLFQM. The uncertainties are the same as those listed in Table \ref{tab:1}.
}\label{tab:2}
\begin{center}
	\scalebox{1.1}{
\begin{tabular}{|cccccc|}
\hline\hline
 & $F$ & $F(0)$ &$F(q^2_{\rm {max}})$ & $a$ & $b$ \\
 \hline
 & $V^{B_{c}^{*} D}$ & $0.26^{+0.01+0.01}_{-0.01-0.02}$  & $0.42^{+0.08+0.16}_{-0.07-0.14}$   &  $1.58^{+0.30+0.25}_{-0.27-0.25}$ & $1.65^{+0.07+0.07}_{-0.06-0.07}$
 \\
 & $A_0^{B_{c}^{*} D}$ & $0.15^{+0.01+0.01}_{-0.01-0.01}$  & $0.24^{+0.06+0.09}_{-0.05-0.08}$   & $1.18^{+0.30+0.27}_{-0.27-0.27}$  & $0.82^{+0.06+0.08}_{-0.07-0.06}$
 \\
 & $A_1^{B_{c}^{*} D}$ & $0.16^{+0.01+0.01}_{-0.01-0.01}$  & $0.26^{+0.02+0.05}_{-0.01-0.04}$   & $1.10^{+0.28+0.24}_{-0.25-0.23}$  & $0.70^{+0.06+0.06}_{-0.04-0.05}$
 \\
 & $A_2^{B_{c}^{*} D}$ & $0.13^{+0.01+0.00}_{-0.01-0.01}$  & $0.20^{+0.01+0.07}_{-0.01-0.06}$   & $1.33^{+0.26+0.23}_{-0.24-0.23}$  & $1.21^{+0.04+0.06}_{-0.04-0.05}$\\
 \hline
 & $V^{B_{c}^{*} B}$ & $3.31^{+0.10+0.18}_{-0.10-0.22}$  & $3.91^{+0.07+0.16}_{-0.07-0.14}$   &  $6.22^{+0.23+0.21}_{-0.21-0.21}$ & $24.49^{+0.04+0.06}_{-0.04-0.06}$
 \\
 & $A_0^{B_{c}^{*} B}$ & $0.43^{+0.01+0.01}_{-0.02-0.00}$  & $0.48^{+0.06+0.08}_{-0.05-0.08}$   & $3.93^{+0.23+0.22}_{-0.21-0.23}$  & $10.86^{+0.06+0.06}_{-0.04-0.06}$\\
 & $A_1^{B_{c}^{*} B}$ & $0.44^{+0.00+0.01}_{-0.00-0.01}$  & $0.49^{+0.01+0.04}_{-0.01-0.04}$   & $4.32^{+0.21+0.19}_{-0.19-0.19}$  & $11.26^{+0.03+0.04}_{-0.03-0.04}$
 \\
 & $A_2^{B_{c}^{*} B}$ & $0.28^{+0.15+0.15}_{-0.18-0.16}$  & $0.27^{+0.01+0.06}_{-0.02-0.06}$   & $-0.65^{+0.19+0.18}_{-0.18-0.18}$  & $9.85^{+0.03+0.04}_{-0.04-0.04}$\\
 \hline
 & $V^{B_{c}^{*} B_{s}}$ & $3.61^{+0.09+0.11}_{-0.10-0.14}$  & $4.12^{+0.03+0.09}_{-0.04-0.08}$   &  $5.78^{+0.13+0.10}_{-0.13-0.10}$ & $17.20^{+0.02+0.03}_{-0.03-0.03}$
 \\
 & $A_0^{B_{c}^{*} B_{s}}$ & $0.50^{+0.01+0.01}_{-0.02-0.01}$  & $0.54^{+0.02+0.07}_{-0.02-0.07}$   & $3.70^{+0.13+0.11}_{-0.13-0.11}$  & $7.75^{+0.02+0.02}_{-0.03-0.02}$\\
 & $A_1^{B_{c}^{*} B_{s}}$ & $0.52^{+0.00+0.00}_{-0.00-0.01}$  & $0.57^{+0.00+0.05}_{-0.01-0.05}$   & $4.01^{+0.12+0.12}_{-0.12-0.12}$  & $7.98^{+0.01+0.02}_{-0.02-0.02}$
 \\
 &$A_2^{B_{c}^{*} B_{s}}$ & $0.27^{+0.17+0.16}_{-0.21-0.16}$  & $0.27^{+0.02+0.05}_{-0.03-0.04}$   & $-0.90^{+0.11+0.11}_{-0.10-0.11}$  & $7.57^{+0.01+0.02}_{-0.01-0.02}$\\
  \hline
 & $V^{B_{c}^{*} \eta_{c}}$ & $0.88^{+0.01+0.04}_{-0.01-0.06}$  & $0.87^{+0.03+0.09}_{-0.04-0.08}$   &  $1.10^{+0.13+0.10}_{-0.13-0.10}$ & $4.09^{+0.02+0.03}_{-0.03-0.03}$
 \\
 & $A_0^{B_{c}^{*} \eta_{c}}$ & $0.56^{+0.01+0.01}_{-0.02-0.02}$  & $0.59^{+0.02+0.07}_{-0.02-0.07}$   & $0.77^{+0.13+0.11}_{-0.13-0.11}$  & $2.15^{+0.02+0.02}_{-0.03-0.02}$\\
 & $A_1^{B_{c}^{*} \eta_{c}}$ & $0.59^{+0.01+0.02}_{-0.02-0.03}$  & $0.62^{+0.00+0.05}_{-0.01-0.05}$   & $0.72^{+0.12+0.12}_{-0.12-0.12}$  & $1.94^{+0.01+0.02}_{-0.02-0.02}$
 \\
 &$A_2^{B_{c}^{*} \eta_{c}}$ & $0.45^{+0.02+0.01}_{-0.02-0.02}$  & $0.46^{+0.02+0.05}_{-0.03-0.04}$   & $0.93^{+0.11+0.11}_{-0.10-0.11}$  & $3.11^{+0.01+0.02}_{-0.01-0.02}$\\
  \hline
 & $V^{B_{c}^{*} \eta_{c}(2S)}$ & $0.56^{+0.01+0.01}_{-0.01-0.03}$  & $0.58^{+0.03+0.09}_{-0.04-0.08}$   &  $0.77^{+0.13+0.10}_{-0.13-0.10}$ & $3.00^{+0.02+0.03}_{-0.03-0.03}$
 \\
 & $A_0^{B_{c}^{*} \eta_{c}(2S)}$ & $0.31^{+0.00+0.05}_{-0.00-0.07}$  & $0.30^{+0.02+0.07}_{-0.02-0.07}$   & $0.18^{+0.13+0.11}_{-0.13-0.11}$  & $1.51^{+0.02+0.02}_{-0.03-0.02}$\\
 & $A_1^{B_{c}^{*} \eta_{c}(2S)}$ & $0.34^{+0.00+0.02}_{-0.00-0.10}$  & $0.34^{+0.00+0.05}_{-0.01-0.05}$   & $0.27^{+0.12+0.12}_{-0.12-0.12}$  & $1.42^{+0.01+0.02}_{-0.02-0.02}$
 \\
 &$A_2^{B_{c}^{*} \eta_{c}(2S)}$ & $0.22^{+0.01+0.04}_{-0.01-0.15}$  & $0.20^{+0.02+0.05}_{-0.03-0.04}$   & $-0.32^{+0.11+0.11}_{-0.10-0.11}$  & $2.08^{+0.01+0.02}_{-0.01-0.02}$\\
 \hline\hline
\end{tabular}
}
\end{center}
\end{table}
In Tables \ref{tab:4}-\ref{tab:5}, we compare the form factor values at maximum recoil ($q^{2}=0$) with those obtained within other different models. The form factors of the transitions $B^{*}_{c}\to B_{(s)}, D, \eta_{c}$ and  $B^{*}_{s}\to D_{s}, K$  were investigated with the WSB model in Refs. \cite{R:2019uyb, Wang:2012hu, Chang:2016cdi}. Recently, the authors in Ref. \cite{Yang:2022jqu} investigated the form factors of the  transitions $B^{*}_{c}\to  B_{(s)}, \eta_{c}(1S,2S)$ within the light-front quark model. Certainly, the form factors of the $B^{*} \to D$ and $B_{s}^{*} \to D_{s}, K$ were analyzed within the
 model independent approach \cite{Ray:2019gkv} as well. For the form factors related to the $B^*_{(s)}$ transitions, one can find that, even in the same WSB model, the values of $A_2$ for the transitions $B^*_s\to D_s, K$ given by different authors \cite{R:2019uyb,Chang:2016cdi} have opposite signs.
Besides the multiple sign differences 
 among these predictions, most of the results given by different approaches are consistent with each other.
Regarding the form factors related to the $B^*_c$ transitions, 
significant differences predicted by the BSW approach exist for different values of $\omega$, which refers to the average transverse quark momentum. For the form factors of the transitions $B^{*}_{c}\to  B_{(s)}, \eta_{c}(1S,2S)$, our predictions are comparable with the previous light-front quark model calculations \cite{Yang:2022jqu} within errors except for the $A_2$ values.
These differences can be clarified in future experiments. 

\begin{table}[H]
\caption{Form factors of the transitions $B^{*}\to D,\pi, B_{s}^{*}\to D_{s},K$ at $q^{2}= 0$ in the
	CLFQM, along with other theoretical results.}
\begin{center}
	\scalebox{1.0}{
\begin{tabular}{cccccc}
\hline\hline
Transitions  &References&$V(0)$&$A_{0}(0)$&$A_{1}(0)$&$A_{2}(0)$\\
\hline
$B^{*}\rightarrow D $&This work&$0.75$&$0.63$&$0.66$&$0.57$\\
\hline
$ $&\cite{Ray:2019gkv}&$0.76$&$0.71$&$0.75$&$0.62$\\
$ $&\cite{Chang:2016cdi}&$0.70$&$0.63$&$0.66$&$0.56$\\
\hline
$B^{*}\rightarrow \pi $&This work&$0.29$&$0.25$&$0.30$&$0.21$\\
\hline
$ $&\cite{Chang:2016cdi}&$0.34$&$0.34$&$0.38$&$0.29$\\
\hline
$B^{*}_{s}\rightarrow D_{s} $&This work&$0.76$&$0.63$&$0.66$&$0.56$\\
\hline
$ $&\cite{R:2019uyb}\footnotemark[1]&$0.68$&$0.61$&$0.63$&$-0.55$\\
$ $&\cite{R:2019uyb}\footnotemark[2]&$0.77$&$0.67$&$0.72$&$-0.56$\\
$ $&\cite{Ray:2019gkv}&$0.72$&$0.66$&$0.69$&$0.59$\\
$ $&\cite{Chang:2016cdi}&$0.67$&$0.59$&$0.61$&$0.54$\\
\hline
$B^{*}_{s}\rightarrow K $&This work&$0.34$&$0.27$&$0.31$&$0.22$\\
\hline
$ $&\cite{R:2019uyb}\footnotemark[1]&$0.30$&$0.27$&$0.28$&$-0.25$\\
$ $&\cite{R:2019uyb}\footnotemark[2]&$0.27$&$0.24$&$0.26$&$-0.21$\\
$ $&\cite{Ray:2019gkv}&$0.30$&$0.28$&$0.29$&$0.26$\\
$ $&\cite{Chang:2016cdi}&$0.32$&$0.29$&$0.31$&$0.28$\\
\hline\hline
\end{tabular}\label{tab:4}}
\end{center}
{\footnotesize $^1$ The form factors are computed with the parameter $\omega=0.4$ GeV in the WSB model.\\
               $^2$ The form factors are computed with flavor dependent parameter $\omega$ in the WSB model.}
\end{table}

\begin{table}[H]
\caption{Form factors of the transitions $B^{*}_c\to D, B_{(s)},\eta_c(1S, 2S)$ at $q^{2}= 0$, along with other theoretical results.}
\begin{center}
\scalebox{0.9}{
\begin{tabular}{|c|c|cccc|c|}
\hline\hline
Transition  &References&$V(0)$&$A_{0}(0)$&$A_{1}(0)$&$A_{2}(0)$\\
\hline
$B^{*}_{c}\rightarrow D $&This work&$0.26$&$0.15$&$0.16$&$0.13$\\
\hline
$ $&\cite{R:2019uyb}\footnotemark[1]&$0.020$&$0.013$&$0.012$&$-0.016$\\
$ $&\cite{R:2019uyb}\footnotemark[2]&$0.084$&$0.045$&$0.052$&$-0.033$\\
$ $&\cite{R:2019uyb}\footnotemark[3]&$0.60$&$0.29$&$0.37$&$-0.15$\\
\hline
$B^{*}_{c}\rightarrow B $&This work&$3.31$&$0.43$&$0.44$&$0.28$\\
\hline
$ $&\cite{R:2019uyb}\footnotemark[1]&$2.62$&$0.31$&$0.35$&$0.13$\\
$ $&\cite{R:2019uyb}\footnotemark[2]&$5.09$&$0.43$&$0.67$&$2.19$\\
$ $&\cite{R:2019uyb}\footnotemark[3]&$6.93$&$0.90$&$0.91$&$-0.83$\\
$ $&\cite{Sun:2017lup}\footnotemark[1]&$1.95$&$0.54$&$0.29$&$3.29$\\
$ $&\cite{Sun:2017lup}\footnotemark[4]&$2.88$&$0.78$&$0.43$&$4.70$\\
$ $&\cite{Sun:2017lup}\footnotemark[3]&$3.61$&$0.94$&$0.54$&$5.41$\\
$ $&\cite{Yang:2022jqu}&$3.08$&$0.60$&$0.65$&$0.91$\\
\hline
$B^{*}_{c}\rightarrow B_{s} $&This work&$3.61$&$0.50$&$0.52$&$0.27$\\
\hline
$ $&\cite{R:2019uyb}\footnotemark[1]&$2.89$&$0.37$&$0.43$&$0.43$\\
$ $&\cite{R:2019uyb}\footnotemark[2]&$5.56$&$0.58$&$0.83$&$2.45$\\
$ $&\cite{R:2019uyb}\footnotemark[3]&$6.52$&$0.92$&$0.98$&$-0.24$\\
$ $&\cite{Sun:2017lup}\footnotemark[1]&$1.87$&$0.61$&$0.36$&$3.62$\\
$ $&\cite{Sun:2017lup}\footnotemark[4]&$2.55$&$0.82$&$0.49$&$4.79$\\
$ $&\cite{Sun:2017lup}\footnotemark[3]&$3.10$&$0.95$&$0.60$&$5.27$\\
$ $&\cite{Yang:2022jqu}&$3.40$&$0.69$&$0.75$&$0.96$\\
\hline
$B^{*}_{c}\rightarrow \eta_{c} $&This work&$0.88$&$0.56$&$0.59$&$0.45$\\
\hline
$ $&\cite{R:2019uyb}\footnotemark[1]&$0.21$&$0.15$&$0.15$&$-0.15$\\
$ $&\cite{R:2019uyb}\footnotemark[2]&$0.79$&$0.53$&$0.57$&$-0.40$\\
$ $&\cite{R:2019uyb}\footnotemark[3]&$1.21$&$0.73$&$0.87$&$-0.33$\\
$ $&\cite{Yang:2022jqu}&$0.91$&$0.66$&$0.69$&$0.59$\\
\hline
$B^{*}_{c}\rightarrow \eta_{c}(2S) $&This work&$0.56$&$0.31$&$0.34$&$0.22$\\
\hline
$ $&\cite{Yang:2022jqu}&$0.59$&$0.43$&$0.41$&$0.51$\\
\hline\hline
\end{tabular}\label{tab:5}
}
\end{center}
{\footnotesize $^1$ The form factors are computed with the parameter $\omega=0.4$ GeV in the WSB model.\\
               $^2$ The form factors are computed with the flavor dependent parameter $\omega$ in the WSB model.\\
               $^3$ The form factors are computed with the QCD inspired parameter $\omega = m_{B^*_c}\alpha_{s}$ in the WSB model\\
               $^4$ The form factors are computed with the flavor dependent parameter $\omega=0.6$ GeV in the WSB model.}
\end{table}

\begin{figure}[H]
\vspace{0.80cm}
  \centering
  \subfigure[]{\includegraphics[width=0.31\textwidth]{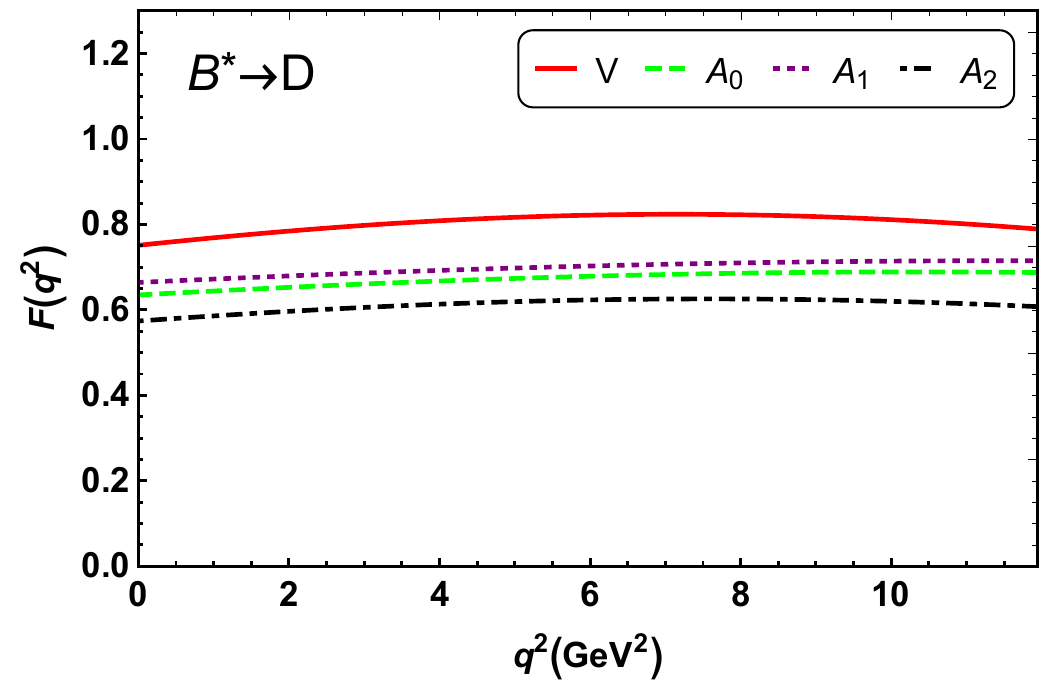}\quad}
  \subfigure[]{\includegraphics[width=0.31\textwidth]{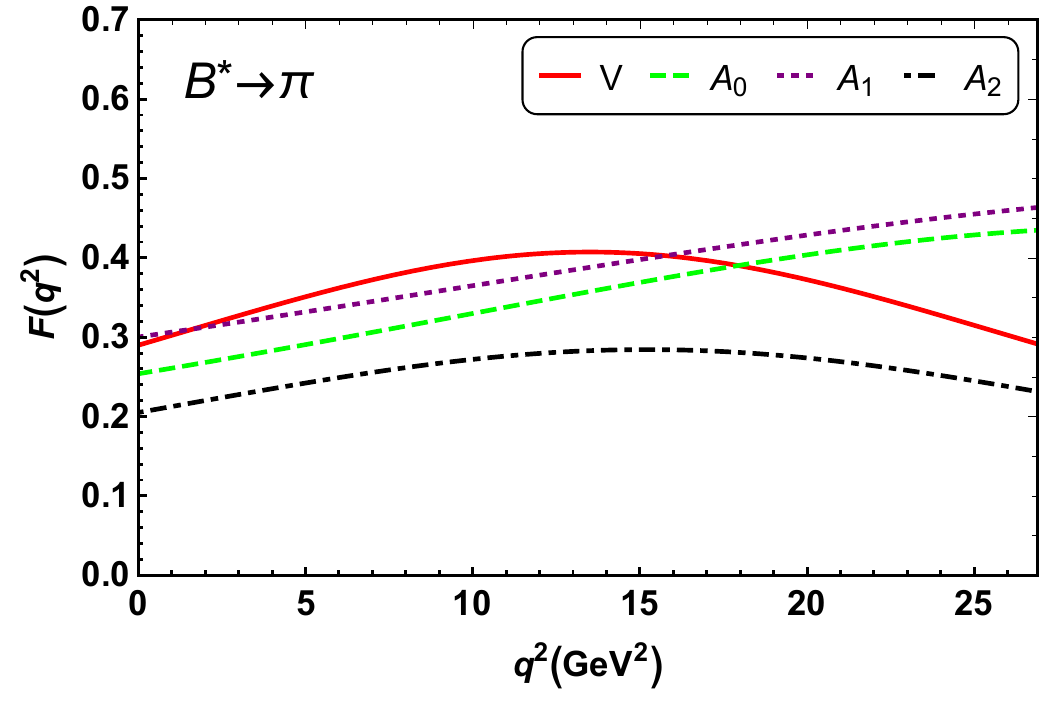}\quad}
  \subfigure[]{\includegraphics[width=0.31\textwidth]{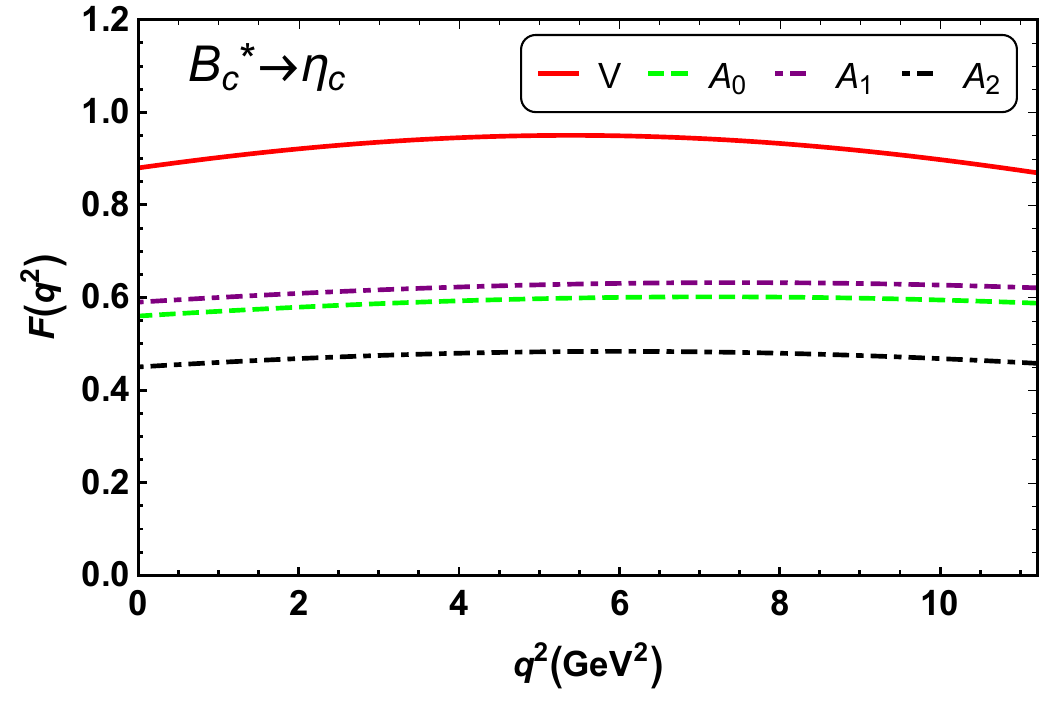}}\\
  \subfigure[]{\includegraphics[width=0.31\textwidth]{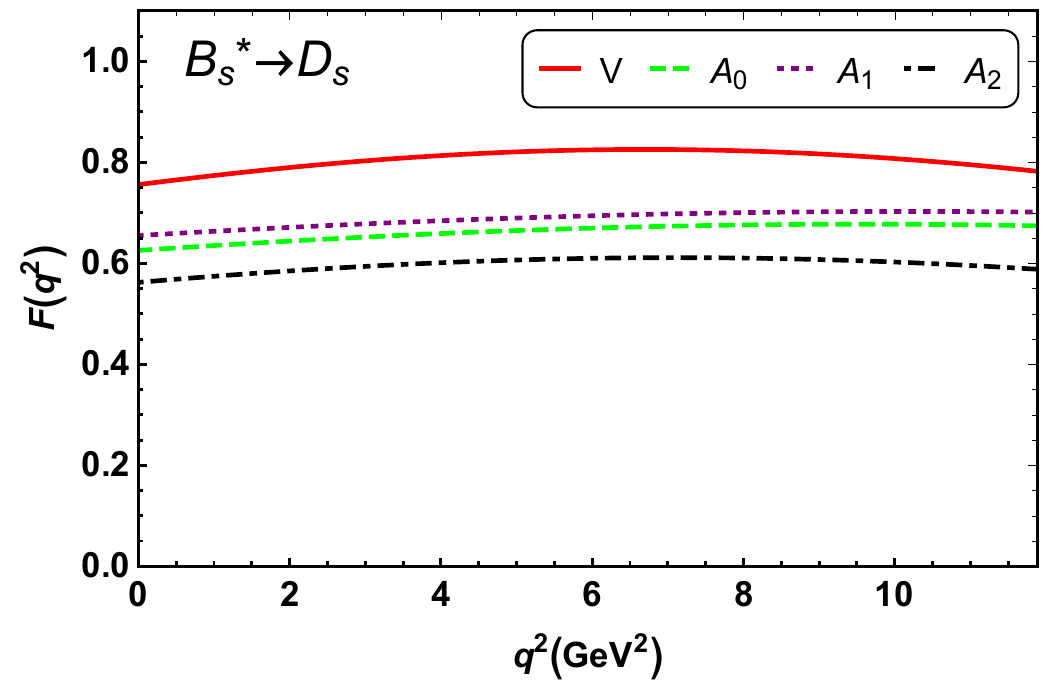}\quad}
  \subfigure[]{\includegraphics[width=0.31\textwidth]{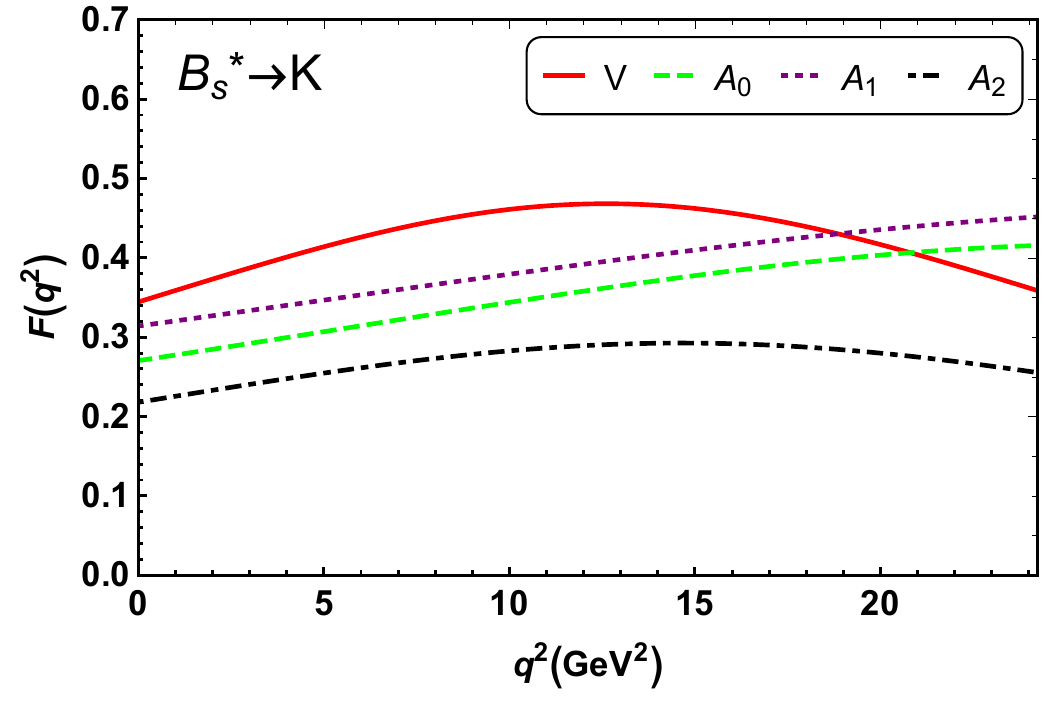}\quad}
  \subfigure[]{\includegraphics[width=0.31\textwidth]{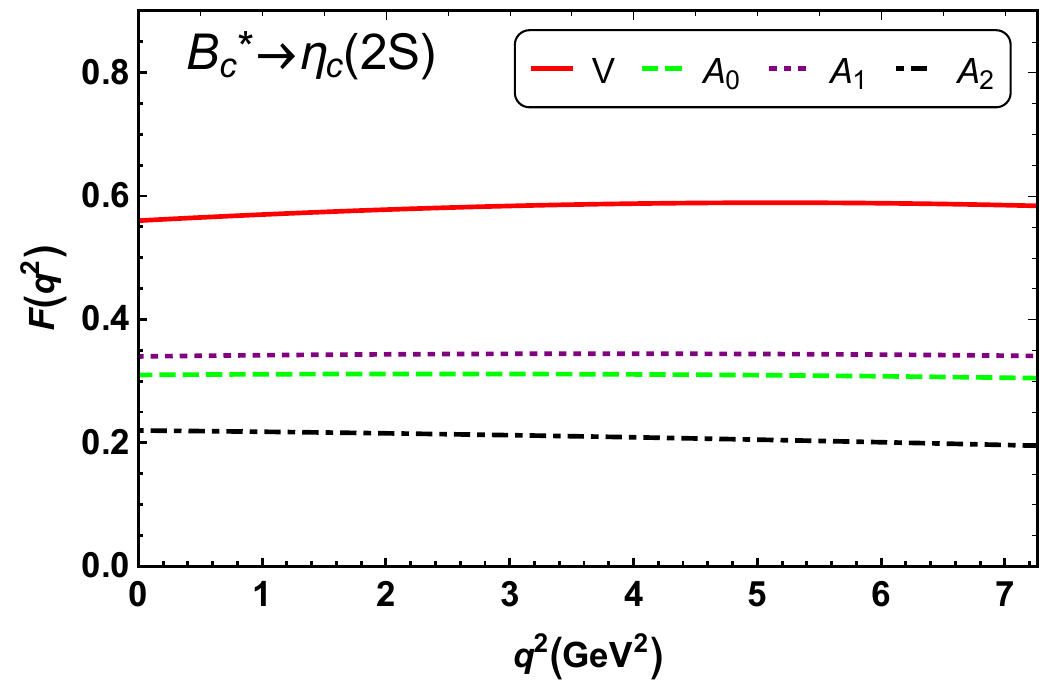}}\\
  \subfigure[]{\includegraphics[width=0.31\textwidth]{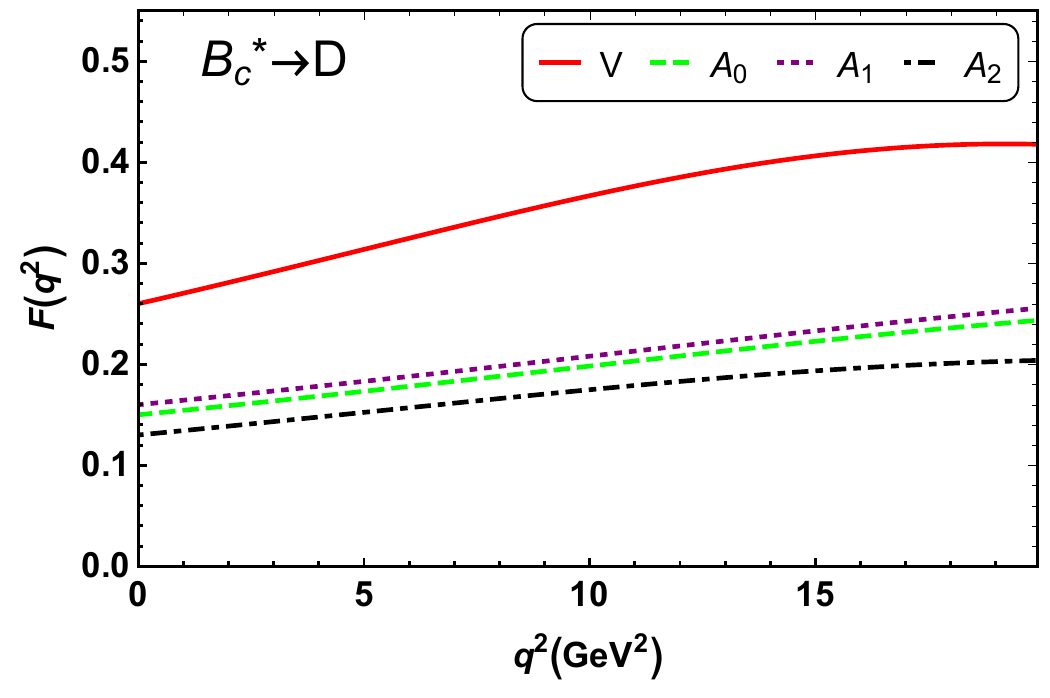}\quad}
   \subfigure[]{\includegraphics[width=0.31\textwidth]{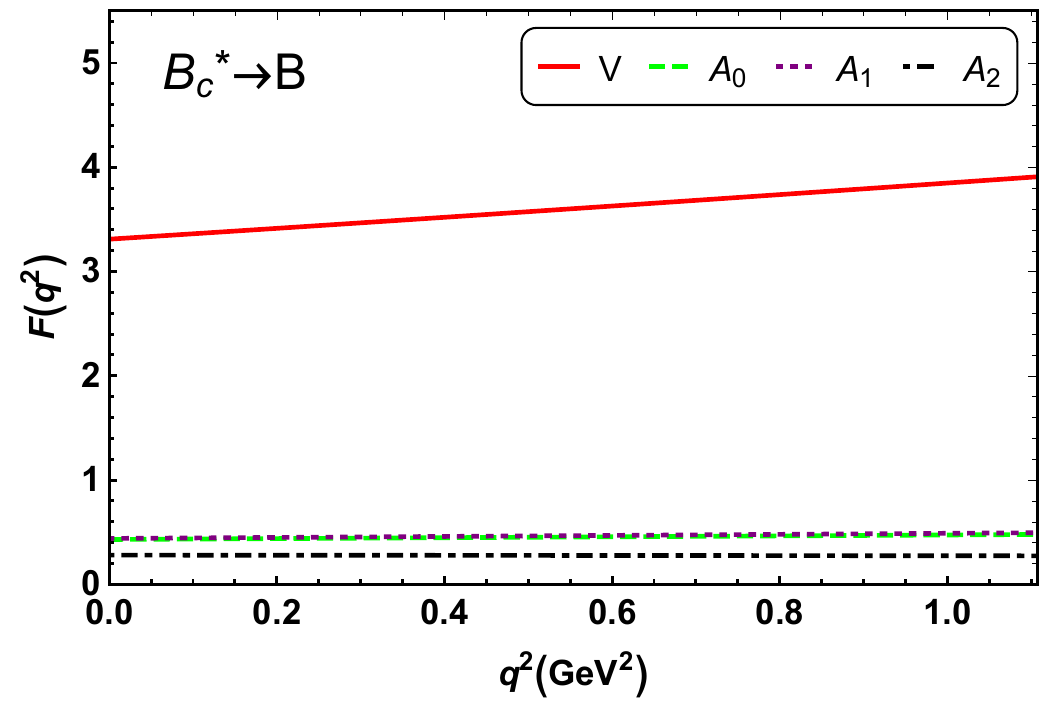} \quad}
  \subfigure[]{\includegraphics[width=0.31\textwidth]{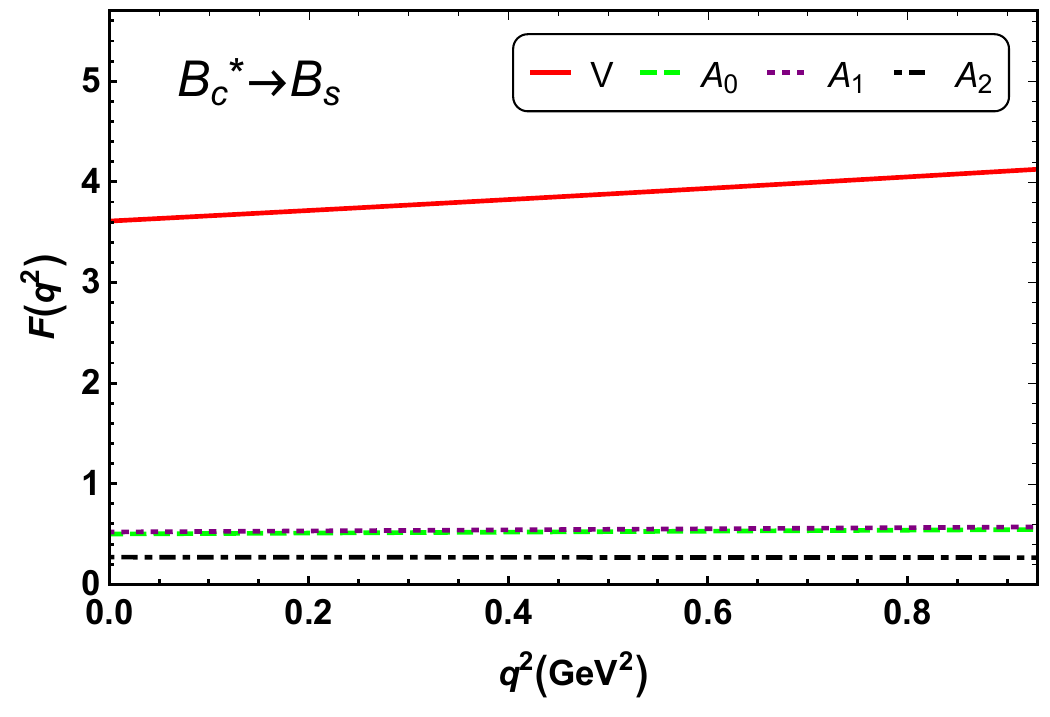}}
\caption{Form factors $V(q^2)$ and $A_{0,1,2}(q^2)$ for the transitions $B^*\to D, \pi$, $B^*_s\to D_{s}, K$, and $B^*_c\to B_{(s)}, D, \eta_c(1S,2S)$.}\label{fig:T5}
\end{figure}

We plot the $q^2$-dependencies of the form factors of the transitions $B^*\to D, \pi$, $B^*_s\to D_{s}, K $ and $B^*_c\to B_{(s)}, D, \eta_c(1S,2S)$, as shown in Figure \ref{fig:T5}. Compared with the form factors of the transitions $B^*_{(s)}\to D_{(s)}$, those of the transitions $B^*\to \pi$ and $B^*_{s}\to K$ are more sensitive to changes in $q^2$. 
Compared with the single pole expression, the double-pole approximation makes the predictions less
model dependent. Nonetheless, when a significant mass difference exists between the initial and final states, for example, the final state is $\pi$ or $K$, the model dependent, non-monotonous $q^2$ behaviors still exist. This is why the transition form factors predicted by the CLFQM in small $q^2$ regions are more reliable than those given in large $q^2$ regions. This is contrary to the case of lattice QCD predictions. If the transitions, such as $B^*_c\to B_{(s)}$, have a small phase space, their corresponding form factors can indeed display monotonous $q^2$ behaviours.
Among the form factors of the transitions $B^*_c\to B_{(s)}$, the values of $V(q^2)$ are much larger than those of $A_{0,1,2}(q^2)$.
\subsection{Semileptonic Decays}
As the flavor changing processes in the Standard Model, the semileptonic heavy flavor meson decays are crucial for extracting the Cabibbo-Kobayashi-Maskawa (CKM) matrix elements. The form factors involving the dynamic information are essential in these semileptonic decays. Based on the form factors and helicity amplitudes provided in the previous section, the branching ratios of the semileptonic $B^{*+}$,  $B_{s}^{*0}$, and $B_{c}^{*+}$ decays are presented in Table \ref{Tab:7}.We also define two other physical observables, that is the forward-backward asymmetry $A_{FB}$ and the longitudinal polarization fraction $f_{L}$, to study the impact of lepton mass and provide a more detailed physical picture for the semileptonic decays. The corresponding results are listed in Tables \ref{AFB} and \ref{FL}, respectively. The theoretical uncertainties in our calculations are caused by 
the decay widths of the initial mesons and the decay
constants of the initial and final state mesons, respectively. Further details are discussed below:
\begin{table}[H]
	\caption{Branching ratios of the semileptonic decays $B^{*}_{u,s,c}\to P \ell^{+}{\nu}_{\ell}$. For each entry, the first error is from the decay width of the initial meson, and the second and third errors are caused by the decay constants of the initial and final mesons, respectively.}
	\begin{center}
		\scalebox{0.74}{
			\begin{tabular}{|cccc|}
				\hline\hline
				Channels&This work& BS \cite{Wang:2018ryc}& WSB \cite{Chang:2016cdi} \\
				\hline
				$B^{*+} \to \bar{D}^{0}e^{+}{\nu}_{e}$&$(5.19^{+1.36+0.17+0.19}_{-0.89-0.02-0.04})\times10^{-8}$&$3.29\times10^{-8}$&\;\;\;\;$2.29\times10^{-8}$\\
				$B^{*+} \to \bar{D}^{0}\mu^{+}{\nu}_{\mu}$&$(5.16^{+1.35+0.17+0.18}_{-0.89-0.02-0.04})\times10^{-8}$&$3.27\times10^{-8}$&\;\;\;\;$2.29\times10^{-8}$\\
				$B^{*+} \to \bar{D}^{0}\tau^{+}{\nu}_{\tau}$&$(1.30^{+0.34+0.04+0.04}_{-0.22-0.01-0.00})\times10^{-8}$&$0.82\times10^{-8}$&\;\;\;\;$0.68\times10^{-8}$\\
				\hline
				$B_{s}^{*0} \to D_{s}^{-}e^{+}{\nu}_{e}$&$(3.25^{+0.77+0.01+0.11}_{-0.52-0.11-0.01})\times10^{-7}$&$2.04\times10^{-7}$&\;\;\;\;$1.39\times10^{-7}$\\
				$B_{s}^{*0} \to D_{s}^{-}\mu^{+}{\nu}_{\mu}$&$(3.23^{+0.77+0.01+0.11}_{-0.52-0.11-0.01})\times10^{-7}$&$2.03\times10^{-7}$&\;\;\;\;$1.39\times10^{-7}$\\
				$B_{s}^{*0} \to D_{s}^{-}\tau^{+}{\nu}_{\tau}$&$(8.09^{+1.92+0.03+0.21}_{-1.30-0.24-0.03})\times10^{-8}$&$5.35\times10^{-8}$&\;\;\;\;$4.08\times10^{-8}$\\
				$B_{s}^{*0} \to K^{-}e^{+}{\nu}_{e}$&$(6.06^{+1.43+0.56+0.31}_{-0.98-0.44-0.17})\times10^{-9}$&$-$&\;\;\;\;$1.67\times10^{-9}$\\
				$B_{s}^{*0} \to K^{-}\mu^{+}{\nu}_{\mu}$&$(6.03^{+1.43+0.56+0.30}_{-0.97-0.44-0.17})\times10^{-9}$&$-$&\;\;\;\;$1.67\times10^{-9}$\\
				$B_{s}^{*0} \to K^{-}\tau^{+}{\nu}_{\tau}$&$(2.56^{+0.61+0.18+0.45}_{-0.41-0.13-0.08})\times10^{-9}$&$-$&\;\;\;\;$1.07\times10^{-9}$\\
				\hline
				$B_{c}^{*+} \to D^{0} e^{+}{\nu}_{e}$&$(1.36^{+0.35+0.13+0.20}_{-0.23-0.13-0.15})\times10^{-9}$&$1.60\times10^{-9}$&\\
				$B_{c}^{*+} \to D^{0} \mu^{+}{\nu}_{\mu}$&$(1.36^{+0.35+0.13+0.20}_{-0.23-0.13-0.15})\times10^{-9}$&$1.60\times10^{-9}$&$-$\\
				$B_{c}^{*+} \to D^{0} \tau^{+}{\nu}_{\tau}$&$(0.65^{+0.17+0.07+0.08}_{-0.11-0.06-0.07})\times10^{-9}$&$1.08\times10^{-9}$&$-$\\
				$B_{c}^{*+} \to B^{0}_{s} e^{+}{\nu}_{e}$&$(7.51^{+1.92+0.17+0.10}_{-1.27-0.08-0.37})\times10^{-7}$&$9.43\times10^{-7}$&$-$\\
				$B_{c}^{*+} \to B^{0}_{s} \mu^{+}{\nu}_{\mu}$&$(7.06^{+1.80+0.14+0.06}_{-1.19-0.10-0.32})\times10^{-7}$&$8.96\times10^{-7}$&$-$\\
				$B_{c}^{*+} \to B^{0} e^{+}{\nu}_{e}$&$(4.37^{+1.11+0.14+0.27}_{-0.74-0.12-0.27})\times10^{-8}$&$5.78\times10^{-8}$&$-$\\
				$B_{c}^{*+} \to B^{0} \mu^{+}{\nu}_{\mu}$&$(4.15^{+1.06+0.11+0.24}_{-0.70-0.10-0.24})\times10^{-8}$&$5.57\times10^{-8}$&$-$\\
				$B_{c}^{*+} \to \eta_{c} e^{+}{\nu}_{e}$&$(4.48^{+1.14+0.28+0.40}_{-0.76-0.10-0.50})\times10^{-7}$&$4.20\times10^{-7}$&$-$\\
				$B_{c}^{*+} \to \eta_{c} \mu^{+}{\nu}_{\mu}$&$(4.45^{+1.14+0.27+0.39}_{-0.75-0.10-0.49})\times10^{-7}$&$4.19\times10^{-7}$&$-$\\
				$B_{c}^{*+} \to \eta_{c} \tau^{+}{\nu}_{\tau}$&$(1.03^{+0.26+0.07+0.07}_{-0.17-0.04-0.10})\times10^{-7}$&$1.26\times10^{-7}$&$-$\\
				$B_{c}^{*+} \to \eta_{c}(2S) e^{+}{\nu}_{e}$&$(5.65^{+1.44+0.02+0.59}_{-0.96-0.03-0.29})\times10^{-8}$&$-$&$-$\\
				$B_{c}^{*+} \to \eta_{c}(2S) \mu^{+}{\nu}_{\mu}$&$(5.60^{+1.43+0.02+0.60}_{-0.95-0.03-0.29})\times10^{-8}$&$-$&$-$\\
				$B_{c}^{*+} \to \eta_{c}(2S) \tau^{+}{\nu}_{\tau}$&$(4.92^{+1.26+0.11+0.12}_{-0.83-0.11-0.23})\times10^{-9}$&$-$&$-$\\
				\hline\hline
			\end{tabular}\label{Tab:7}}
	\end{center}
\end{table}
\begin{enumerate}
\item
From Table \ref{Tab:7}, the branching ratios of the decays  $B^{*}_{s}\to K\ell{\nu}_{\ell}$ are significantly $1\sim2$ orders smaller compared to those of the decays $B^{*}_{(s)}\to D_{(s)}\ell{\nu}_{\ell}$. This is mainly due to the smaller CKM matrix element $V_{ub}=0.00382$ compared to $V_{cb}=0.0408$.  Furthermore, the form factor $A_0^{B^{*}_{s}K}$  is only approximately 0.43 times $A_0^{B^{*}_{(s)}D_{(s)}}$. In addition,
$\mathcal{B}r(B_{s}^{*0} \to D_{s}^{-}\ell^{+}{\nu}_{\ell})$ is approximately six times $\mathcal{B}r(B^{*+} \to \bar{D}^{0}\ell^{+}{\nu}_{\ell})$, which is attributed to the small total decay width $\Gamma_{tot}(B_{s}^{*0})$, which is only approximately one sixth of $\Gamma_{tot}(B^{*+})$. Among these $B^*_{(s)}$ decays, the channels $B_{s}^{*0} \to D_{s}^{-}\ell^{\prime+}{\nu}_{\ell^\prime}$ with $\ell^{\prime}=e, \mu$ \footnote{In the following we will use $\ell^\prime$ to represent $e,\mu$  for simplicity. } have the largest branching fractions of the $10^{-7}$ order, making them a top priority for observation. The branching ratios of the decays $B^{*+}\to \bar{D}^{0}\ell^{+}{\nu}_{\ell}$ and $B^{*0}_{s}\to D^{-}_s\tau^{+}{\nu}_{\tau}$ are of the order of $10^{-8}$, which can be in the measurement scopes of the LHC and SuperKEKB in the future. However, observing the decays $B^{*0}_{s}\to K^{-}\ell^{+}{\nu}_{\ell} $ in experiments may be challenging, owing to their small branching fractions of the $10^{-9}$ order. 
\item
For the branching ratios of the $B^{*+}$ and $B_{s}^{*0}$ decays, our results are at least twice as large as those given by the WSB model  \cite{Chang:2016cdi}. Although the form factors calculated using the CLFQM and the WSB are consistent with at $q^{2}=0$, their dependencies on $q^2$ between these two approaches are very different, which is owing to the difference in the branching ratios. Conversely, the Bethe-Salpeter (BS) equation approach \cite{Wang:2018ryc} resulted in intermediate values obtained by solving the instantaneous BS equation with a Cornell-like potential.  
\item
The branching ratios of the $B^{*}_{c}$ decays induced by the $\bar b\to \bar c(\bar u)$ and $c\to d(s)$ transitions are calculated, which are also shown in Table \ref{Tab:7}. Our predictions are consistent with the results obtained in the BS method \cite{Wang:2018ryc}. The decays $B_{c}^{*+} \to B^{0}_{s} \ell^{\prime+}{\nu}_{\ell^\prime}$ and $B_{c}^{*+} \to \eta_c \ell^{+}{\nu}_{\ell}$, which possess larger branching ratios compared to other semileptonic $B^{*}_{c}$ decays, can serve as golden channels to search for the $B^*_c$ meson in experiments. The branching ratios of the decays $B_{c}^{*+} \to D^{0}\ell^{+}{\nu}_{\ell}$ are much smaller than those of other channels, because the CKM matrix element $V_{ub}=3.82\times10^{-3}$ involved in these decays is very small. Comparing the branching ratios of the semileptonic decays $B_{c}^{*+} \to \eta_{c}\ell^{+}{\nu}_{\ell}$ and $B_{c}^{*+} \to \eta_{c}(2S)\ell^{+}{\nu}_{\ell}$, we can find that the latter ratios are about one order smaller than the former ratios owing to the smaller form factor of the transition $B_{c}^{*} \to \eta_{c}(2S)$  and the smaller phase space for the decays  $B_{c}^{*+} \to \eta_{c}(2S)\ell^{+}{\nu}_{\ell}$. 
\item
Some branching fraction ratios for the $B_{(s)}$ and $B_{c}$ meson decays have garnered significant attention owing to their deviations between theoretical predictions and experimental data, such as the $R_{J/\Psi}$ anomaly \cite{LHCb:2015gmp}, which may indicate possible new physics beyond the SM. As such, checking whether similar cases exist in their vector partner decays is important. In view of this purpose, we define the following quantities:
\be
\mathcal{R}_P=\frac{\mathcal{B}r(B^*_{u,s,c} \rightarrow P \tau {\nu}_{\tau})}{\mathcal{B}r(B^*_{u,s,c} \rightarrow P e {\nu}_{e})},
\en
where $P$ represents a 
pseudoscalar meson, such as $K, D_{(s)}, \eta_c(1S,2S)$. 

The corresponding values are listed as
\begin{tiny}
	\begin{equation}
		\begin{aligned}
	\mathcal{R}_{\bar D_{0}}&=\frac{\mathcal{B}r(B^{*+} \to \bar{D}^{0}\tau^{+}{\nu}_{\tau})}{\mathcal{B}r(B^{*+} \to \bar{D}^{0}e^{+}{\nu}_{e})}=0.250\pm0.093,\;\;\;
	\mathcal{R}_{D_{s}}=\frac{\mathcal{B}r(B_{s}^{*0} \to D_{s}^{-}\tau^{+}{\nu}_{\tau})}{\mathcal{B}r(B_{s}^{*0} \to D_{s}^{-}e^{+}{\nu}_{e})}=0.249\pm0.059 ,\\
	\mathcal{R}_{K}&=\frac{\mathcal{B}r(B_{s}^{*0} \to K^{-}\tau^{+}{\nu}_{\tau})}{\mathcal{B}r(B_{s}^{*0} \to K^{-}e^{+}{\nu}_{e})}=0.425\pm0.143,\;\;\;
	\mathcal{R}_{D_{0}}=\frac{\mathcal{B}r(B_{c}^{*+} \to D^{0} \tau^{+}{\nu}_{\tau})}{\mathcal{B}r(B_{c}^{*+} \to D^{0} e^{+}{\nu}_{e})}=0.478\pm0.175 ,\\
	\mathcal{R}_{\eta_{c}}&=\frac{\mathcal{B}r(B_{c}^{*+} \to \eta_{c} \tau^{+}{\nu}_{\tau})}{\mathcal{B}r(B_{c}^{*+} \to \eta_{c} e^{+}{\nu}_{e})}=0.229\pm0.059,\;\;\;
	\mathcal{R}_{\eta_{c}(2S)}=\frac{\mathcal{B}r(B_{c}^{*+} \to \eta_{c}(2S) \tau^{+}{\nu}_{\tau})}{\mathcal{B}r(B_{c}^{*+} \to \eta_{c}(2S) e^{+}{\nu}_{e})}=0.087\pm0.022.
	\end{aligned}
\end{equation}
\end{tiny}
From the above ratios, we can find that our predictions are consistent with the results calculated by using the BS method in Ref. \cite{Wang:2018ryc} within errors, where $\mathcal{R}_{\bar D_{0}}$= 0.249, $\mathcal{R}_{D_{s}}$= 0.262,   $\mathcal{R}_{{D}_{0}}$= 0.675, and $\mathcal{R}_{\eta_{c}}$= 0.300. Our results are also comparable with those obtained by the BSW model in Ref. \cite{Chang:2016cdi}, where $\mathcal{R}_{\bar D_{0}}$= 0.297, $\mathcal{R}_{D_{s}}$= 0.294, $\mathcal{R}_{K}$= 0.641. Since the numerator and denominator involving the same initial and final hadrons for each ratio share the same CKM matrix element and form factor, some uncertainties in the calculations can be canceled. Therefore, these ratios are less model-dependent and can be checked in future experiments.
\begin{table}[H]
\caption{Forward-backward asymmetries $A_{FB}$ for the decays $B^*_{u,s,c} \to P\ell^{+}\nu_{\ell}$.}
\begin{center}
\scalebox{0.8}{
\begin{tabular}{|c|c|c|c|}
\hline\hline
 Channels  &$B^{*+} \to \bar{D}^{0}e^{+}\nu_{e}$&$B^{*+} \to \bar{D}^{0}\mu^{+}\nu_{\mu}$&$B^{*+} \to \bar{D}^{0}\tau^{+}\nu_{\tau}$\\
 \hline
 $A_{FB}$&$-0.19^{+0.00+0.04+0.04}_{-0.01-0.04-0.04}$&$-0.19^{+0.00+0.04+0.04}_{-0.01-0.04-0.04}$&$-0.14^{+0.02+0.00+0.00}_{-0.04-0.00-0.00}$\\
 \hline
  Channel  &$B_{s}^{*0} \to D^{-}_{s}e^{+}\nu_{e}$&$B_{s}^{*0} \to D^{-}_{s}\mu^{+}\nu_{\mu}$&$B_{s}^{*0} \to D^{-}_{s}\tau^{+}\nu_{\tau}$\\
   \hline
 $ A_{FB}$&$-0.19^{+0.03+0.00+0.00}_{-0.04-0.00-0.00}$&$-0.18^{+0.03+0.01+0.00}_{-0.04-0.01-0.00}$&$-0.13^{+0.02+0.00+0.00}_{-0.03-0.00-0.00}$\\
\hline\hline
Channels&$B_{s}^{*0} \to K^{-}e^{+}\nu_{e}$&$B_{s}^{*0} \to K^{-}\mu^{+}\nu_{\mu}$&$B_{s}^{*0} \to K^{+}\tau^{+}\nu_{\tau}$\\
\hline
$A_{FB}$&$-0.10^{+0.02+0.00+0.01}_{-0.02-0.00-0.01}$&$-0.10^{+0.02+0.00+0.01}_{-0.02-0.00-0.01}$&$-0.09^{+0.02+0.00+0.01}_{-0.02-0.00-0.01}$\\
\hline
Channels&$B_{c}^{*+} \to D^{0}e^{+}\nu_{e}$&$B_{c}^{*+} \to D^{0}\mu^{+}\nu_{\mu}$&$B_{c}^{*+} \to D^{0}\tau^{+}\nu_{\tau}$\\
\hline
$A_{FB}$&$-0.28^{+0.05+0.03+0.04}_{-0.07-0.02-0.03}$&$-0.28^{+0.05+0.03+0.04}_{-0.07-0.02-0.03}$&$-0.24^{+0.04+0.02+0.03}_{-0.06-0.02-0.02}$\\
\hline
Channels&$B_{c}^{*+} \to \eta_{c}e^{+}\nu_{e}$&$B_{c}^{*+} \to \eta_{c}\mu^{+}\nu_{\mu}$&$B_{c}^{*+} \to \eta_{c}\tau^{+}\nu_{\tau}$\\
\hline
$A_{FB}$&$-0.19^{+0.03+0.01+0.02}_{-0.05-0.01-0.01}$&$-0.19^{+0.03+0.01+0.02}_{-0.05-0.01-0.01}$&$-0.14^{+0.02+0.00+0.02}_{-0.04-0.01-0.01}$\\
\hline
Channels&$B_{c}^{*+} \to \eta_{c}(2S)e^{+}\nu_{e}$&$B_{c}^{*+} \to \eta_{c}(2S)\mu^{+}\nu_{\mu}$&$B_{c}^{*+} \to \eta_{c}(2S)\tau^{+}\nu_{\tau}$\\
\hline
$A_{FB}$&$-0.16^{+0.03+0.00+0.03}_{-0.04-0.00-0.02}$&$-0.16^{+0.03+0.00+0.03}_{-0.04-0.00-0.02}$&$-0.11^{+0.02+0.00+0.02}_{-0.03-0.00-0.05}$\\
\hline
Channels&$B_{c}^{*+} \to B^{0}_{s}e^{+}\nu_{e}$&$B_{c}^{*+} \to B^{0}_{s}\mu^{+}\nu_{\mu}$&$-$\\
\hline
$A_{FB}$&$-0.22^{+0.04+0.00+0.01}_{-0.06-0.01-0.01}$&$-0.22^{+0.04+0.00+0.01}_{-0.06-0.01-0.01}$&$-$\\
\hline
Channels&$B_{c}^{*+} \to B^{0}e^{+}\nu_{e}$&$B_{c}^{*+} \to B^{0}\mu^{+}\nu_{\mu}$&$-$\\
\hline
$A_{FB}$&$-0.26^{+0.04+0.01+0.02}_{-0.07-0.01-0.02}$&$-0.26^{+0.04+0.01+0.02}_{-0.07-0.01-0.02}$&$-$\\
\hline\hline
\end{tabular}\label{AFB}}
\end{center}
\end{table}

\begin{table}[H]
\caption{Longitudinal polarization fractions $f_{L}$ for the decays $B^*_{u,s,c} \to P\ell^{+}\nu_{\ell}$ in Region 1,  Region 2, and the total physical 
region.}
\begin{center}
\scalebox{0.8}{
\begin{tabular}{|c|c|c|c|c|c|c|c|}
\hline\hline
Observables&Region 1&Region 2&Total&Observables&Region 1&Region 2&Total\\
\hline\hline
$f_{L}(B^{*+} \to \bar{D}^{0}l^{'+}\nu_{l^{'}})$&$0.69$&$0.43$&$0.57^{+0.15+0.02+0.02}_{-0.10-0.00-0.01}$&$f_{L}(B_{s}^{*0} \to D^{-}_{s}l^{'+}\nu_{l^{'}})$&$0.69$&$0.43$&$0.57^{+0.14+0.00+0.02}_{-0.09-0.02-0.00}$\\
\hline
$f_{L}(B^{*+} \to \bar{D}^{0}\tau^{+}\nu_{\tau})$&$0.56$&$0.42$&$0.47^{+0.12+0.02+0.02}_{-0.08-0.00-0.00}$&$f_{L}(B_{s}^{*0} \to D^{-}_{s}\tau^{+}\nu_{\tau})$&$0.56$&$0.42$&$0.47^{+0.11+0.00+0.00}_{-0.08-0.02-0.01}$\\
\hline\hline
Observables&Region 1&Region 2&Total&Observables&Region 1&Region 2&Total\\
\hline\hline
$f_{L}(B_{s}^{*0} \to K^{-}l^{'+}\nu_{l^{'}})$&$0.88$&$0.70$&$0.82^{+0.18+0.06+0.18}_{-0.14-0.08-0.03}$&$f_{L}(B_{c}^{*+} \to D^{0}l^{'+}\nu_{l^{'}})$&$0.65$&$0.42$&$0.53^{+0.14+0.05+0.09}_{-0.09-0.05-0.05}$\\
\hline
$f_{L}(B_{s}^{*0} \to K^{-}\tau^{+}\nu_{\tau})$&$0.83$&$0.68$&$0.76^{+0.17+0.05+0.13}_{-0.13-0.04-0.03}$&$f_{L}(B_{c}^{*+} \to D^{0}\tau^{+}\nu_{\tau})$&$0.57$&$0.42$&$0.48^{+0.12+0.05+0.07}_{-0.08-0.04-0.05}$\\
\hline\hline
Observables&Region 1&Region 2&Total&Observables&Region 1&Region 2&Total\\
\hline\hline
$f_{L}(B_{c}^{*+} \to \eta_{c}l^{'+}\nu_{l^{'}})$&$0.68$&$0.42$&$0.57^{+0.14+0.03+0.06}_{-0.10-0.10-0.07}$&$f_{L}(B_{c}^{*+} \to \eta_{c}(2S)l^{'+}\nu_{l^{'}})$&$0.70$&$0.43$&$0.58^{+0.15+0.00+0.02}_{-0.10-0.00-0.32}$\\
\hline
$f_{L}(B_{c}^{*+} \to \eta_{c}\tau^{+}\nu_{\tau})$&$0.55$&$0.41$&$0.46^{+0.12+0.03+0.03}_{-0.08-0.01-0.05}$&$f_{L}(B_{c}^{*+} \to \eta_{c}(2S)\tau^{+}\nu_{\tau})$&$0.50$&$0.39$&$0.43^{+0.11+0.01+0.09}_{-0.07-0.01-0.20}$\\
\hline\hline
Observables&Region 1&Region 2&Total&Observables&Region 1&Region 2&Total\\
\hline\hline
$f_{L}(B_{c}^{*+} \to B^{0}_{s}e^{+}\nu_{e})$&$0.65$&$0.40$&$0.53^{+0.13+0.00+0.01}_{-0.09-0.02-0.03}$&$f_{L}(B_{c}^{*+} \to B^{0}e^{+}\nu_{e})$&$0.63$&$0.39$&$0.51^{+0.13+0.03+0.04}_{-0.19-0.02-0.04}$\\
\hline
$f_{L}(B_{c}^{*+} \to B^{0}_{s}\mu^{+}\nu_{\mu})$&$0.64$&$0.40$&$0.52^{+0.13+0.00+0.01}_{-0.09-0.02-0.03}$&$f_{L}(B_{c}^{*+} \to B^{0}\mu^{+}\nu_{\mu})$&$0.62$&$0.39$&$0.51^{+0.13+0.03+0.04}_{-0.19-0.02-0.04}$\\
\hline\hline
\end{tabular}\label{FL}}
\end{center}
\end{table}
\item
The calculations showed that the longitudinal polarization fractions $f_{L}$ between the decays $B^*_{u,s,c} \to P e\nu_{e}$ and $B^*_{u,s,c} \to P\mu\nu_{\mu}$ are very close to each other, while  $f_L(B^*_{u,s,c} \to P \tau\bar{\nu}_{\tau})$ is up to $15$ \% smaller, as shown in Table \ref{FL}, that is
\be
f_{L}(B^*_{u,s,c} \to P e{\nu}_{e})\sim f_{L}(B^*_{u,s,c} \to P \mu{\nu}_{\mu})> f_{L}(B^*_{u,s,c} \to P \tau{\nu}_{\tau}),
\en
which reflects the lepton flavor universality (LFU).
To determine the dependence of the polarization on $q^2$, we calculate the longitudinal polarization fractions
by dividing the full energy region into two regions for each decay. Region 1 is defined as $m_{\ell}^{2}<q^{2}<\frac{(m_{B^*}-m_{P})^{2}+m_{\ell}^{2}}{2} $ and Region 2 is $\frac{(m_{B^*}-m_{P})^{2}+m_{\ell}^{2}}{2} <q^{2}<(m_{B^*}-m_{P})^{2}$. Interestingly, these decays $B^*_{u,s,c} \to P \ell^{+}{\nu}_{\ell}$ ,except for the channels $B_{s}^{*0} \to K^{-}\ell^{+}\nu_{\ell}$, are dominated by the transverse polarization in Region 2. Furthermore, among these decays $B^*_{u, s, c}\to P \tau^{+}{\nu}_{\tau}$, only the channel $B_{s}^{*0} \to K^{-}\tau^{+}\nu_{\tau}$ is dominated by the longitudinal polarization in the entire physical region. The special polarization for the decay $B_{s}^{*0} \to K^{-}\ell^{+}\nu_{\ell}$ can be tested in future high-luminosity experiments.
\begin{figure}[H]
\vspace{0.40cm}
  \centering
  \subfigure[]{\includegraphics[width=0.22\textwidth]{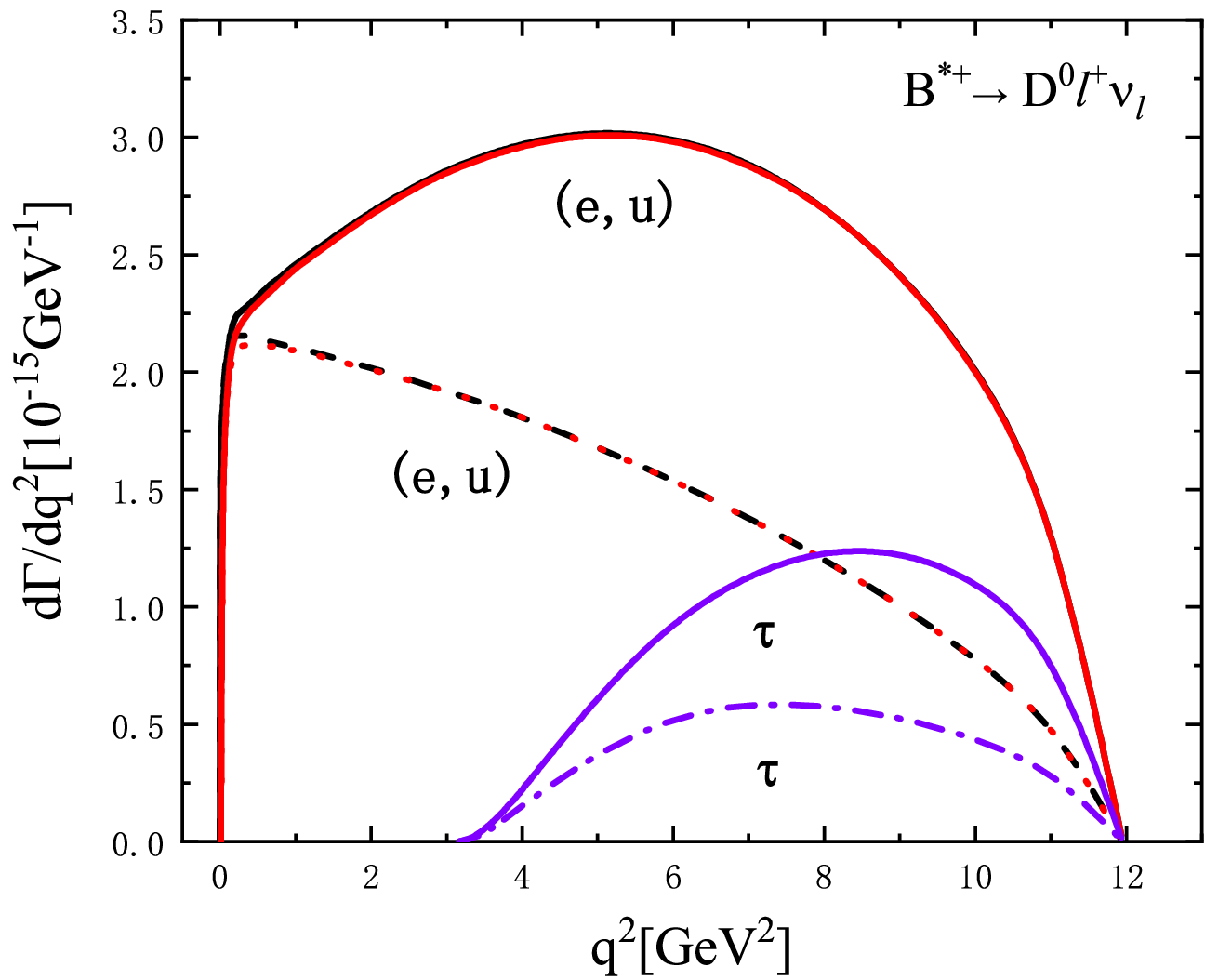}\quad}
  \subfigure[]{\includegraphics[width=0.22\textwidth]{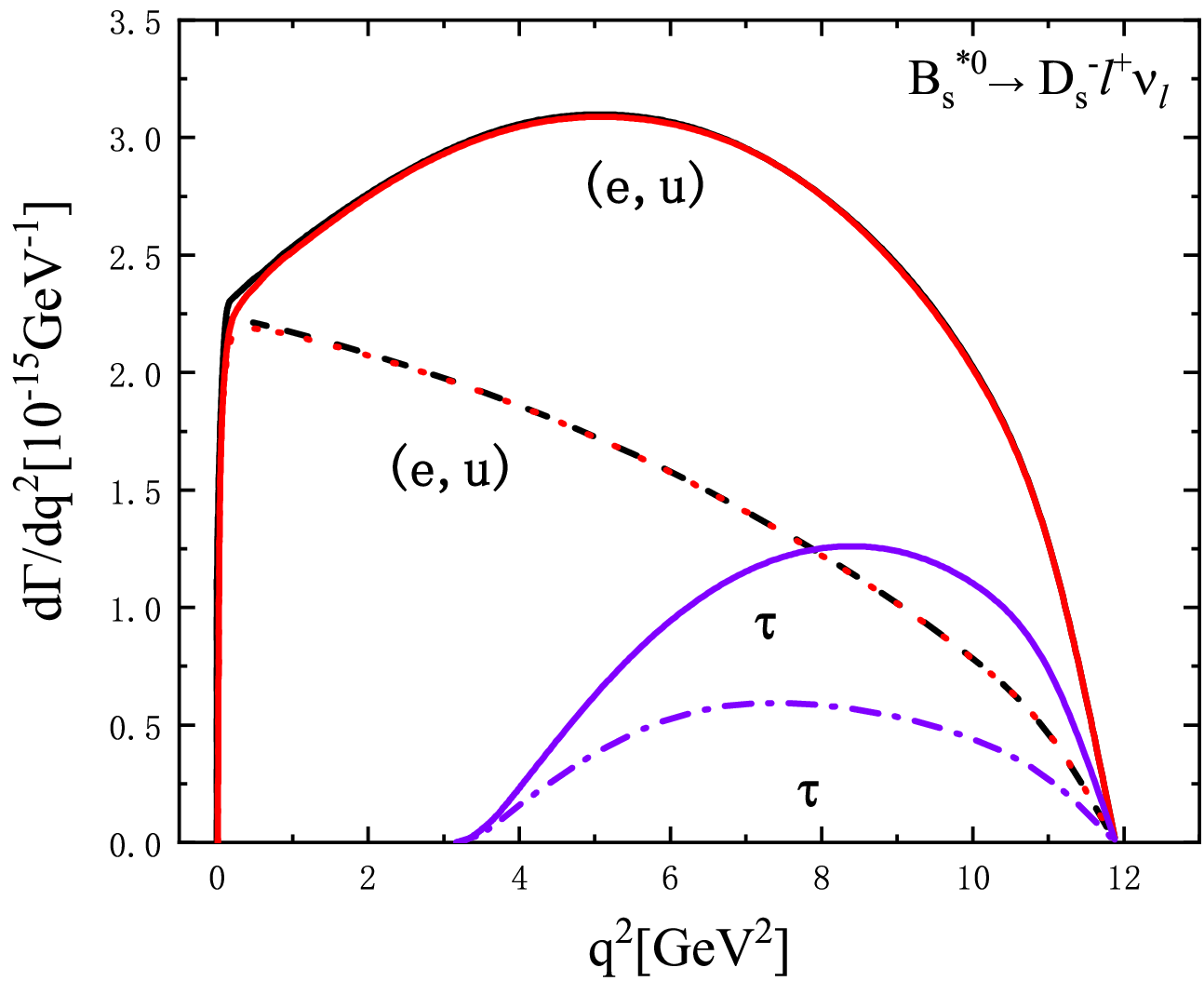}\quad}
  \subfigure[]{\includegraphics[width=0.22\textwidth]{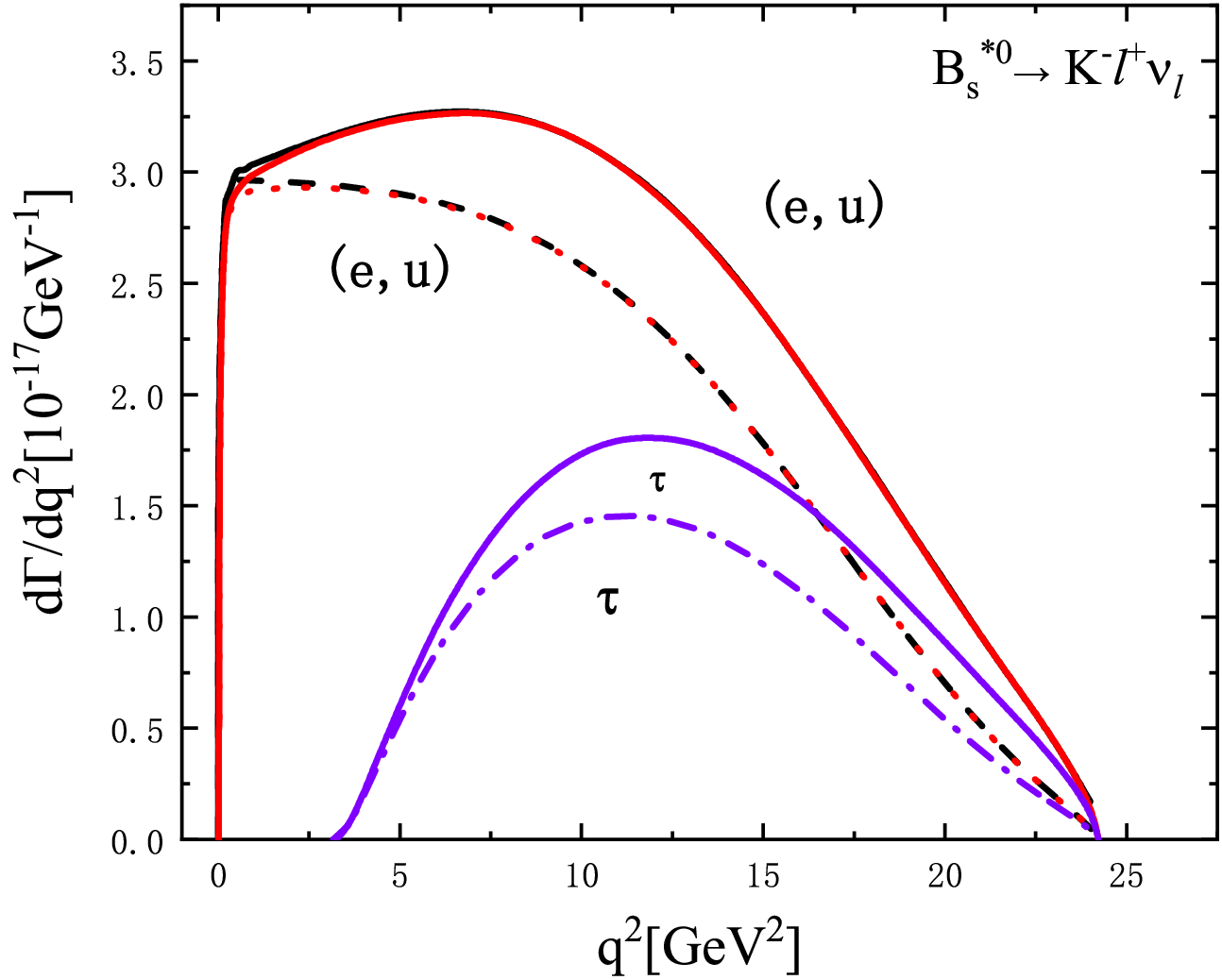}\quad}
  \subfigure[]{\includegraphics[width=0.22\textwidth]{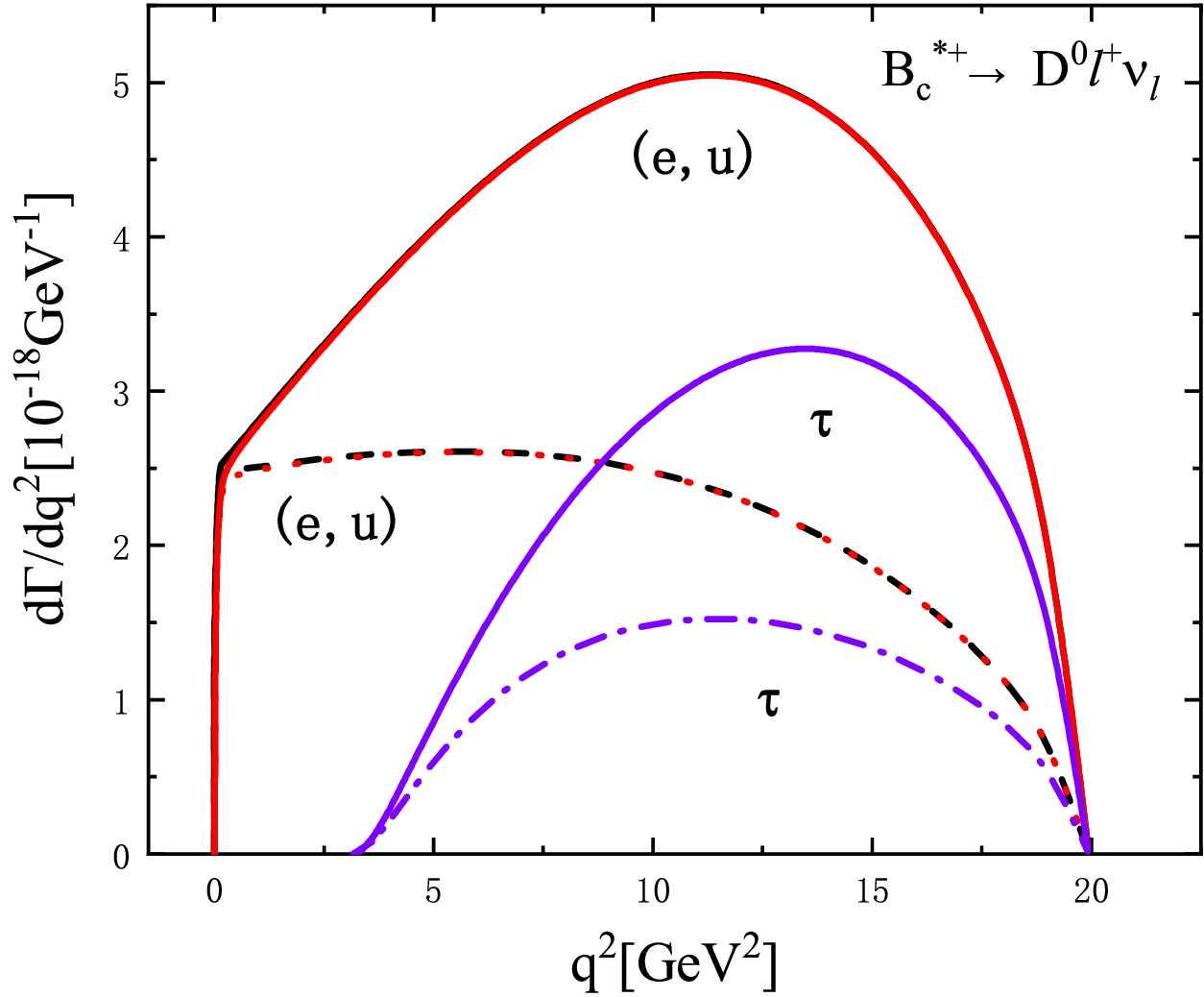}}\\
  \subfigure[]{\includegraphics[width=0.22\textwidth]{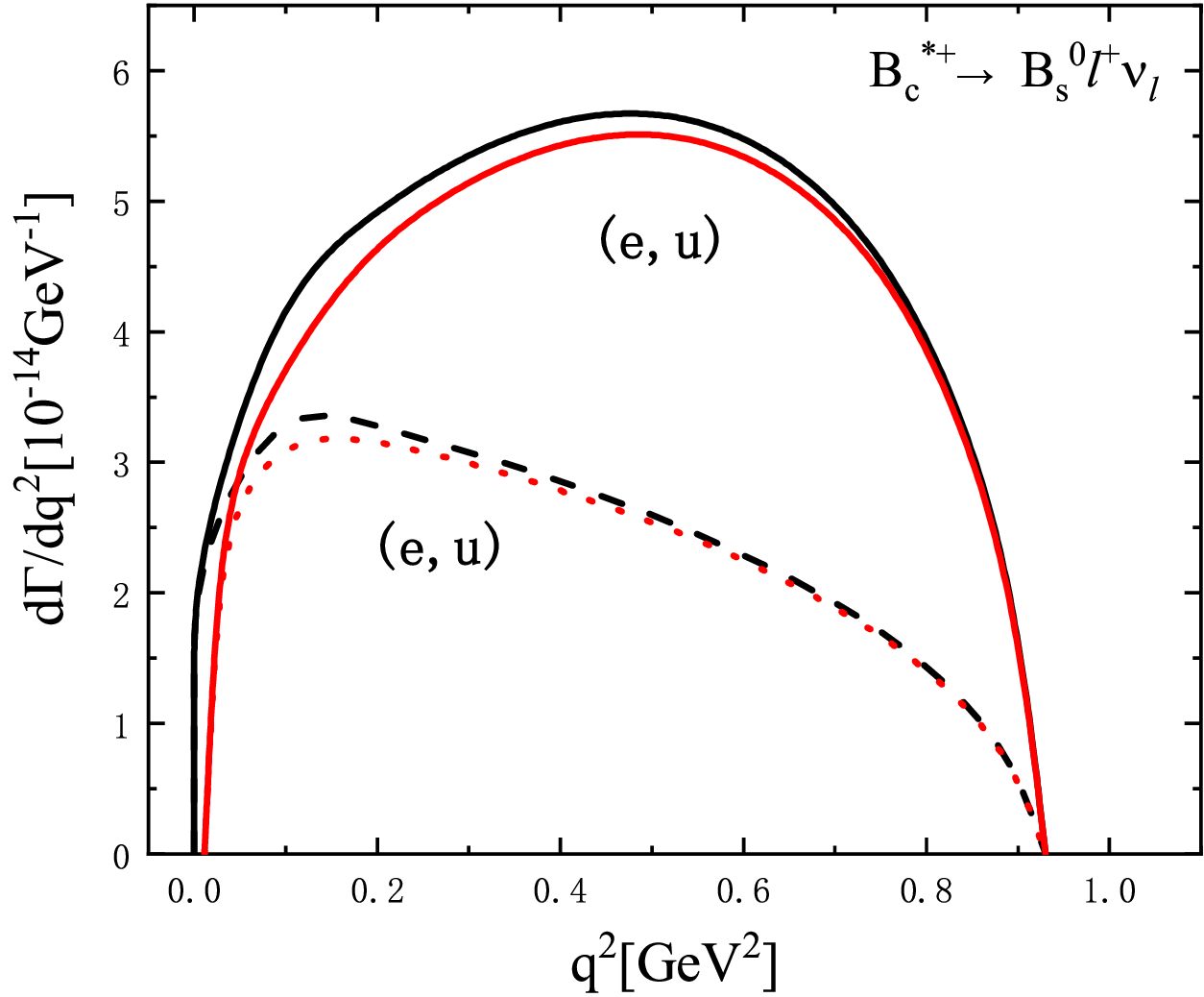}\quad}
  \subfigure[]{\includegraphics[width=0.22\textwidth]{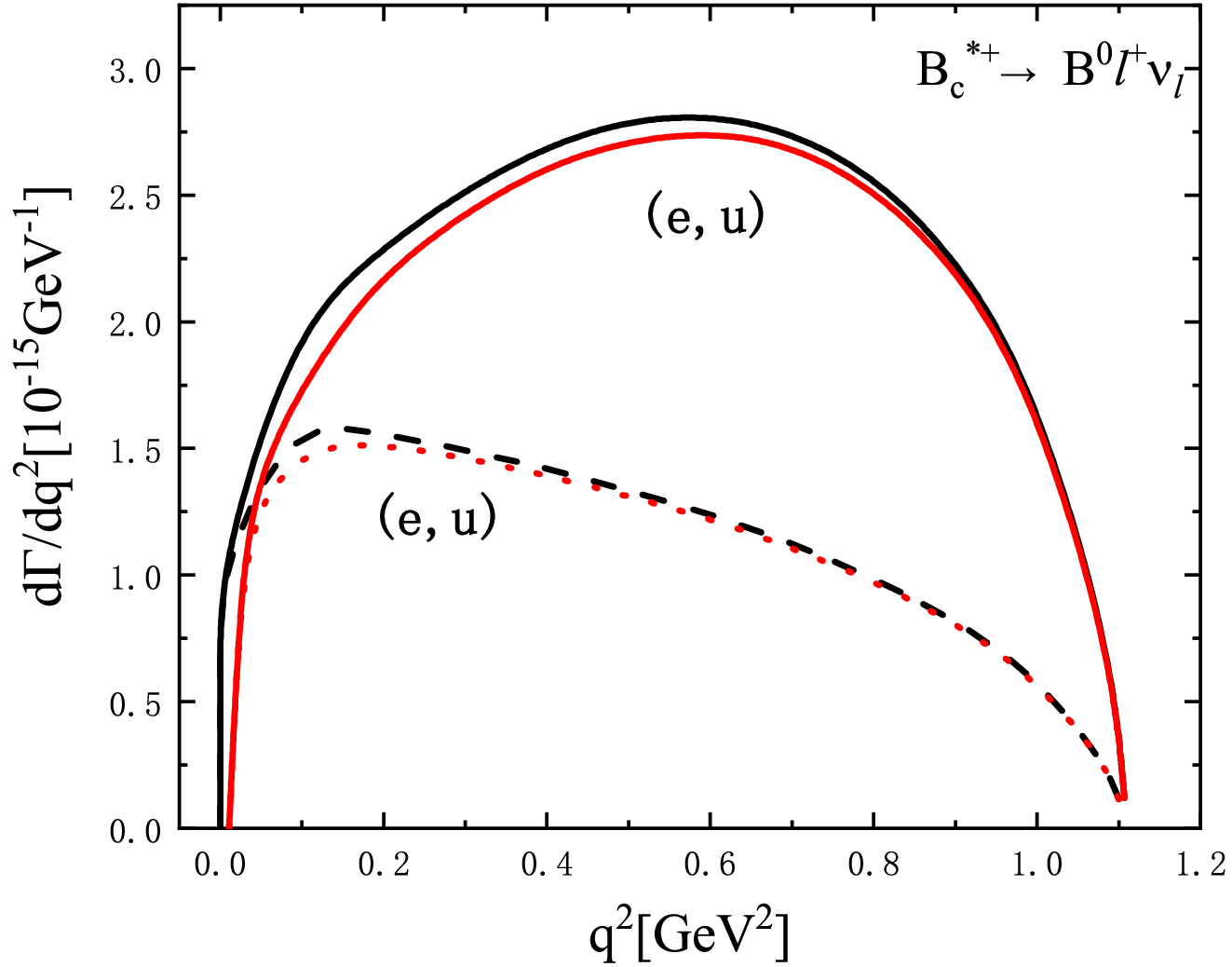}\quad}
  \subfigure[]{\includegraphics[width=0.22\textwidth]{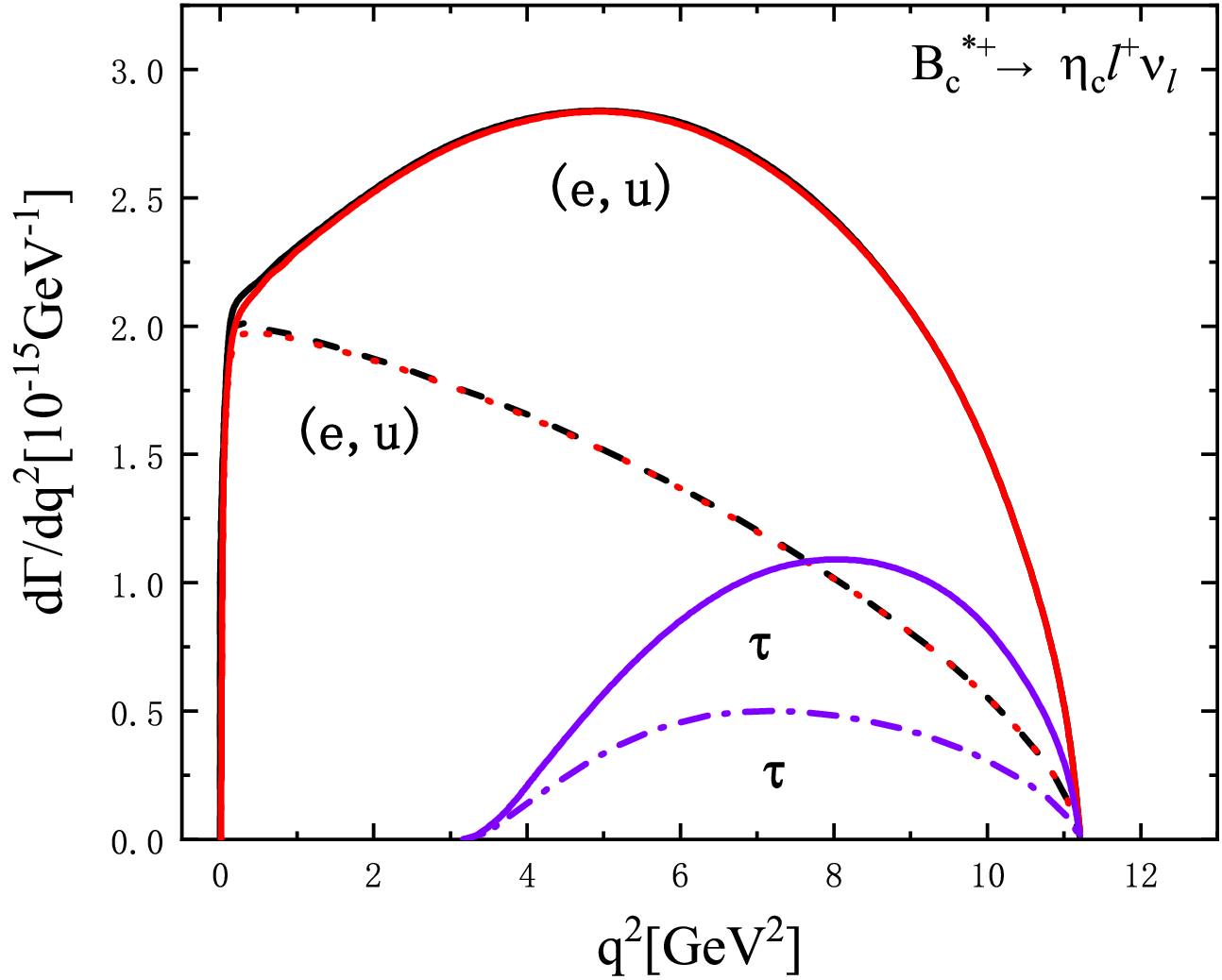}\quad}
  \subfigure[]{\includegraphics[width=0.22\textwidth]{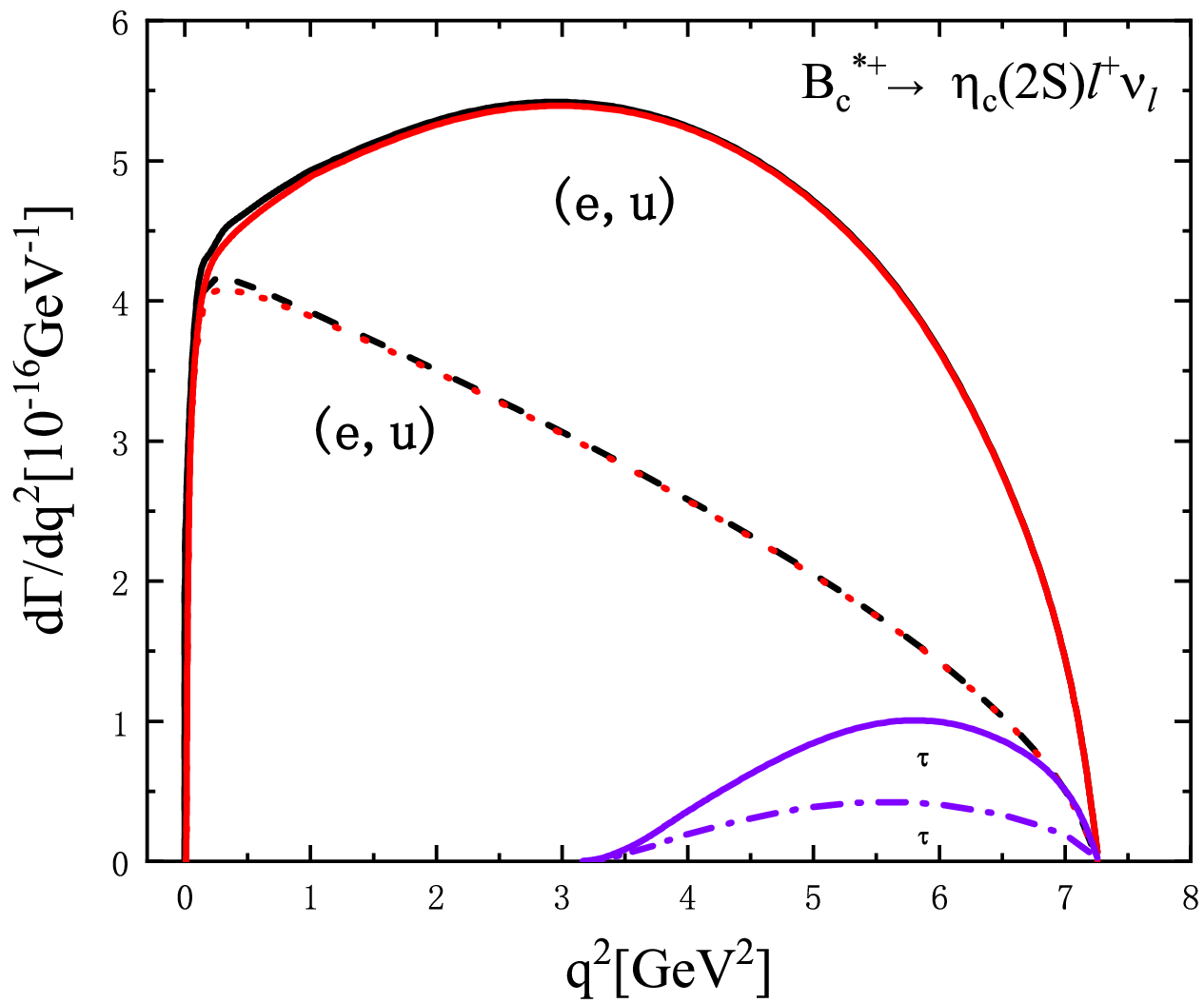}}
    \caption{$q^2$ dependencies of the differential decay rates $d\Gamma/dq^2$ (solid lines) and $d\Gamma^{L}/dq^2$ (the dashed lines refer to the decays with $e^+\nu_e$ involved, the dotted lines represent the decays with $\mu^+\nu_\mu$ involved, and  the dash-dotted lines are for the decays with $\tau^+\nu_\tau$ involved) for the decays $B^*_{u,s,c} \to P\ell^{+}\nu_{\ell}$. }\label{fig:T2}
\end{figure}
\item
In Figures \ref{fig:T2} and \ref{fig:T4}, we also display the  $q^{2}$-dependencies of the differential decay rates $d\Gamma_{(L)}/dq^{2}$ and the forward-backward asymmetries $A_{FB}$, respectively.  In Figure \ref{fig:T2}, one can find that the $q^2$ dependencies of the differential decay rates $d\Gamma/dq^2$ ($d\Gamma^{L}/dq^2$) for the decays 
$B^*\to P e\nu_e$  and $B^*\to P \mu\nu_\mu$ almost coincide with each other. The longitudinal polarizations are dominant in the decays $B_{s}^{*0} \to K^{-}\ell^{+}\nu_{\ell}$, which are obviously different from those of other decays. In terms of the forward-backward asymmetries $A_{FB}$ for the decays $B_{s}^{*0} \to K^{-}\ell^{\prime+}\nu_{\ell^\prime}$ and $B_{s}^{*0} \to K^{-}\tau^{+}\nu_{\tau}$, their extreme values are closer to each other, as shown in Figure \ref{fig:T4}. 
\end{enumerate}
\begin{figure}[H]
\vspace{0.40cm}
  \centering
  \subfigure[]{\includegraphics[width=0.22\textwidth]{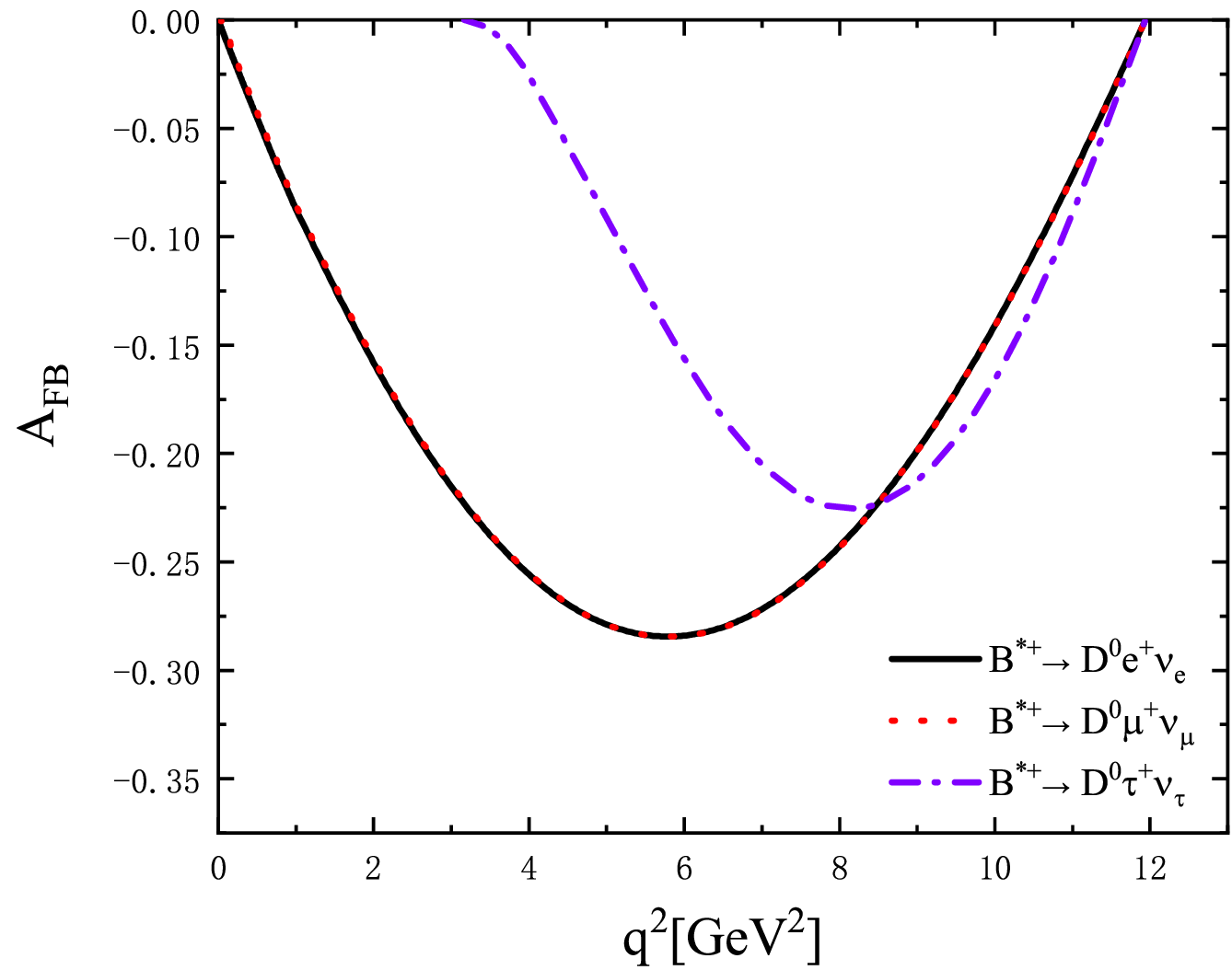}\quad}
  \subfigure[]{\includegraphics[width=0.22\textwidth]{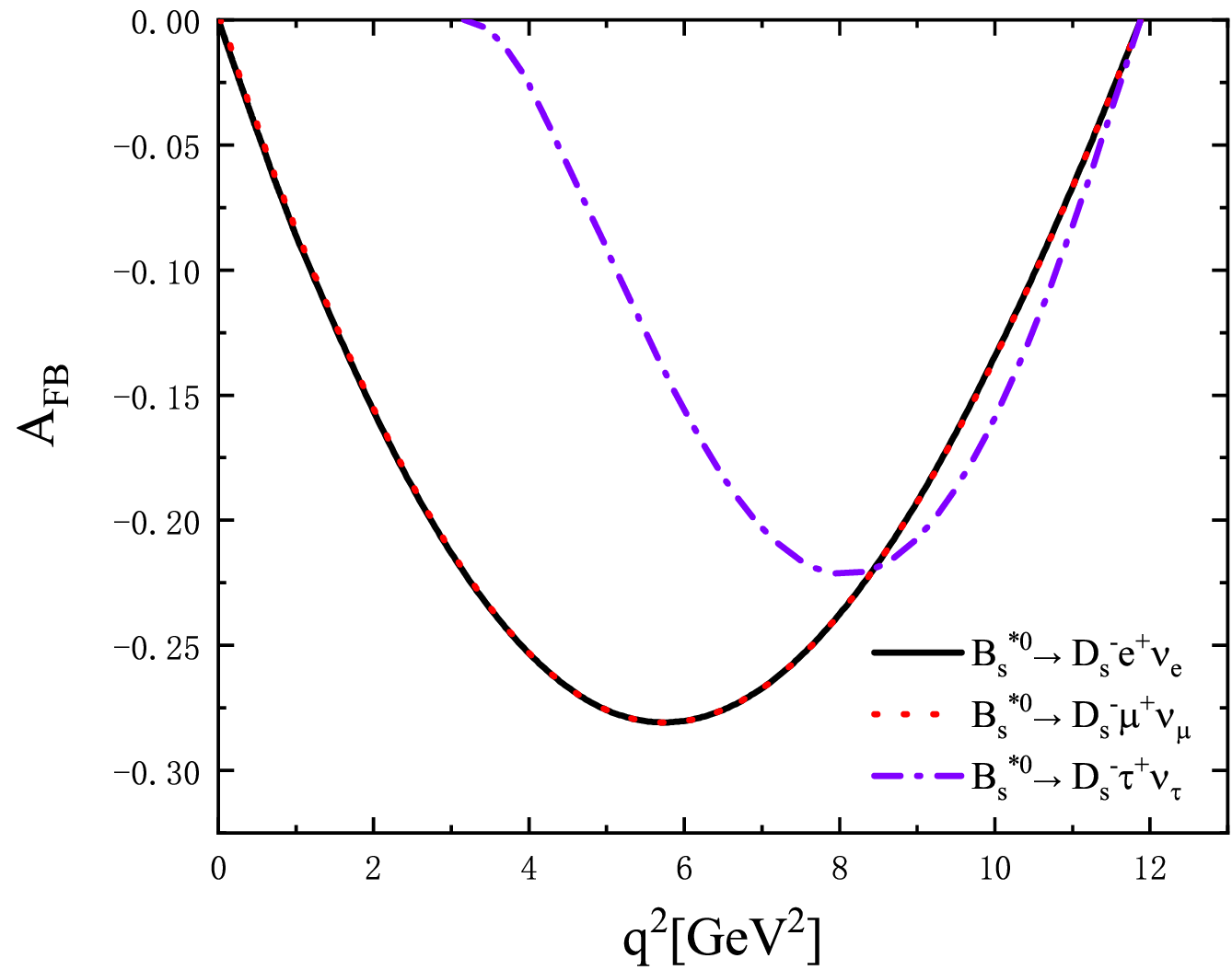}\quad}
  \subfigure[]{\includegraphics[width=0.22\textwidth]{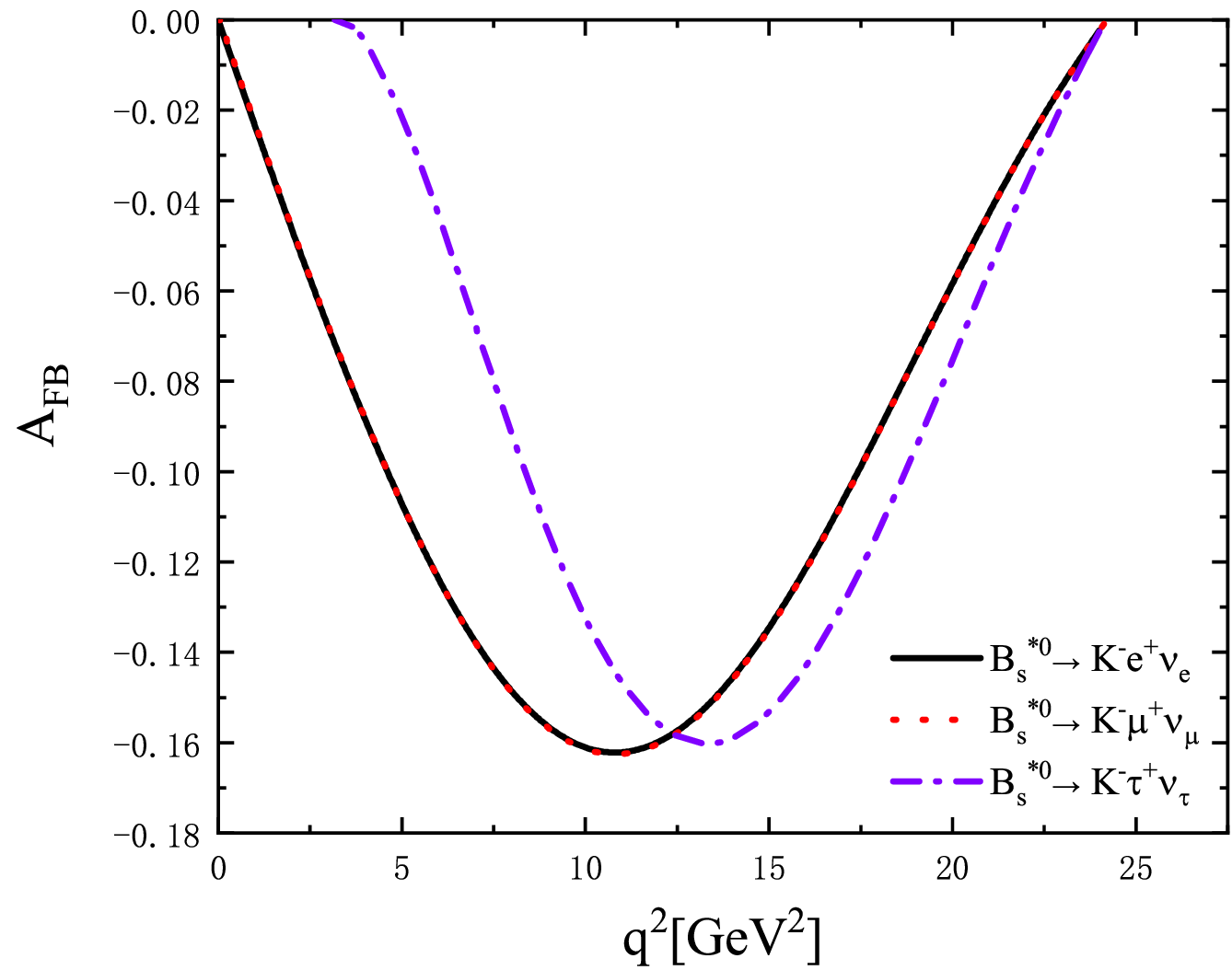}\quad}
  \subfigure[]{\includegraphics[width=0.22\textwidth]{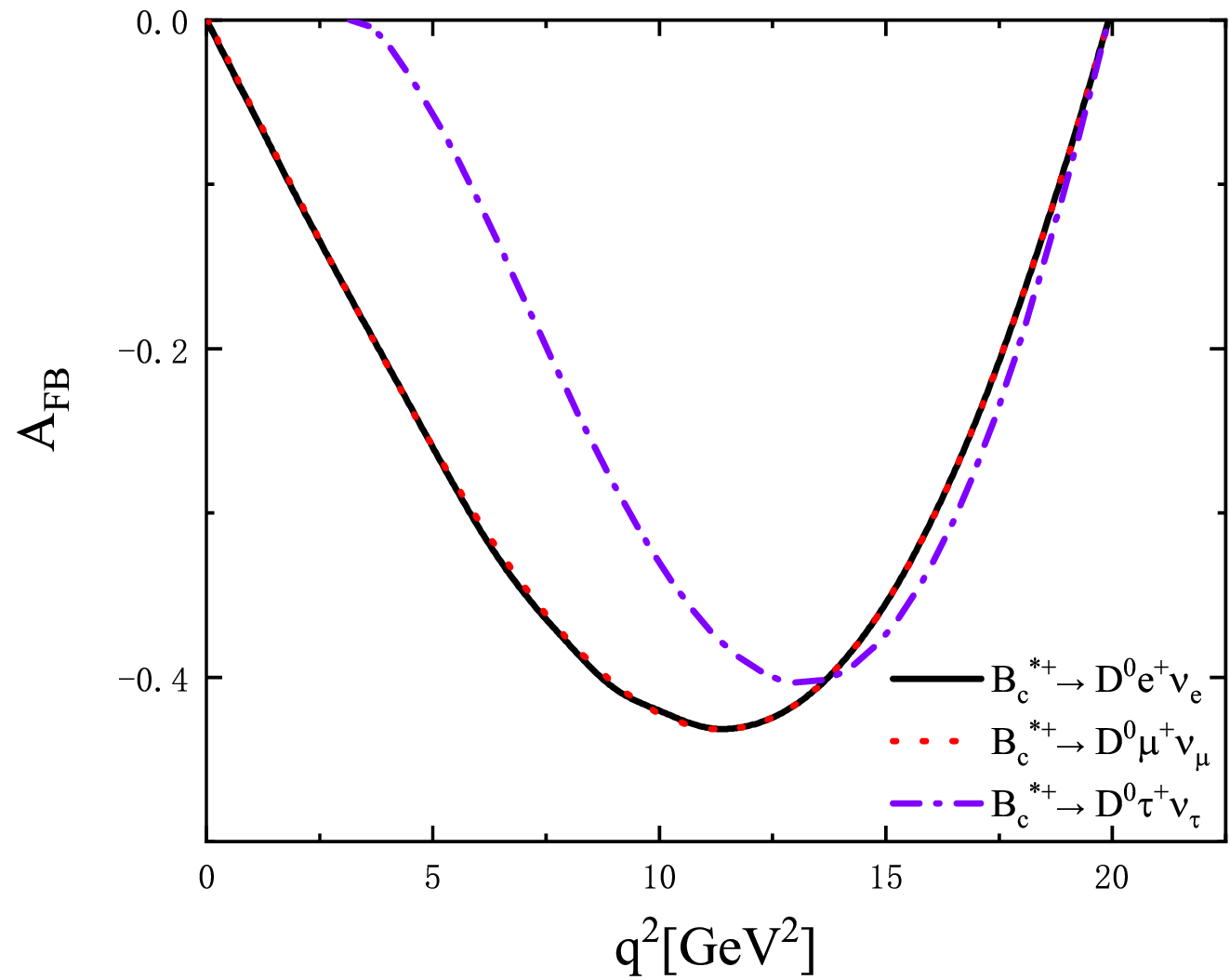}}\\
  \subfigure[]{\includegraphics[width=0.22\textwidth]{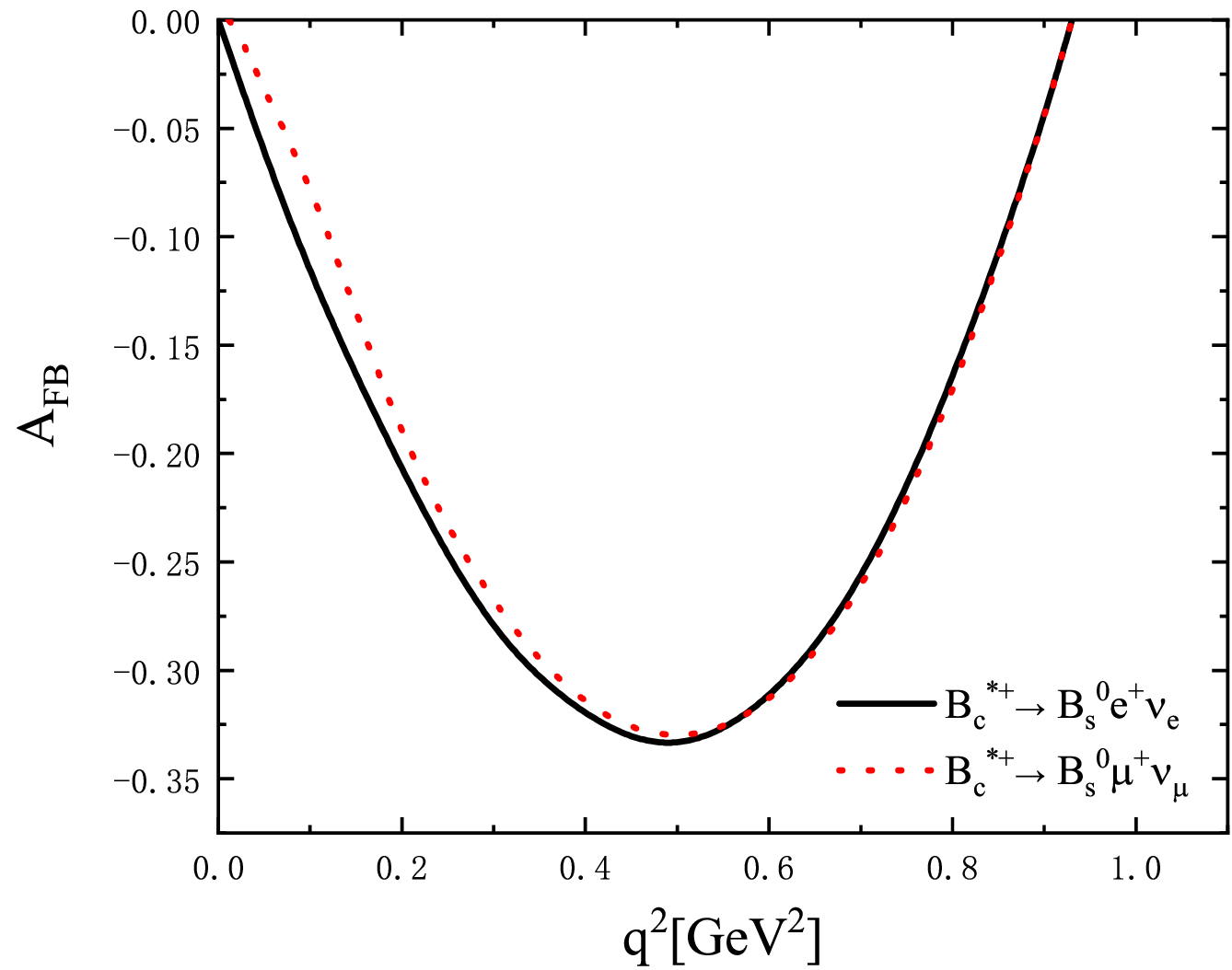}\quad}
  \subfigure[]{\includegraphics[width=0.22\textwidth]{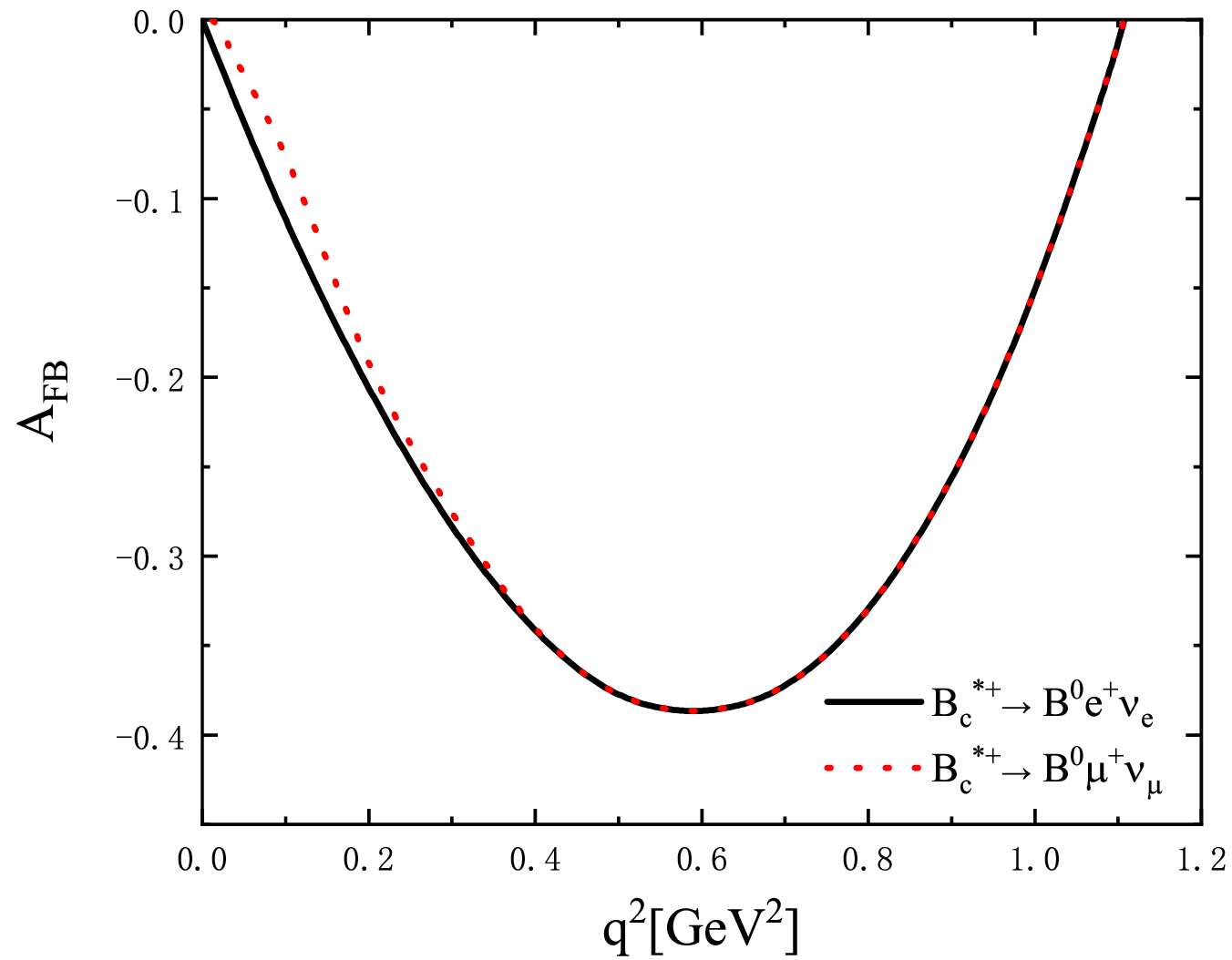}\quad}
  \subfigure[]{\includegraphics[width=0.22\textwidth]{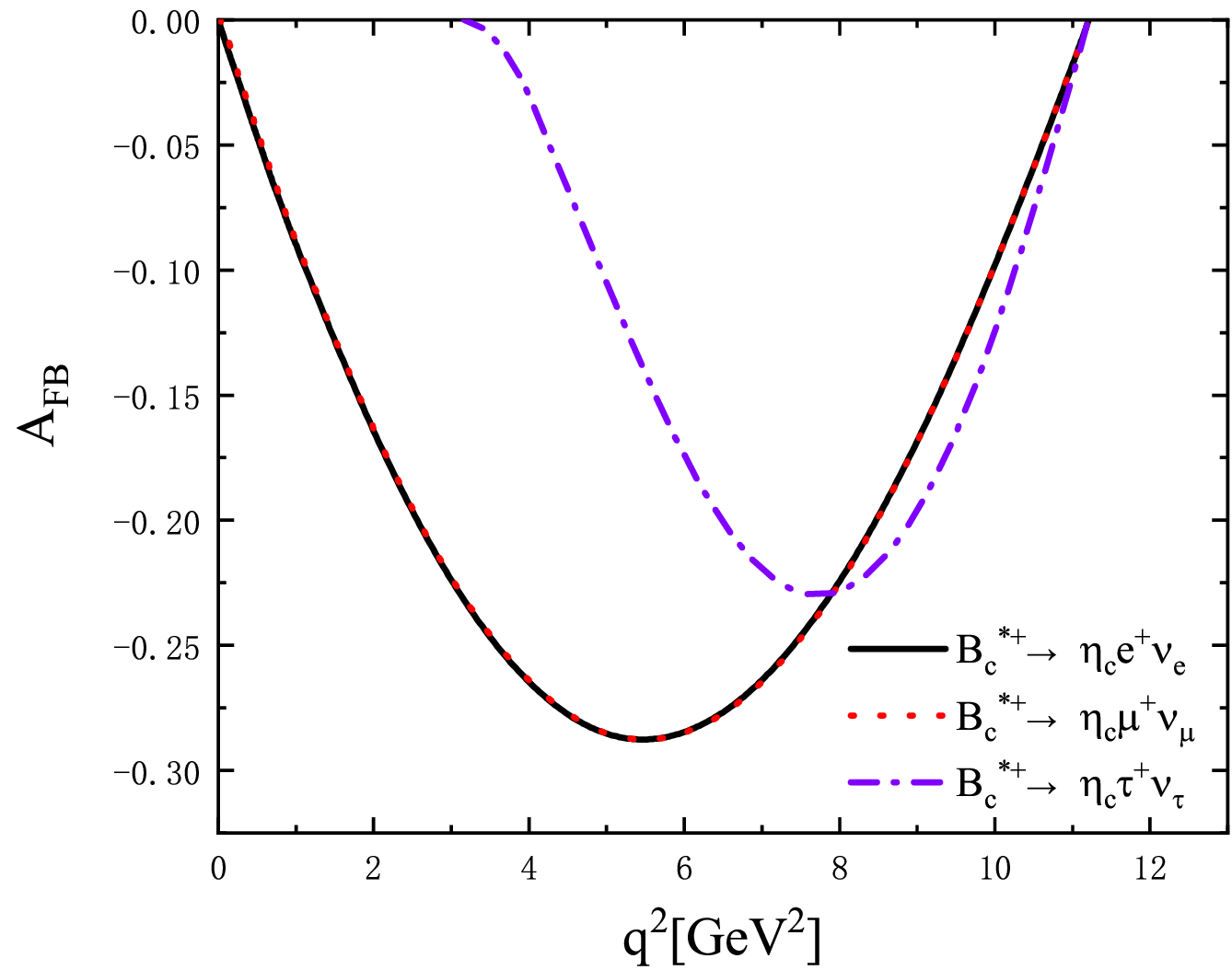}\quad}
  \subfigure[]{\includegraphics[width=0.22\textwidth]{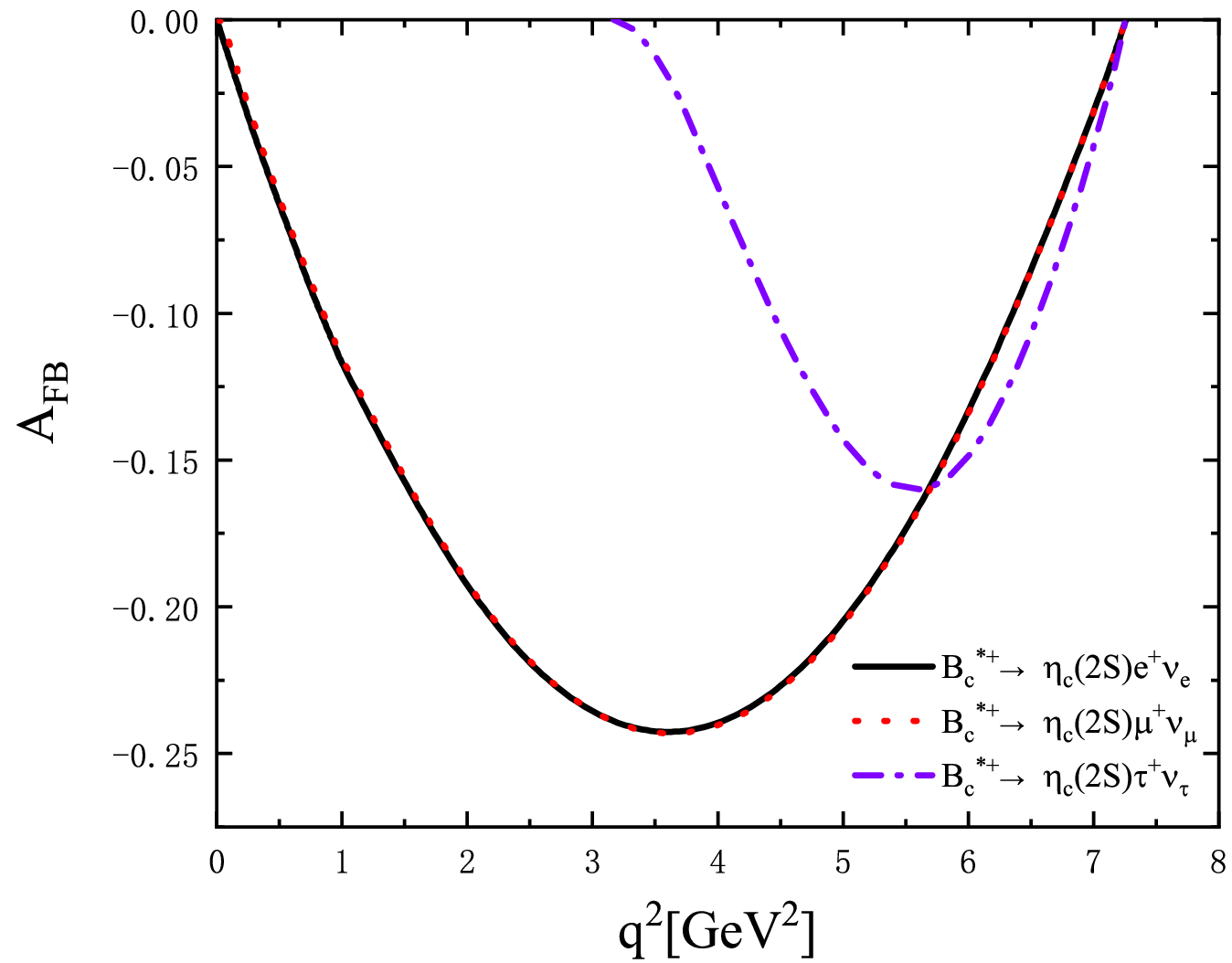}}

\caption{$q^{2}$-dependencies of the forward-backward asymmetries $A_{FB}$ for the decays $B^*_{u,s,c} \to P\ell^{+}\nu_{\ell}$.}\label{fig:T4}
\end{figure}
\subsection{Nonleptonic Decays}
The decay rates of the nonleptonic $B^{*}_{u,d,s,c}$ decays can be written as
\be
\mathcal{B} r(B^{*}\to M_{1}M_{2})=\frac{p_{\mathrm{cm}}}{24 \pi m_{ B^{*}}^{2} \Gamma_{ B^{*}}}|\mathcal{A}(B^{*}\to M_{1}M_{2})|^{2},
\en
where $p_{cm}$ represents the three-momentum of the final states and is defined as
\be
p_{cm}=\frac{\sqrt{\left[m_{B^{*}}^{2}-\left(m_{M_{1}}+m_{M_{2}}\right)^{2}\right]\left[m_{B^{*}}^{2}-
\left(m_{M_{1}}-m_{M_{2}}\right)^{2}\right]}}{2 m_{B^{*}}}.
\en
Using the input parameters given in Table \ref{tab:constant}, we can calculate the branching ratios of the nonleptonic decays $B^{*}_{u,d,s,c}\to M_{1} M_{2}$, which are listed in Tables \ref{BR1}, \ref{BR2}, and \ref{BR3} with the values obtained from Refs. \cite{Wang:2018ryc,Sun:2017lup,Yang:2022jqu,R:2019uyb,Q.L,Sun:2017lla} for comparison. The uncertainties in our results arise from the decay widths of the vector mesons $B^{*}_{u,d,s,c}$ and the decay constants of the initial and final state mesons. The decay modes considered here are dominated by the color-favored factorizable contributions and are insensitive to the nonfactorizable contributions. Numerically, we adopt the Wilson coefficients $a_{1}=1.2$ and $a_{2}=-0.5$. Further comments are detailed below:
\begin{enumerate}
\item
For  the hierarchical relations between the CKM matrix elements $|V_{ud}|>|V_{us}|$, the branching ratios of the decays $B^{*+}_{c}\to\eta_{c}\pi^{+}(\rho^{+})$ are at least an order of magnitude larger than those of the decays $B^{*+}_{c}\to\eta_{c}K^{+} (K^{*+})$. This is the similar to the cases where $\eta_c$ is replaced by $\eta_c(2S)$ in these decays.
Owing to the smaller decay constant $f_{\eta_{c}(2S)}$ compared with $f_{\eta_c}$ and the compacter phase space for the final states $\eta_{c}(2S)P(V)$ compared with the $\eta_c P(V)$,  the branching ratios of the decays $B^*_c\to \eta_c(2S)P(V)$ are smaller than those of the corresponding decays $B^*_c\to \eta_cP(V)$ by approximately a factor of 5. 
In addition, the CKM-suppressed decays $B_{c}^{*}\to \eta_{c}(1S,2S)K^{(*)}$  have relatively small branching ratios, which only reach up to the $10^{-10}\sim10^{-9}$ order. Our predictions are comparable with the naive factorization (NF) approach \cite{Yang:2022jqu}, the perturbative QCD (PQCD) approach \cite{Sun:2017lla} and the BSW model \cite{R:2019uyb,Q.L}. Our results for the decays $B_{c}^{*+} \to \eta_{c}\rho^{+}(K^{*+})$ are thus consistent with the previous LFQM calculations \cite{Q.L}, with differences attributed to the different input values.
\begin{table}[H]
	\caption{Branching ratios of the decays $B^{*+}_{c}\to \eta_c(1S,2S)P$ with $P$ referring to $\pi, K, \rho, K^*$.}
	\begin{center}
		\scalebox{0.85}{
			\begin{tabular}{|c|c|c|c|cc|c|c|}
				\hline\hline
				 & && &\;\;\;\;\;\;Ref. \cite{Q.L}&&&\\ \cline{5-6}Decay modes&This work& NF \cite{Yang:2022jqu}&PQCD \cite{Sun:2017lla}&BSW&LFQM& BSW \cite{R:2019uyb}&Unit\\
				\hline
				$\mathcal{B}r(B_{c}^{*+} \to \eta_{c}  \pi^{+})$&$1.14^{+0.29+0.06+0.03}_{-0.19-0.08-0.07}$&$1.5$&$2.2$&$-$&$-$&$0.77$&$10^{-8}$\\
				$\mathcal{B}r(B_{c}^{*+} \to \eta_{c} K^{+})$&$8.69^{+2.21+0.46+0.25}_{-1.47-0.59-0.57}$&$11$&$17$&$-$&$-$&$-$&$10^{-10}$\\
				$\mathcal{B}r(B_{c}^{*+} \to \eta_{c}  \rho^{+})$&$3.21^{+0.82+0.16+0.12}_{-0.54-0.21-0.23}$&$4.3$&$-$&$3.02$&$2.45$&$7.72$&$10^{-8}$\\
				$\mathcal{B}r(B_{c}^{*+} \to \eta_{c} K^{*+})$&$1.93^{+0.49+0.10+0.07}_{-0.33-0.12-0.14}$&$2.3$&$-$&$1.7$&$1.4$&$4.18$&$10^{-9}$\\
				\hline
				$\mathcal{B}r(B_{c}^{*+} \to \eta_{c}(2S) \pi^{+})$&$2.32^{+0.59+0.04+0.67}_{-0.39-0.05-0.89}$&$4.1$&$2.4$&$-$&$-$&$-$&$10^{-9}$\\
				$\mathcal{B}r(B_{c}^{*+} \to \eta_{c}(2S) K^{+})$&$1.75^{+0.45+0.03+0.52}_{-0.29-0.04-0.67}$&$2.9$&$3.4$&$-$&$-$&$-$&$10^{-10}$\\
				$\mathcal{B}r(B_{c}^{*+} \to \eta_{c}(2S) \rho^{+})$&$6.91^{+1.76+0.09+0.17}_{-1.17-0.13-0.29}$&$12$&$-$&$-$&$-$&$-$&$10^{-9}$\\
				$\mathcal{B}r(B_{c}^{*+} \to \eta_{c}(2S) K^{*+})$&$3.89^{+0.99+0.05+0.14}_{-0.66-0.07-0.26}$&$6.1$&$-$&$-$&$-$&$-$&$10^{-10}$\\
				\hline\hline
			\end{tabular}\label{BR3}
		}
	\end{center}
\end{table}
\item
Owing to the large CKM factors $V_{ud}V_{cs}=0.95$, the Cabibbo-favored decays $B_{c}^{*+}\to B_{s}\pi^+(\rho^+)$ have the largest branching ratios, which can reach up to the $10^{-7}$ order, especially for the decay$B_{c}^{*+}\to B_{s} \rho^+$, which has a branching ratio that is very close to the $10^{-6}$ order. These two decays might be observable in future LHCb and Super KEKB experiments and can serve as golden channels to search for the $B^*_c$ meson. However, the decays $B_{c}^{*}\to B\pi(\rho), B_{s}K^{(*)}$ have relatively smaller branching ratios, which are only of the order of $10^{-8}$ owing to the smaller CKM factors $V_{ud}V_{cd}\approx V_{us}V_{cs}\approx0.22$. The Cabibbo-double-suppressed decays $B_{c}^{*}\to BK^{(*)}$ have the smallest branching ratios within the range of  $10^{-10}\sim10^{-9}$ for the smallest CKM factors $V_{us}V_{cd}\approx0.050$. A clear hierarchical relationship between their branching ratios can be obtained as follows: 
\begin{footnotesize}
	\begin{equation}
		\begin{aligned}
			\mathcal{B}r(B_{c}^{*+} \to B_{s} \pi(\rho^{+}))\gg \mathcal{B}r(B_{c}^{*+} \to B_{s} K^{(*)+})\sim 
			\mathcal{B}r(B_{c}^{*+} \to B \pi^+(\rho^{+}))\gg \mathcal{B}r(B_{c}^{*+} \to B K^{(*)+}).\\
		\end{aligned}
	\end{equation}
\end{footnotesize}
Furthermore, three partial polarization amplitudes exist for the decays $B^{*}_c\to B_{(s)}\rho(K^*)$, where the transverse polarizations are dominant and will promote the branching ratios; however, only the p-wave amplitude is involved for the decays $B^{*}_c\to B_{(s)}\pi(K)$. Therefore, the branching ratios of the decays $B^{*}_c\to B_{(s)}\rho(K^*)$ are approximately $3\sim 4$ times larger than those of the corresponding decays $B^{*}_c\to B_{(s)}\pi(K)$. This is similar to the cases with $B_{(s)}$ replaced by $\eta_c(1S,2S)$ in these decays; $\mathcal{B}r(B_{c}^{*} \to \eta_{c}(1S,2S)\rho(K^*))$ are approximately $2\sim3$ times larger than $\mathcal{B}r(B_{c}^{*} \to \eta_{c}(1S,2S)\pi(K))$.
\begin{table}[H]
	\caption{Branching ratios of the decays $B^{*}_{c} \to B_{(s)}\pi(K)$ and  $B^{*}_{c} \to B_{(s)}\rho(K^*)$.}
	\begin{center}
		\scalebox{0.85}{
			\begin{tabular}{|c|c|c|ccc|c|}
				\hline\hline
				&\;\; \;\;& &\;\;\;\;\;\;\;\;\;\;\;\;\;\;\;\;Ref. \cite{Sun:2017lup}&$$&$$&Unit \\ \cline{4-6} Decay modes&This work&  Ref. \cite{Yang:2022jqu}&$\omega=0.4$GeV &$\omega=0.6$ GeV &$\omega=m\alpha_{s}$&\\
				\hline
				$B_{c}^{*+} \to B \pi^{+}$&$1.12^{+0.29+0.07+0.03}_{-0.19-0.09-0.04}$&$2.1$&$2.14$&$4.50$&$6.53$&$10^{-8}$\\
				$B_{c}^{*+} \to B K^{+}$&$6.28^{+1.60+0.38+0.02}_{-1.06-0.52-0.21}$&$11$&$11.5$&$24.1$&$35.0$&$10^{-10}$\\
				$B_{c}^{*+} \to B \rho^{+}$&$4.76^{+1.22+0.03+0.13}_{-0.80-0.09-0.26}$&$9.7$&$3.5$&$7.5$&$11.5$&$10^{-8}$\\
				$B_{c}^{*+} \to B K^{*+}$&$2.47^{+0.63+0.01+0.08}_{-0.42-0.04-0.14}$&$4.4$&$1.45$&$3.13$&$4.84$&$10^{-9}$\\
				$B_{c}^{*+} \to B_{s} \pi^{+}$&$2.34^{+0.60+0.13+0.08}_{-0.39-0.18-0.01}$&$4.4$&$4.00$&$7.26$&$9.82$&$10^{-7}$\\
				$B_{c}^{*+} \to B_{s} K^{+}$&$1.21^{+0.31+0.07+0.04}_{-0.20-0.09-0.01}$&$2.2$&$1.98$&$3.60$&$4.87$&$10^{-8}$\\
				$B_{c}^{*+} \to B_{s} \rho^{+}$&$9.36^{+2.39+0.04+0.01}_{-1.58-0.18-0.23}$&$19$&$6.3$&$11.7$&$16.8$&$10^{-7}$\\
				$B_{c}^{*+} \to B_{s} K^{*+}$&$3.99^{+1.02+0.01+0.01}_{-0.67-0.06-0.11}$&$7.2$&$2.21$&$4.12$&$6.01$&$10^{-8}$\\
				\hline\hline
			\end{tabular}\label{BR2}
		}
	\end{center}
\end{table}
\item
We provide the following SU(3) symmetry breaking relations $R^{K^*/\rho}_M$ and $R^{K/\pi}_M$, with $M$ being $B_{(s)},\eta_c(1S,2S)$.
\begin{tiny}
\begin{equation}
	\begin{aligned}
R^{K^*/\rho}_{B_{s}}&\equiv\frac{\mathcal{B}r(B_{c}^{*+} \to B_{s} K^{*+})}{\mathcal{B}r(B_{c}^{*+} \to B_{s} \rho^{+})}=0.042\pm 0.015,\;\;\;\;\;\;R^{K/\pi}_{B_{s}}\equiv\frac{\mathcal{B}r(B_{c}^{*+} \to B_{s} K^{+})}{\mathcal{B}r(B_{c}^{*+} \to B_{s} \pi^{+})}=0.052\pm0.019,\\
R^{K^*/\rho}_{B}&\equiv \frac{\mathcal{B}r(B_{c}^{*+} \to B K^{*+})}{\mathcal{B}r(B_{c}^{*+} \to B \rho^{+})}=0.052\pm0.018,\;\;\;\;\;\;R^{K/\pi}_{B}\equiv\frac{\mathcal{B}r(B_{c}^{*+} \to B K^{+})}{\mathcal{B}r(B_{c}^{*+} \to B \pi^{+})}=0.056\pm0.020,\\
R^{K^*/\rho}_{\eta_{c}}&\equiv\frac{\mathcal{B}r(B_{c}^{*+} \to \eta_{c} K^{*+})}{\mathcal{B}r(B_{c}^{*+} \to \eta_{c}  \rho^{+})}=0.060\pm0.015 ,\;\;\;\;\;\;R^{K/\pi}_{\eta_{c}}\equiv\frac{\mathcal{B}r(B_{c}^{*+} \to \eta_{c} K^{+})}{\mathcal{B}r(B_{c}^{*+} \to \eta_{c}  \pi^{+})}=0.076\pm0.027,\\
R^{K^*/\rho}_{\eta_{c}(2S)}&\equiv\frac{\mathcal{B}r(B_{c}^{*+} \to \eta_{c}(2S)  K^{*+})}{\mathcal{B}r(B_{c}^{*+} \to \eta_{c}(2S)  \rho^{+})}=0.056\pm0.020 ,\;\;\;\;\;\;R^{K/\pi}_{\eta_{c}(2S)}\equiv\frac{\mathcal{B}r(B_{c}^{*+} \to \eta_{c}(2S) K^{+})}{\mathcal{B}r(B_{c}^{*+} \to \eta_{c}(2S)  \pi^{+})}=0.075\pm0.027,
\end{aligned}
\end{equation}
\end{tiny}
which are consistent with the estimations $R^{K^*/\rho}_M=\left|\dfrac{V_{us}f_{K^{*}}}{V_{ud}f_{\rho}}\right|^{2}\approx0.057$ and $R^{K/\pi}_M=\left|\dfrac{V_{us}f_{K}}{V_{ud}f_{\pi}}\right|^{2}\approx0.078$ obtained from the factorization assumption within errors. Furthermore, these ratios $R^{K^*/\rho}_{B_{(s)}}$ and $R^{K/\pi}_{B_{(s)}}$are also consistent with the results given by the naive factorization approach \cite{Yang:2022jqu}, the WSB model \cite{Sun:2017lup, Q.L}, and the PQCD approach \cite{Sun:2017lla}.
\begin{table}[H]
	\caption{Branching ratios of the decays $B^{*}_{(s)} \to \pi(K) D_{(s)}, \rho(K^*)D_{(s)}$.}
	\begin{center}
		\scalebox{1}{
			\begin{tabular}{|c|c|c|c|c|c|c|}
				\hline\hline
				Channel&\;\;This work & \cite{Chang:2015jla}&\cite{Chang:2016eto}&Unit \\
				\hline
				$B^{*+} \to \pi^{+} {D}^{0}$&$3.14^{+0.82+0.06+0.12}_{-0.54-0.08-0.10}$&$[0.5,3.2]$&$-$&$10^{-14}$\\
				$B^{*+} \to \pi^{+} \bar D^{0}$      &$0.76^{+0.20+0.03+0.08}_{-0.13-0.09-0.09}$&$[0.6,3.9]$&$-$&$10^{-9}$\\
				$B^{*0} \to \pi^{-} D^{+}$&$0.86^{+0.20+0.02+0.03}_{-0.14-0.02-0.03}$&$[0.4,3.0]$&$-$&$10^{-12}$\\
				$B^{*0} \to \pi^{-} D_{s}^{+}$&$2.46^{+0.57+0.05+0.10}_{-0.39-0.07-0.08}$&$[1.3,8.4]$&$-$&$10^{-11}$\\
				$B^{*0} \to \pi^{+} D^{-}$&$6.36^{+1.47+0.05+0.03}_{-1.00-0.09-0.05}$&$[2,13]$&$-$&$10^{-9}$\\
				$B^{*0} \to K^{+} D^{-}$&$4.81^{+1.11+0.04+0.02}_{-0.76-0.07-0.04}$&$[1.5,9.8]$&$-$&$10^{-10}$\\
				$B^{*0} \to \rho^{+} D^{-}$&$1.77^{+0.41+0.01+0.01}_{-0.28-0.03-0.01}$&$-$&$1.51$&$10^{-8}$\\
				$B^{*0} \to K^{*+} D^{-}$&$1.05^{+0.24+0.02+0.01}_{-0.17-0.01-0.01}$&$-$&$0.87$&$10^{-9}$\\
				\hline
				$B_{s}^{*0} \to K^{-} D^{+}$&$12.58^{+2.98+0.34+0.14}_{-2.02-0.38-0.14}$&$[3,21]$&$-$&$10^{-12}$\\
				$B_{s}^{*0} \to \pi^{+} D_{s}^{-}$&$7.85^{+1.86+0.06+0.14}_{-1.26-0.11-0.19}$&$-$&$-$&$10^{-9}$\\
				$B_{s}^{*0} \to K^{+} D_{s}^{-}$&$5.92^{+1.40+0.05+0.11}_{-0.95-0.08-0.15}$&$[1.4,8.7]$&$-$&$10^{-10}$\\
				$B_{s}^{*0} \to K^{-} D_{s}^{+}$&$3.62^{+0.86+0.09+0.04}_{-0.58-0.11-0.04}$&$[0.9,5.9]$&$-$&$10^{-11}$\\
				$B_{s}^{*0} \to \rho^{+} D_{s}^{-}$&$2.18^{+0.52+0.02+0.05}_{-0.35-0.03-0.06}$&$-$&$2.89$&$10^{-8}$\\
				$B_{s}^{*0} \to K^{*-} D_{s}^{-}$&$1.30^{+0.31+0.01+0.03}_{-0.21-0.02-0.04}$&$-$&$1.66$&$10^{-9}$\\
				\hline\hline
			\end{tabular}\label{BR1}
		}
	\end{center}
\end{table}
\item 
In Ref. \cite{Chang:2015jla}, a range of the total decay width for the initial meson $B^{*}_{(s)}$ was used in the numerical calculations, so the ranges of the branching ratios  were obtained, as listed in Table \ref{BR1}. Our predictions all fall within the ranges given in the BSW model \cite{Chang:2015jla}. In addition, for the branching ratios of the decays $B_{(s)}^{*} \to \rho D_{(s)}, K^{*} D_{(s)}$, our results are consistent with the QCDF calculations \cite{Chang:2016eto} within errors.
In these decays, the channels $B_{(s)}^{*} \to \rho^+ D_{(s)}^-$ have the largest branching ratios, which can reach up to the $10^{-8}$ order and may be observed in future high-luminosity LHC experiments. 
\item 
 We present the results of the polarization fractions in Table \ref{BR10}. The decays $B_{c}^{*+} \to \eta_{c}(1S,2S)  \rho^{+}, \eta_{c}(1S,2S) K^{*+}$ and $B^{*0}_{(s)} \to \rho^{+} D^{-}_{(s)}, K^{*+} D^{-}_{(s)}$ are clearly dominated by the longitudinal polarization, while the channels $B_{c}^{*+} \to B_{(s)} \rho^{+}, B_{(s)} K^{*+}$ are clearly dominated by the transverse polarizations.
 \begin{table}[H]
	\caption{Polarization fractions $f_L, f_{\|}$ for the decays $B^*\to P\rho(K^*)$ with P referring to $\eta_c(1S,2S), B_{(s)}, D_{(s)}$.}
	\begin{center}
		\scalebox{1}{
			\begin{tabular}{|c|c|c|c|c|}
				\hline\hline
				Channels&$B_{c}^{*+} \to \eta_{c}  \rho^{+}$&$B_{c}^{*+} \to \eta_{c} K^{*+}$&$B_{c}^{*+} \to \eta_{c}(2S) \rho^{+}$ &$B_{c}^{*+} \to \eta_{c}(2S) K^{*+}$ \\
				\hline\hline
				$f_{L}[\%]$&$86.60$&$83.07$&$81.64$&$77.19$\\
				BSW \cite{Q.L}&$88.0$&$84.6$&$-$&$-$\\
				LFQM \cite{Q.L}&$86.6$&$83.1$&$-$&$-$\\
				$f_{\|}[\%]$&$10.58$&$13.42$&$15.52$&$19.37$\\
				BSW \cite{Q.L}&$10.0$&$12.7$&$-$&$-$\\
				LFQM \cite{Q.L}&$10.2$&$13.0$&$-$&$-$\\
				\hline\hline
				Channel&$B_{c}^{*+} \to B^0 \rho^{+}$&$B_{c}^{*+} \to B^0 K^{*+}$&$B_{c}^{*+} \to B^0_{s} \rho^{+}$ &$B_{c}^{*+} \to B^0_{s} K^{*+}$ \\
				\hline\hline
				$f_{L}[\%]$&$42.49$&$37.69$&$40.30$&$35.56$\\
				$f_{\|}[\%]$&$47.43$&$55.12$&$53.43$&$61.49$\\
				\hline\hline
				Channels&$B^{*0} \to D^{+}\rho^{-} $&$B^{*0} \to D^{-}K^{*+} $&$B_{s}^{*0} \to D_{s}^{+}\rho^{-} $ &$B_{s}^{*0} \to D_{s}^{-}K^{*+} $ \\
				\hline\hline
				$f_{L}[\%]$&$87.65$&$84.33$&$87.62$&$82.29$\\
				$f_{\|}[\%]$&$9.67$&$12.33$&$9.73$&$12.40$\\
				\hline\hline
			\end{tabular}\label{BR10}
		}
	\end{center}
\end{table}
\end{enumerate}
\section{Summary}\label{sum}
In this study, we used the covariant light-front method to comprehensively investigate the semileptonic and nonleptonic rare weak decays of the vector heavy mesons $B^{*}_{u,d,s,c}$ to provide an important reference for future experiments. Here, the transition form factors $B^*\to D,\pi$, $B^*_s\to D_s, K$, and $B^*_c\to B_{(s)},D,\eta_c(1S,2S)$ were calculated, as they play a crucial role in our considered decays. Notably, the total decay widths of these b-flavored vector mesons were estimated by the partial widths of their corresponding single-photon decay channels. The following are some notable findings:
\begin{enumerate}
\item 
In these semileptonic decays, the channels $B_{s}^{*0}\to D_{s}^{-}\ell^{\prime+}{\nu}_{\ell^\prime}$ and $B_{c}^{*+}\to B_{s}^{0}\ell^{\prime+}{\nu}_{\ell^\prime}, \eta_{c}\ell^{\prime+}{\nu}_{\ell^\prime}$ have the largest branching ratios, which can reach up to the $10^{-7}$ order. Therefore, they are likely to be observed in future LHCb experiments or can serve as golden channels to search for the $B^*_c$ meson in experiments. 
\item 
The branching fraction ratio is an interesting quantity in experiments. We present some branching fraction ratios for the vector meson decays
\begin{tiny}
	\begin{equation}
		\begin{aligned}
			\mathcal{R}_{D_{0}}&=\frac{\mathcal{B}r(B^{*+} \to \bar{D}^{0}\tau^{+}\bar{\nu}_{\tau})}{\mathcal{B}r(B^{*+} \to \bar{D}^{0}e^{+}\bar{\nu}_{e})}=0.250\pm0.093,\;\;\;
			\mathcal{R}_{D_{s}}=\frac{\mathcal{B}r(B_{s}^{*0} \to D_{s}^{-}\tau^{+}\bar{\nu}_{\tau})}{\mathcal{B}r(B_{s}^{*0} \to D_{s}^{-}e^{+}\bar{\nu}_{e})}=0.249\pm0.059 ,\\
			\mathcal{R}_{K}&=\frac{\mathcal{B}r(B_{s}^{*0} \to K^{-}\tau^{+}\bar{\nu}_{\tau})}{\mathcal{B}r(B_{s}^{*0} \to K^{-}e^{+}\bar{\nu}_{e})}=0.425\pm0.143,\;\;\;
			\mathcal{R}_{\bar{D}_{0}}=\frac{\mathcal{B}r(B_{c}^{*+} \to D^{0} \tau^{+}\bar{\nu}_{\tau})}{\mathcal{B}r(B_{c}^{*+} \to D^{0} e^{+}\bar{\nu}_{e})}=0.478\pm0.175 ,\\
			\mathcal{R}_{\eta_{c}}&=\frac{\mathcal{B}r(B_{c}^{*+} \to \eta_{c} \tau^{+}\bar{\nu}_{\tau})}{\mathcal{B}r(B_{c}^{*+} \to \eta_{c} e^{-}\bar{\nu}_{e})}=0.229\pm0.059,\;\;\;
			\mathcal{R}_{\eta_{c}(2S)}=\frac{\mathcal{B}r(B_{c}^{*+} \to \eta_{c}(2S) \tau^{+}\bar{\nu}_{\tau})}{\mathcal{B}r(B_{c}^{*+} \to \eta_{c}(2S) e^{+}\bar{\nu}_{e})}=0.087\pm0.022,
		\end{aligned}
	\end{equation}
\end{tiny}
which are consistent with other theoretical predictions.
\item
In the nonleptonic decays, the Cabibbo-favored channels $B_{c}^{*+}\to B^0_{s} \pi^{+}, B^0_{s} \rho^{+}$ have the largest branching ratios, which can reach up 
to the $10^{-7}$ order. Evidently, these two decays are also golden channels to search for the $B_c^*$ meson in experiments. 
\item
Using the obtained branching ratios, we can consider the SU(3) symmetry breaking parameters $R^{V}_M$ and $R^{K/\pi}_M$, with $M$ being $B_{(s)}$ and $\eta_c(1S,2S)$,
\begin{tiny}
	\begin{equation}
		\begin{aligned}
			R^{K^*/\rho}_{B_{s}}&\equiv\frac{\mathcal{B}r(B_{c}^{*+} \to B_{s} K^{*+})}{\mathcal{B}r(B_{c}^{*+} \to B_{s} \rho^{+})}=0.042\pm 0.015,\;\;\;\;\;\;R^{K/\pi}_{B_{s}}\equiv\frac{\mathcal{B}r(B_{c}^{*+} \to B_{s} K^{+})}{\mathcal{B}r(B_{c}^{*+} \to B_{s} \pi^{+})}=0.052\pm0.019,\\
			R^{K^*/\rho}_{B}&\equiv \frac{\mathcal{B}r(B_{c}^{*+} \to B K^{*+})}{\mathcal{B}r(B_{c}^{*+} \to B \rho^{+})}=0.052\pm0.018,\;\;\;\;\;\;R^{K/\pi}_{B}\equiv\frac{\mathcal{B}r(B_{c}^{*+} \to B K^{+})}{\mathcal{B}r(B_{c}^{*+} \to B \pi^{+})}=0.056\pm0.020,\\
			R^{K^*/\rho}_{\eta_{c}}&\equiv\frac{\mathcal{B}r(B_{c}^{*+} \to \eta_{c} K^{*+})}{\mathcal{B}r(B_{c}^{*+} \to \eta_{c}  \rho^{+})}=0.060\pm0.015 ,\;\;\;\;\;\;R^{K/\pi}_{\eta_{c}}\equiv\frac{\mathcal{B}r(B_{c}^{*+} \to \eta_{c} K^{+})}{\mathcal{B}r(B_{c}^{*+} \to \eta_{c}  \pi^{+})}=0.076\pm0.027,\\
			R^{K^*/\rho}_{\eta_{c}(2S)}&\equiv\frac{\mathcal{B}r(B_{c}^{*+} \to \eta_{c}(2S)  K^{*+})}{\mathcal{B}r(B_{c}^{*+} \to \eta_{c}(2S)  \rho^{+})}=0.056\pm0.020 ,\;\;\;\;\;\;R^{K/\pi}_{\eta_{c}(2S)}\equiv\frac{\mathcal{B}r(B_{c}^{*+} \to \eta_{c}(2S) K^{+})}{\mathcal{B}r(B_{c}^{*+} \to \eta_{c}(2S)  \pi^{+})}=0.075\pm0.027,
		\end{aligned}
	\end{equation}
\end{tiny}
which are consistent with the estimations $R^{K^*/\rho}_M=\left|\dfrac{V_{us}f_{K^{*}}}{V_{ud}f_{\rho}}\right|^{2}\approx0.057$ and $R^{K/\pi}_M=\left|\dfrac{V_{us}f_{K}}{V_{ud}f_{\pi}}\right|^{2}\approx0.078$.

In the future, the self-consistency of the CLFQM needs to be improved systematically. A known example is the vector meson decay constant, that is, the $f_V$ inconsistency puzzle \cite{Choi,Chang}. It is resolved by replacing the physical mass $M$ with the invariant mass $M_0$. In the past, authors usually investigated the distribution amplitudes (DAs) up to the leading-twist, which describes the longitudinal momentum distribution of valence quarks inside the hadrons. Self-consistency is needed to further check carefully when the higher-twist DAs are involved \cite{Arifi}. Recognizing the higher-twist effects to the hadron structures is important. Furthermore, it is worth examining such effects in detail when radially and orbitally excited states are involved.
\end{enumerate}
\section*{Acknowledgment}
This work is partly supported by the National Natural Science
Foundation of China under Grant No. 11347030, by the Program of
Science and Technology Innovation Talents in Universities of Henan
Province 14HASTIT037, and the Natural Science Foundation of Henan
Province under grant no. 232300420116.
\appendix
\section{Some specific rules under the $p^-$ integration}
When preforming the integration, we need to include the zero-mode contribution. This involves properly performing the integration in the CLFQM. Specifically, we
use the following rules given in Refs. \cite{Y. Cheng,41,44,60}:
\begin{footnotesize}
	\begin{eqnarray}
\hat{p}_{1 \mu}^{\prime} &\doteq &   P_{\mu}
A_{1}^{(1)}+q_{\mu} A_{2}^{(1)},\\
\hat{p}_{1 \mu}^{\prime}
\hat{p}_{1 \nu}^{\prime}  &\doteq & g_{\mu \nu} A_{1}^{(2)} +P_{\mu}
P_{\nu} A_{2}^{(2)}+\left(P_{\mu} q_{\nu}+q_{\mu} P_{\nu}\right)
A_{3}^{(2)}+q_{\mu} q_{\nu} A_{4}^{(2)},\\
Z_{2}&=&\hat{N}_{1}^{\prime}+m_{1}^{\prime 2}-m_{2}^{2}+\left(1-2
x_{1}\right) M^{\prime 2} +\left(q^{2}+q \cdot P\right)
\frac{p_{\perp}^{\prime} \cdot q_{\perp}}{q^{2}},\\
\hat{p}_{1 \mu}^{\prime} \hat{N}_{2} & \rightarrow & q_{\mu}\left[A_{2}^{(1)} Z_{2}+\frac{q \cdot P}{q^{2}} A_{1}^{(2)}\right],
\en
\be
\hat{p}_{1 \mu}^{\prime} \hat{p}_{1 \nu}^{\prime} \hat{N}_{2} & \rightarrow &g_{\mu \nu} A_{1}^{(2)} Z_{2}+q_{\mu} q_{\nu}\left[A_{4}^{(2)} Z_{2}+2 \frac{q \cdot P}{q^{2}} A_{2}^{(1)} A_{1}^{(2)}\right],\\
A_{1}^{(1)}&=&\frac{x_{1}}{2}, \quad A_{2}^{(1)}=
A_{1}^{(1)}-\frac{p_{\perp}^{\prime} \cdot q_{\perp}}{q^{2}},\quad A_{3}^{(2)}=A_{1}^{(1)} A_{2}^{(1)},\\
A_{4}^{(2)}&=&\left(A_{2}^{(1)}\right)^{2}-\frac{1}{q^{2}}A_{1}^{(2)},\quad A_{1}^{(2)}=-p_{\perp}^{\prime 2}-\frac{\left(p_{\perp}^{\prime}
\cdot q_{\perp}\right)^{2}}{q^{2}}, \quad A_{2}^{(2)}=\left(A_{1}^{(1)}\right)^{2}.  
\end{eqnarray}
\end{footnotesize}
\section{EXPRESSIONS OF $B^* \to P $ FORM FACTORS}
The following are the analytical expressions of the form factors of the transitions $B^* \to P $ in the CLFQM:
\begin{footnotesize}
\begin{eqnarray}
V^{B^* P}(q^{2})&=&\frac{N_{c}(M^{'}+M^{''})}{16 \pi^{3}} \int d x_{2} d^{2} p_{\perp}^{\prime} \frac{2 h_{B^*}^{\prime}
 h_{P}^{\prime \prime}}{x_{2} \hat{N}_{1}^{\prime} \hat{N}_{1}^{\prime \prime}}\left\{x_{2} m_{1}^{\prime}
 +x_{1} m_{2}+\left(m_{1}^{\prime}-m_{1}^{\prime \prime}\right) \frac{p_{\perp}^{\prime} \cdot q_{\perp}}{q^{2}}\right.\non &&\left.
 +\frac{2}{w_{V}^{\prime \prime}}\left[p_{\perp}^{\prime 2}+\frac{\left(p_{\perp}^{\prime} \cdot q_{\perp}\right)^{2}}{q^{2}}\right]\right\},\\
 A_0^{B^* P}(q^{2})&=& \frac{M^{'}+M^{''}}{2M^{''}}A_1^{B^* P}(q^{2})-\frac{M^{'}-M^{''}}{2M^{''}}A_2^{B^* P}(q^{2})-\frac{q^2}{2M^{''}}\frac{N_{c}}{16 \pi^{3}} \int d x_{2} d^{2} p_{\perp}^{\prime} \frac{h_{B^*}^{\prime} h_{P}^{\prime \prime}}{x_{2} \hat{N}_{1}^{\prime}
\hat{N}_{1}^{\prime \prime}}\left\{2\left(2 x_{1}-3\right)\right.\non &&\left.\left(x_{2} m_{1}^{\prime}+x_{1} m_{2}\right)-8\left(m_{1}^{\prime}-m_{2}\right)
\times\left[\frac{p_{\perp}^{\prime 2}}{q^{2}}
+2 \frac{\left(p_{\perp}^{\prime} \cdot q_{\perp}\right)^{2}}{q^{4}}\right]-\left[\left(14-12 x_{1}\right) m_{1}^{\prime}\right.\right. \non &&\left.\left.-2 m_{1}^{\prime \prime}-\left(8-12 x_{1}\right) m_{2}\right] \frac{p_{\perp}^{\prime} \cdot q_{\perp}}{q^{2}}
+\frac{4}{w_{V}^{\prime \prime}}\left(\left[M^{\prime 2}+M^{\prime \prime 2}-q^{2}+2\left(m_{1}^{\prime}-m_{2}\right)\left(m_{1}^{\prime \prime}
+m_{2}\right)\right]\right.\right.\non &&\left.\left.\times\left(A_{3}^{(2)}+A_{4}^{(2)}-A_{2}^{(1)}\right)
+Z_{2}\left(3 A_{2}^{(1)}-2 A_{4}^{(2)}-1\right)+\frac{1}{2}\left[x_{1}\left(q^{2}+q \cdot P\right)
-2 M^{\prime 2}-2 p_{\perp}^{\prime} \cdot q_{\perp}\right.\right.\right.\non &&\left.\left.\left.-2 m_{1}^{\prime}\left(m_{1}^{\prime \prime}+m_{2}\right)
-2 m_{2}\left(m_{1}^{\prime}-m_{2}\right)\right]\left(A_{1}^{(1)}+A_{2}^{(1)}-1\right) q \cdot p\left[\frac{p_{\perp}^{\prime 2}}{q^{2}}
+\frac{\left(p_{\perp}^{\prime} \cdot q_{\perp}\right)^{2}}{q^{4}}\right]\right.\right.\non &&\left.\left.\times\left(4 A_{2}^{(1)}-3\right)\right)\right\},\;\;\;\\
A_1^{B^* P}(q^{2})&=& -\frac{1}{M^{'}+M^{''}}\frac{N_{c}}{16 \pi^{3}} \int d x_{2} d^{2} p_{\perp}^{\prime} \frac{h_{B^*}^{\prime} h_{P}^{\prime \prime}}{x_{2}
\hat{N}_{1}^{\prime}
\hat{N}_{1}^{\prime \prime}}\left\{2 x_{1}\left(m_{2}-m_{1}^{\prime}\right)\left(M_{0}^{\prime 2}+M_{0}^{\prime \prime 2}\right)
-4 x_{1} m_{1}^{\prime \prime} M_{0}^{\prime 2}\right.\non
&&\left.+2 x_{2} m_{1}^{\prime} q \cdot P+2 m_{2} q^{2}-2 x_{1} m_{2}\left(M^{\prime 2}+M^{\prime \prime 2}\right)+2\left(m_{1}^{\prime}-m_{2}\right)\left(m_{1}^{\prime}
+m_{1}^{\prime \prime}\right)^{2}+8\left(m_{1}^{\prime}-m_{2}\right) \right.\non &&
\left. \times\left[p_{\perp}^{\prime 2}+\frac{\left(p_{\perp}^{\prime}
\cdot q_{\perp}\right)^{2}}{q^{2}}\right]+2\left(m_{1}^{\prime}+m_{1}^{\prime \prime}\right)\left(q^{2}+q \cdot p\right) \frac{p_{\perp}^{\prime} \cdot q_{\perp}}{q^{2}}
-4 \frac{q^{2} p_{\perp}^{\prime 2}+\left(p_{\perp}^{\prime} \cdot q_{\perp}\right)^{2}}{q^{2} w_{V}^{\prime \prime}}
\right.\non && \left.\times\left[2 x_{1}\left(M^{\prime 2}+M_{0}^{\prime 2}\right)-q^{2}-q \cdot p-2\left(q^{2}+q \cdot p\right) \frac{p_{\perp}^{\prime} \cdot q_{\perp}}{q^{2}}-2\left(m_{1}^{\prime}-m_{1}^{\prime \prime}\right)\left(m_{1}^{\prime}-m_{2}\right)\right]\right\},\;\;\;\;\;
\en
\be
A_2^{B^* P}(q^{2})&=& \frac{N_{c}(M^{'}+M^{''})}{16 \pi^{3}} \int d x_{2} d^{2} p_{\perp}^{\prime} \frac{2 h_{B^*}^{\prime} h_{P}^{\prime \prime}}{x_{2} \hat{N}_{1}^{\prime}
\hat{N}_{1}^{\prime \prime}}\left\{\left(x_{1}-x_{2}\right)\left(x_{2} m_{1}^{\prime}+x_{1} m_{2}\right)-\frac{p_{\perp}^{\prime} \cdot q_{\perp}}{q^{2}}\left[2 x_{1} m_{2}
+m_{1}^{\prime \prime} \right.\right.\non &&
\left.\left.+\left(x_{2}-x_{1}\right) m_{1}^{\prime}\right]-2 \frac{x_{2} q^{2}+p_{\perp}^{\prime} \cdot q_{\perp}}{x_{2} q^{2} w_{V}^{\prime \prime}}\left[p_{\perp}^{\prime} \cdot p_{\perp}^{\prime \prime}
+\left(x_{1} m_{2}+x_{2} m_{1}^{\prime}\right)\left(x_{1} m_{2}-x_{2} m_{1}^{\prime \prime}\right)\right]\right\}.
\end{eqnarray}
\end{footnotesize}

\end{document}